\documentclass[twocolumn,iop]{emulateapj}

\usepackage{apjfonts}
\usepackage{amsmath,amssymb}
\usepackage{mathrsfs}
\usepackage{bm}

\newcommand{\beq}{\begin{equation}}
\newcommand{\eeq}{\end{equation}}
\newcommand{\intd}{{\rm d}}
\newcommand{\nf}{\ensuremath{N_{\rm F}}}
\newcommand{\prior}{\ensuremath{^{\rm prior}}}
\newcommand{\ri}{\ensuremath{\mathscr{R}_i}}
\newcommand{\rv}{\ensuremath{\mathscr{R}_V}}
\newcommand{\mh}{\ensuremath{\rm{[M/H]}}}
\newcommand{\feh}{\ensuremath{\rm{[Fe/H]}}}
\newcommand{\actaa}{Acta Astronomica}

\shorttitle{Global model for Cepheids}
\shortauthors{Pejcha \& Kochanek}

\begin{document}

\title{A Global Physical Model for Cepheids}

\author{ Ond\v{r}ej Pejcha\altaffilmark{1} and Christopher S. Kochanek\altaffilmark{1,2} }
\altaffiltext{1}{Department of Astronomy, The Ohio State University, 140 West 18th Avenue, Columbus, OH 43210, USA}
\altaffiltext{2}{Center for Cosmology and Astroparticle Physics, The Ohio State University, 191 West Woodruff Avenue, Columbus, OH 43210, USA}
\email{pejcha,ckochanek@astronomy.ohio-state.edu}

\begin{abstract}
We perform a global fit to $\sim\! 5,000$ radial velocity and $\sim\!177,000$ magnitude measurements in $29$ photometric bands covering $0.3\,\mu$m to $8.0\,\mu$m distributed among $287$ Galactic, LMC, and SMC Cepheids with $P>10$\,days. We assume that the Cepheid light curves and radial velocities are fully characterized by distance, reddening, and time-dependent radius and temperature variations. We construct phase curves of radius and temperature for periods between $10$ and $100$\,days, which yield light curve templates for all our photometric bands and can be easily generalized to any additional band. With only $4$ to $6$ parameters per Cepheid, depending on the existence of velocity data and the amount of freedom in the distance, the models have typical rms light and velocity curve residuals of $0.05$\,mag and $3.5$\,km\,s$^{-1}$. The model derives the mean Cepheid spectral energy distribution and its derivative with respect to temperature, which deviate from a black body in agreement with metal-line and molecular opacity effects. We determine a mean reddening law towards the Cepheids in our sample, which is not consistent with standard assumptions in either the optical or near-IR. Based on stellar atmosphere models we predict the biases in distance, reddening, and temperature determinations due to the metallicity and we quantify the metallicity signature expected for our fit residuals. The observed residuals as a function of wavelength show clear differences between the individual galaxies, which are compatible with these predictions. In particular, we find that metal-poor Cepheids appear hotter. Finally, we provide a framework for optimally selecting filters that yield the smallest overall errors in Cepheid parameter determination, or filter combinations for suppressing or enhancing the metallicity effects on distance determinations. We make our templates publicly available.
\end{abstract}

\section{Introduction}

The classical Cepheids play a key role in the extragalactic distance determination, both as a direct way of obtaining distances of nearby galaxies and as a calibrator for other methods (e.g.\ the review by \citealp{freedman10a}, and \citealp{madore91,feast97,freedman01,benedict07,riess09a,riess09b,riess11a,riess11b}). This is enabled by the relatively tight relation between the pulsational period of a Cepheid and its luminosity  (the $PL$ relation). Accurate distance estimates using Cepheids require increasingly better understandings of systematic effects such as extinction, composition (metallicity) and blending.

For example, the light from Cepheids is typically extinguished by dust, which can have different properties in different galaxies \citep[e.g.][]{cardelli89,laney93,falco99,misselt99,motta02,gordon03,indebetouw05,marshall06,laney07}. Cepheid observations must be corrected for extinction before any subsequent analysis. This usually makes use of an assumption that Cepheids with a given period have a specific color and that there is a known, universal extinction law. This holds for the Baade-Wesselink and related methods as well as purely photometric analyses. To alleviate these shortcomings, recent efforts have shifted from the optical (Hubble Key Project) to the infrared wavelengths ($H$, $I$ and Spitzer IRAC), where the effects of reddening are reduced \citep[e.g.][]{mcgonegal82,gieren05,ref167,freedman08,madore09,ngeow10,storm11a,storm11b}. There are, however, differences in the near/mid-infrared extinction law for different sightlines in the Galaxy \citep[e.g.][]{roman07,nishiyama09}.

Cepheid properties must also change as a function of chemical composition, but the associated observational signatures have been a matter of considerable debate. While the general conclusion of observational studies is that metal-rich Cepheids are brighter than their metal-poor counterparts \citep[e.g.][]{stothers88,kochanek97,sakai04,macri06,valle09,shappee11}, results from stellar pulsation models and some observations yield the opposite dependence \citep[e.g.][]{fiorentino02,fiorentino07,marconi05,bono08,romaniello08,freedman11}. Estimates of the metallicity correction vary by almost an order of magnitude, as can be seen in Figure~14 of \citet{gerke11}. It is also now clear, at least in the extragalactic context, that uncertainties in the appropriate metallicities and metallicity gradients are nearly as important as the actual metallicity dependence of Cepheid parameters \citep[see][and references therein]{bresolin11,gerke11,shappee11}. Furthermore, studies find different trends in different filters \citep{bono08,bono10,ngeow11,storm11b}. Many of these problems arise because the effects of metallicity have filter-dependent degeneracies with reddening and distance, making it crucial to fully recover all covariances in the final results, as emphasized by \citet{gould94}, \citet{kochanek97}, \citet{sasselov97}, and \citet{riess11a}. As a result, the question of metallicity effects on the Cepheid distance scale remains a controversial issue. The problem of blending then adds further complications for more distant galaxies \citep{mochejska00}.

In many circumstances, light curve templates for the individual observational bands are needed to accurately determine the mean magnitudes of extragalactic Cepheids because of their sparse light curves. \citet{stetson96} constructed $V$- and $I_{\rm C}$-band templates for the LMC, SMC, and Galactic Cepheids with periods between $7$ and $100$\,days by fitting Fourier series to the light curves. He assumed that the amplitudes and phases of individual modes vary continuously with period, thus forming the well-known Hertzsprung sequence \citep{hertzsprung26}. \citet{hendry99}, \citet{ngeow03}, \citet{tanvir05} and \citet{yoachim09} expanded on these models using Principal Component Analysis (PCA) techniques to build new templates for analysis of HST Cepheid data. All of these templates are limited in the sense that they are defined only for a small set of filters (typically $V$ and $I_{\rm C}$), and there is no well-defined means of shifting them to other, similar filters, let alone to very different wavelengths. \citet{freedman10b} show that all light curves are related by a linear transformation which provides a path towards addressing this problem.

In this work, we address these issues by constructing a global model of Cepheid light curves and radial velocity curves in the Galaxy, LMC, and SMC. We self-consistently determine distances, reddenings, radii, and temperatures of individual Cepheids along with a global reddening law, the mean SED of a Cepheid and its dependence on temperature, and the phase variations of the radii and temperatures as a function of period. We self-consistently determine the uncertainties in {\it all} parameters including {\it all} their covariances. Our model enables us to construct a light curve template for an arbitrary filter given just a single calculable parameter for the relative contribution of radius and temperature variations to that filter. We evaluate metallicity effects on Cepheid observations first from a theoretical point of view, and then we search for metallicity effects in the estimated parameters and residuals of the fit. 

The approach to the problem is closely related to the Baade-Wesselink method \citep{baade26,wesselink46} and assumes that the magnitude of a Cepheid at a given time depends only its distance, reddening, and instantaneous radius and temperature. Rather than carrying out a study of individual Cepheids, we simply fit a global model to all the available photometric and velocity data. A simpler variation on this idea was recently proposed by \citet{freedman10b}. This paper is organized as follows. In Section~\ref{sec:model_data}, we outline our physical model, priors, dataset and the method of fitting. In Section~\ref{sec:results}, we describe results of the fit, discussing in turn the fit residuals, global quantities and individual properties of the Cepheids. In Section~\ref{sec:discussion}, we discuss our results and search for signs of additional physics, principally metallicity effects. We review our findings and outline future directions in Section~\ref{sec:conclusions}. An Appendix outlines a general quantitative approach to selecting filters for Cepheid studies.

\section{Model and Data}
\label{sec:model_data}

In this Section we describe our physical model of Cepheid light curves and radial velocity curves in detail (Section~\ref{sec:phys_model}) along with its connection to traditional methods (Section~\ref{sec:connect}). The priors needed to address degeneracies in our master equation, such as the zero point of extinction, are described in \S\ref{sec:priors}. The data that are used for the fit are described in \S\ref{sec:dataset} and the fitting method is outlined in Section~\ref{sec:fit_method}.

\subsection{Physical Model}
\label{sec:phys_model}

For each Cepheid, the radial velocity $v(t)$ and magnitude $m_i(t)$ in filter $i$ at a time $t$ are given as
\begin{eqnarray}
m_i(t) &=& \overline{M}_i + \mu + \ri E(B-V) - 5 \log\!\! \left(\!\!\frac{R(t)}{R_0}\!\!\right)\! - 2.5\beta_i\! \log\!\! \left(\!\!\frac{T(t)}{T_0}\!\!\right)\!,\label{eq:phys_model_m}\\
v(t) &=& \bar{v}  -\frac{1}{p}\frac{\intd R(t)}{\intd t},
\label{eq:phys_model_v}
\end{eqnarray}
where $\overline{M}_i$ is the absolute magnitude of a ``mean'' Cepheid with radius $R_0 \equiv 10$\,R$_\odot$ and temperature $T_0 \equiv 5400$\,K, $\mu$ is the distance modulus, $\mathscr{R}_i$ is the ratio of total to selective extinction in filter $i$, $E(B-V)$ is the reddening, $R(t)$ and $T(t)$ are the radius and temperature of the Cepheid, $\bar{v}$ is the mean radial velocity, and $p\equiv 1.36$ is a projection factor for converting observed radial velocities to photospheric ones \citep[e.g.][]{burki82,nardetto04}. Although the exact value of $p$ and its dependence on the pulsational period is a matter of debate, these uncertainties largely cancel because we homogeneously analyze the three galaxies in our sample with the same value of $p$ \citep{storm11a}. The coefficients $\beta_i$ are the logarithmic derivatives of the spectral energy distribution (SED) with respect to temperature at the reference temperature $T_0$,
\beq
\beta_i = \left.\frac{\partial \log F_i}{\partial \log T}\right|_{T_0},
\label{eq:beta_general}
\eeq
where $F_i$ is the flux in filter $i$. All logarithms in this paper are base $10$.

The radii $R$ and temperatures $T$ depend on the time $t$ through the pulsational phase $\phi = (t-t_0)/P$, where $t_0$ is the reference time and $P$ is the pulsational period. We express the time-dependent parts of Equations~(\ref{eq:phys_model_m})--(\ref{eq:phys_model_v}) as
\begin{eqnarray}
\log\left(\frac{R(t)}{R_0}\right) &=& \bar{\rho} + A^2 \delta\rho(\phi),\label{eq:rho}\\
\log\left(\frac{T(t)}{T_0}\right) &=& \bar{\tau} + A^2 \delta \tau(\phi)\label{eq:tau},
\end{eqnarray}
where $\bar{\rho}$ and $\bar{\tau}$ are the mean logarithmic radii and temperatures of a Cepheid with respect to the ``mean'' Cepheid with radius $R_0$ and temperature $T_0$, $A^2$ is the dimensionless amplitude, and $\delta\rho(\phi)$ and $\delta\tau(\phi)$ are the period and phase dependent changes in the radius and temperature. In the above definitions, quantities with subscript $i$ are different for each filter, while $\mu$, $E(B-V)$,  $\bar{v}$, $\bar{\rho}$, $\bar{\tau}$, $A^2$, $P$, and $t_0$ are different for each Cepheid. Only the radius and temperature changes $\delta\rho$ and $\delta\tau$ are functions of time $t$. 

Not all Cepheids in our sample have enough data to allow for independent estimates of $\bar{\rho}$ and $\bar{\tau}$, but they can still contribute to the determination of global variables. In analogy to the normal period--luminosity ($PL$) relations, we assume that the mean radius and temperature of a Cepheid are functions of the pulsational period,
\begin{subequations}
\label{eqs:rhotau_per}
\begin{eqnarray}
\langle\bar{\rho}(P)\rangle &=& a_{\bar{\rho}} +b_{\bar{\rho}} \log\!\left(\! \frac{P}{10\,{\rm d}}\!\right), \label{eq:rho_per}\\
\langle\bar{\tau}(P)\rangle &=& a_{\bar{\tau}} +b_{\bar{\tau}} \log\!\left(\! \frac{P}{10\,{\rm d}}\!\right), \label{eq:tau_per}
\end{eqnarray}
\end{subequations}
where $\langle\bar{\rho}\rangle$ and $\langle\bar{\tau}\rangle$ are the average radii and temperatures at period $P$, and the coefficients $a_{\bar{\rho}}$, $b_{\bar{\rho}}$, $a_{\bar{\tau}}$, and $b_{\bar{\tau}}$ are estimated during the fit. These relations have corresponding widths, $\sigma_{\bar{\rho}}$ and $\sigma_{\bar{\tau}}$, which in principle could be estimated during the fit as well, but for simplicity we leave them fixed at $\sigma_{\bar{\rho}} = \sigma_{\bar{\tau}} = 0.02$ based on the differences between canonical and non-canonical theoretical pulsational models of \citet{bono98,bono05}. Equations~(\ref{eq:rho_per})--(\ref{eq:tau_per}) also allow us to connect our approach to the traditional methods based on $PL$ relations and to define template light curves at any period or wavelength (Section~\ref{sec:connect}).

We model the time-dependent components as a truncated Fourier series of order $N_F$,
\begin{eqnarray}
\delta \rho &=& \sum\limits_{j=1}^{\nf}  \left[c_{\rho,j}\cos(2\pi j\phi) + s_{\rho,j}\sin(2\pi j\phi)\right],\label{eq:fou_rho}  \\ 
\delta \tau &=& \sum\limits_{j=1}^{\nf}  \left[c_{\tau,j}\cos(2\pi j\phi) + s_{\tau,j}\sin(2\pi j\phi)\right],\label{eq:fou_tau}
\end{eqnarray}
where the coefficients are normalized in phase and amplitude such that $c_{\rho,1}\equiv 1$ and $s_{\rho,1}\equiv 0$. The normalization of the temperature template with respect to the radial template is a part of the solution since the relative amplitudes of the radius and temperature changes are captured in the magnitude of $c_{\tau}$ and $s_{\tau}$. We use a Cartesian representation of the Fourier coefficients $(c_{\rho,j}, s_{\rho,j})$ and $(c_{\tau,j}, s_{\tau,j})$ in order to avoid the coordinate singularities which occur in a ``polar'' representation (i.e.\ $\delta\rho(\phi) \propto a_j \cos[2\pi j\phi+b_j]$) when the amplitude of a Fourier mode vanishes ($a_j \rightarrow 0$) and the phase is degenerate, as is seen in the \citet{stetson96} Cepheid templates. In order to model the period dependence of the radius and temperature variations, we specify $c_{\rho,j}$, $c_{\tau,j}$, $s_{\rho,j}$ and $s_{\tau,j}$ at $N_P$ periods chosen in such a way that the number of Cepheids in a given period bin is approximately constant (except for the longest periods). Values of the coefficients for a particular Cepheid's period are obtained by linear interpolation on the grid of templates in $\log\,P$. We choose $\nf=20$ and $N_P=19$.

\subsection{Connection to Traditional Methods}
\label{sec:connect}

In this Section we relate our approach to traditional $PL$ studies. Within our approach, the mean absolute magnitude $M_i$ in filter $i$ of a Cepheid can be constructed from Equation~(\ref{eq:phys_model_m}) as
\beq
M_i = \overline{M}_i -5\bar{\rho} - 2.5\beta_i\bar{\tau},
\label{eq:abs_m}
\eeq
where $\bar{\rho}$ and $\bar{\tau}$ are different for each Cepheid. This means that the average $PL$ relation is defined by substituting the mean radii and temperatures from Equations~(\ref{eq:rho_per})--(\ref{eq:tau_per}) into Equation~(\ref{eq:abs_m}) to obtain
\begin{eqnarray}
M_i &=& \overline{M}_i - 5\langle\bar{\rho}(P)\rangle - 2.5\beta_i \langle\bar{\tau}(P)\rangle = \nonumber\\
 &=& \overline{M}_i - (5 a_{\bar{\rho}}+2.5\beta_i a_{\bar{\tau}}) - (5b_\rho+2.5\beta_ib_{\bar{\tau}})\, \log\! \left(\!\! \frac{P}{10\,{\rm d}}\!\!\right). 
\label{eq:pl_theo}
\end{eqnarray}
Thus our model leads to a $PL$ relation with a zero point of $\overline{M}_i - 5 a_{\bar{\rho}} -2.5\beta_i a_{\bar{\tau}}$ and a slope of $-(5b_{\bar{\rho}}+2.5\beta_ib_{\bar{\tau}})$. For uncorrelated radius and temperature deviations ($\sigma_{\bar{\rho}}$ and $\sigma_{\bar{\tau}}$) from the mean trends (Eq.~[\ref{eqs:rhotau_per}]), the scatter in the $PL$ is $(5^2\sigma_{\bar{\rho}}^2+2.5^2\beta_i^2\sigma_{\bar{\tau}}^2)^{1/2}$, so there is a strong correlation of wavelength ($\beta_i$), slope and scatter about the $PL$ \citep[see][]{madore11}. Similarly, the mean template light curve for a filter $i$ is
\beq
m_i(\phi) = -5\delta\rho(\phi) - 2.5\beta_i\delta\tau(\phi),
\eeq
which has a mean of zero ($\langle m_i \rangle =0$) and is scaled to an amplitude of $A^2 =1$. \citet{freedman10b} argue that light curves in one band can always be constructed as weighted sums of those in two other bands. This is true by construction for our models, where to produce band $3$ from bands $1$ and $2$ one chooses a scale factor $x$ such that $\beta_3 = x\beta_1 + (1-x)\beta_2$ so that $m_3(\phi) = xm_1(\phi)+(1-x)m_2(\phi)$. In general, this can be an extrapolation since one can choose bands such that $x>1$ or $x<0$.

\subsection{Priors}
\label{sec:priors}

The parameters of the model are obtained by minimizing the master constraint $\mathcal{H}$ defined as
\beq
\mathcal{H} = \chi^2 + S,
\label{eq:master_constraint}
\eeq
where $\chi^2$ is the sum of the squares of differences between the model and the observed magnitudes $m^{\rm obs}$ and radial velocities $v^{\rm obs}$
\beq
\chi^2 = \sum\limits_{\rm all\ data} \left(\frac{m^{\rm obs} - m}{\sigma} \right)^2+\sum\limits_{\rm all\ data} \left(\frac{v^{\rm obs} - v}{\sigma} \right)^2,
\label{eq:chi2}
\eeq
where $\sigma$ is the measurement error, and $S$ includes contributions from all priors, and the sums are over all stars, filters and measurements.

Although we fit Equation~(\ref{eq:chi2}) to a huge dataset, the parameters in Equation~(\ref{eq:chi2}) suffer from degeneracies without the addition of priors on the wavelength dependent vectors $\overline{M}_i$, $\ri$, and $\beta_i$ and the absolute distance and extinction scale. For example, we can simultaneously shift the extinction by $E(B-V) \rightarrow E(B-V) + \Delta E(B-V)$ and the absolute magnitude vector $\overline{M}_i \rightarrow \overline{M}_i - \ri\Delta E(B-V)$ while keeping $m_i$ the same. More generally, $\overline{M}_i$  is well-determined up to adding components proportional to a constant (a change in the distance or radius scale), $\ri$ (a change in the extinction zero point), and $\beta_i$ (a change in the temperature zero point). This is true also for $\ri$. Because we assume that $\beta_i$ is the same for the mean and time-variable effects of temperature, it is not subject to the same degeneracies.

In order to address these degeneracies, we first define priors for the vectors $\overline{M}_i$, $\beta_i$, and $\ri$. We add priors on $\overline{M}_i$, $\beta_i$ and $\ri$ of the form
\beq
S_{\overline{M}_i} = \sum\limits_{\rm i} \left(\frac{\overline{M}_i - \overline{M}_i\prior}{\sigma_{\overline{M}_i}}\right)^2,
\eeq
where $\overline{M}_i\prior$ is obtained by convolving the $T_0=5400$\,K, $R_0 = 10\,R_\sun$, $\log g = 1.5$, $\mh=0.0$ ``mean'' Cepheid based on the \citet{castelli04} model atmospheres with a tophat function defined by the central wavelength and width of each filter. The sum is over all passbands. We choose $\sigma_{\overline{M}_i}=0.1$\,mag for all filters where we know the zero points. For the remaining filters\footnote{We were unable to find zero points for the CTIO $JHK$ filters and the Washington filters $CMT_1T_2$.}, we estimate the conversion factor based on our data and then set $\sigma_{\overline{M}_i}=0.2$\,mag. As we will see in Section~\ref{sec:global}, this assumption is unimportant. We constrain the vector $\beta_i$ by evaluating Equation~(\ref{eq:beta_general}) as a derivative in temperature about this reference model. The width of this prior is $\sigma_{\beta_i} = 0.1\beta_i\prior$. We fix the coefficient $\beta_V\equiv 5.14$ to its \citet{castelli04} prior value in order to prevent an overall shift of $\beta_i$ due to a weak degeneracy with $\bar{\rho}$. Similarly, we add a prior on $\ri$, where $\ri\prior$ is obtained from the \citet{cardelli89} galactic extinction law with $\mathscr{R}_V \equiv 3.3$ and $\mathscr{R}_B \equiv \mathscr{R}_V+1$ is held fixed so that $E(B-V)$ has the standard interpretation. The remaining coefficients are constrained by an ``uncertainty''  $\sigma_{\ri,i} = 0.05\ri\prior$. The value $\mathscr{R}_V = 3.3$ was chosen to agree with the Hubble Key Project \citep{madore91,freedman01}, but is actually an additional source of uncertainty that could be explored. 

The second set of priors is on the mean radii and temperatures (Eqs.~[\ref{eq:rho_per}]--[\ref{eq:tau_per}]). First, we add two priors
\begin{subequations}
\begin{eqnarray}
S_{\bar{\rho}} &=& \sum\limits_{{\rm all\ stars}\ k} \left(\frac{\bar{\rho}_k - \langle\bar{\rho}(P)\rangle}{\sigma_{\bar{\rho}}}\right)^2, \label{eq:rhobar_prior}\\
S_{\bar{\tau}} &=& \sum\limits_{{\rm all\ stars}\ k} \left(\frac{\bar{\tau}_k - \langle\bar{\tau}(P)\rangle}{\sigma_{\bar{\tau}}}\right)^2, \label{eq:taubar_prior}
\end{eqnarray}
\end{subequations}
where $\langle\bar{\rho}(P)\rangle$ and $\langle\bar{\tau}(P)\rangle$ are the mean radii and temperatures at a period $P$, which are given in Equations~(\ref{eq:rho_per})--(\ref{eq:tau_per}), and which are estimated as part of the fit. The widths of these relations are fixed at $\sigma_{\bar{\rho}} = \sigma_{\bar{\tau}} =0.02$ based on differences between the canonical and non-canonical theoretical pulsational models of \citet{bono98,bono05}. These priors drive a Cepheid onto the mean period--radius and period--temperature relations if there are insufficient data to independently constrain its radius and temperatures. The second set of priors is on the coefficients of $\langle\bar{\rho}(P)\rangle$ and $\langle\bar{\tau}(P)\rangle$ in Equations.~(\ref{eq:rho_per})--(\ref{eq:tau_per}). These have the form
\beq
S_{a_{\bar{\rho}}} = \left(\frac{a_{\bar{\rho}}-a_{\bar{\rho}}\prior}{0.02}\right)^2,\quad S_{b_{\bar{\rho}}} = \left(\frac{b_{\bar{\rho}}-b_{\bar{\rho}}\prior}{0.01}\right)^2,
\label{eq:bono_prior}
\eeq
along with a similar set of priors for $a_{\bar{\tau}}$ and $b_{\bar{\tau}}$. The prior values of 
\begin{subequations}
\label{eqs:rhotau_coefs}
\begin{eqnarray}
a_{\bar{\rho}}\prior = 0.843, \quad b_{\bar{\rho}}\prior = 0.655,\label{eq:rho_coefs}\\
a_{\bar{\tau}}\prior =  -0.019,\quad b_{\bar{\tau}}\prior =  -0.080,\label{eq:tau_coefs}
\end{eqnarray}
\end{subequations}
were set to match the ``canonical'' solar-metallicity, period--radius and effective temperature--luminosity relations of \citet{bono98} and \citet{bono05}. We derived the temperature priors assuming the standard relation of luminosity, radius and effective temperature $L=4\pi R^2\sigma T^4$. The prior widths of $0.02$ and $0.01$ were chosen to roughly correspond to the differences between the solar-metallicity ``canonical'' and ``non-canonical'' models of \citet{bono98,bono05}, as an estimate of the systematic uncertainties in the models. However, the specific choices for the prior values and their widths are of little consequence for the actual results. The zero points $a_{\bar{\rho}}$ and $a_{\bar{\tau}}$ are well-constrained by the data. The slopes $b_{\bar{\rho}}$ and $b_{\bar{\tau}}$ are less constrained because of the limited number of long-period Cepheids and our restricted period range. For these coefficients, the width of the prior is somewhat important. 

The third set of priors set the distance and extinction scales. In order to fix the distance scale, we assume that the LMC Cepheids occupy a thin disk with an inclination of $30.7^\circ$ and a position angle of $151.0^\circ$ \citep{nikolaev04} with a distance modulus at the center of $18.50$\,mag. The distances to the individual Cepheids in the SMC are free to vary, but we impose a prior that $\mu_{\rm SMC}\prior = 18.90$\,mag with a scatter of $\sigma_{\mu_{\rm SMC}} = 0.10$\,mag \citep[e.g.][]{hilditch05,keller06}. The LMC distance scale cannot be fixed using a prior that assumes a dispersion around some mean value like the one we use for the SMC Cepheids because it leads to period-dependent residuals in distance moduli and reddenings which absorb any differences between the requirements of the data and the priors on radii and temperatures. Essentially, the $PL$ relation is dictated by the radius and temperature priors rather than the data. This problem would be solved by including a sample of Cepheids that truly lie at a common distance modulus. To fix the extinction scale, we impose a prior that the LMC Cepheids have a mean extinction of $\langle E(B-V)\rangle =0.147$\,mag \citep{udalski99} with a width $\sigma_{E(B-V)} = 0.02$\,mag. 

The effects of most of these priors on the final results are weak, because most of the values are ultimately controlled by the data. The key exceptions are the mean extinction in the LMC, the distance to the center of the LMC, and setting $\mathscr{R}_V \equiv 3.3$ and $\mathscr{R}_B-\mathscr{R}_V \equiv 1$. It is possible to recognize where the data dominate a prior by the final values and uncertainties -- if the final results match the prior and the uncertainty is comparable to the prior width, then the data added no information, while if the values have shifted and the uncertainties are markedly smaller, then the data dominate.

\begin{figure}
\plotone{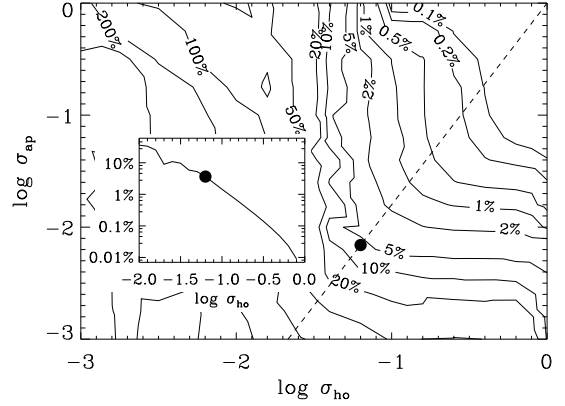}
\caption{Effect of the priors on the high-order Fourier coefficients and templates at adjacent periods on the overall goodness of fit. The contours indicate the $\chi^2$ excess (denoted with numbers at each contour) over a fit without the priors $S_{\rm ho}$ and $S_{\rm ap}$ on the smoothness of the templates. The inset plot shows $\chi^2$ excess along the line of approximately equal contributions of $S_{\rm ho}$ and $S_{\rm ap}$ to the $\chi^2$ ($\log\sigma_{\rm ap} = 1.8\log\sigma_{\rm ho}$, dashed line in the main plot). The black filled circle corresponds to the final choice of $\sigma_{\rm ho}$ and $\sigma_{\rm ho}$, which give a $\chi^2$ excess compared to no smoothing of about $5\%$.}
\label{fig:prior_coef}
\end{figure}

The last set of priors concern the templates themselves. In order to minimize unnecessary ``oscillations'' in the templates, we add a prior to keep the high-order terms of the Fourier series small
\beq
S_{\rm ho} = \frac{1}{\sigma_{\rm ho}^2}\sum\limits_{j=3}^{\nf} j^4 \left( c_{\rho,j}^2+s_{\rho,j}^2+c_{\tau,j}^2+s_{\tau,j}^2\right),
\eeq
and to minimize differences between adjacent period bins we set
\begin{eqnarray}
S_{\rm ap} &=& \frac{1}{\sigma_{\rm ap}^2}\sum\limits_{j=1}^{\nf-1} j^2  \left[ (c_{\rho,j}-c_{\rho,j+1})^2+(s_{\rho,j}-s_{\rho,j+1})^2+ \right. \nonumber\\
 &+ & \left. (c_{\tau,j}-c_{\tau,j+1})^2+(s_{\tau,j}-s_{\tau,j+1})^2 \right].
\end{eqnarray}
The relative strengths of priors $S_{\rm ho}$ and $S_{\rm ap}$ and their relative weights with respect to other components of the $\mathcal{H}$ are determined empirically following Figure~\ref{fig:prior_coef}. We vary the values of $\sigma_{\rm ho}$ and $\sigma_{\rm ap}$, and we record the change in the $\chi^2$ relative to a fit with negligible values for the weights. The final choice of $\sigma_{\rm ho}$ and $\sigma_{\rm ho}$ gives an equal weight to both priors and leads to an increase in the $\chi^2$ of the fit by about $5\%$ compared to using no smoothing. We choose these particular values of $\sigma_{\rm ho}$ and $\sigma_{\rm ap}$ to see some smoothing of the templates while only introducing a small increase in $\chi^2$ compared to having no smoothing. The exact choice is somewhat subjective, but also has no important consequences for the results.

\subsection{Data}
\label{sec:dataset}

In order to obtain well-determined templates, the properties of individual Cepheids and the global parameters of the model solution, we require a sample of Cepheids with a large quantity of photometric and radial velocity measurements. The data include the huge database of OGLE-III $V$ and $I$ measurements from \citeauthor{ref166} (\citeyear{ref166,ref175}), the major databases of \citet{ref079} and \citet{ref119a,ref119b,ref119c}, the large sample of near-IR measurements by \citet{ref167}, and the Spitzer IRAC measurements of \citet{ref176}, \citet{ref177}, and \citet{ref178}. The references for the photometric and radial velocity data are given in Tables~\ref{tab:phot} and \ref{tab:rv}, respectively.

These data include photometry in $29$ filters covering the wavelength range from $0.3\,\mu$m to $8.0\,\mu$m, as summarized in Table~\ref{tab:filters}. In the optical, we include the standard Johnson $UBV$, Cousins $(RI)_{\rm C}$ and Johnson $(RI)_{\rm J}$ bands. We also use Hipparcos and Tycho photometry (bands Hp, $B_{\rm T}$, and $V_{\rm T}$), photometry in the Washington system ($C$, $M$, $T_1$, and $T_2$), and measurements in the Walraven photometric bands ($(WULBV)_{\rm W}$). The Washington system was designed to provide metallicity and temperature estimates for G and K giants \citep{wallerstein66}. The Walraven system was designed for studies of early-type stars \citep{walraven60}. We convert the Walraven data from their default $\log_{10}$ scale to magnitudes by multiplying the data by a factor of $2.5$. In the near-IR we include the $JHK$ filters of the SAAO \citep{glass73} and CTIO \citep{elias82} systems separately. Finally, we include the Spitzer IRAC [3.6], [4.5], [5.8], and [8.0] bands.

Since we are mostly interested in extragalactic uses, we restrict our dataset to fundamental-mode Cepheids with $P\ge 10$\,d. We do not include ultra-long period Cepheids with $P>100$\,days \citep{bird09}. As we proceeded with the fit, we purged the data set of obviously wrong measurements and database errors. We also removed Cepheids with too few data, binary systems, strongly blended systems, and Cepheids with obvious period changes. In the end, we have $177,314$ photometric and $5,031$ radial velocity measurements for $287$ Cepheids in $29$ photometric bands.

\subsection{Fitting Method}
\label{sec:fit_method}

Minimizing Equation~(\ref{eq:master_constraint}) based on the physical model given by Equations~(\ref{eq:phys_model_m})--(\ref{eq:fou_tau}) is somewhat similar to principal component analysis (PCA). Specifically, we are trying to decompose the measurement $m_{ik}$ in filter $i$ for Cepheid $k$ into an unknown set of vectors  $\overline{M}_i$, $\ri$, $\beta_i$, $\delta\rho$, and $\delta\tau$ while simultaneously obtaining the coefficients of the expansion $\mu_k$, $E(B-V)_k$, $\bar{\rho}_k$, $\bar{\tau}_k$, and $A^2_k$. In a PCA, we would decompose a data point $m_i$ as a sum $m_i = \sum_j \alpha_{j} e_{ij}$ of coefficients $\alpha_{j}$ multiplied by orthonormal vectors $\mathbf{e}_j$, $\sum_i e_{ij} e_{ik} = \delta_{jk}$, simultaneously determining $\alpha_j$ and $\mathbf{e}_j$. In this paper, we fit a measurement $m_{ik}$ of a Cepheid $k$ in filter $i$ as a sum $m_{ik} = \sum_j \alpha_{kj} e_{ij}$, where from Equation~(\ref{eq:phys_model_m}) it follows that $\mathbf{e}_i=\{\overline{M}_i, 1,\ri, -5, -5\delta\rho, -2.5\beta_i, -2.5\beta_i\delta\tau\}$ and $\bm{\alpha}_k = \{1, \mu_k, E(B-V)_k, \bar{\rho}_k, A^2_k, \bar{\tau}_k, A^2_k\}$. The basis vectors $\mathbf{e}_j$ are not orthogonal and do not have unit norms in order to maintain their physical meaning. The lack of orthogonality then leads to degeneracies, which we must control by introducing the priors discussed in Section~\ref{sec:priors}. Mathematically, however, the similarity to PCA means that the model is well defined and can be solved by standard iterative or minimization methods to yield a unique solution given the data and priors.

To this end, we have developed a versatile program that allows individual variables to be switched on or off, to reinitialize the physical variables to their prior values, and to alternate fitting all variables simultaneously with fitting just the properties of individual Cepheids. We assign a weight to each measurement calculated as a maximum of the reported measurement error (if available) and a rms scatter of all measurements in the given filter from that particular data source. The fitting procedure itself can proceed in two ways. First, all variables are minimized independently in each iteration using analytic first and second derivatives through the conjugate gradient method \citep{press92}. This method is fast, but does not provide any explicit error estimates. The second option is to construct the full covariance matrix, which is then inverted using the Cholesky decomposition. This method is much slower, but provides error estimates that include {\it all} the correlations of the model as well as the measurement errors. Given any reasonable starting point, the fits are stable and well-behaved for all parameters except the phase reference time $t_0$, where the $\chi^2$ surface is more complex and a manual intervention is sometimes necessary. While we have kept periods $P$ and period derivatives $\dot{P} \equiv 0$ fixed, they can be included without difficulty, but are not as stable because the $\chi^2$ surface in $P$ and $\dot{P}$ is not smooth.

\section{Results}
\label{sec:results}

In the following Sections we present the detailed results of our model. In Section~\ref{sec:residuals}, we discuss residuals to the global fit. In Section~\ref{sec:global}, we discuss the global parameters: the zero point $\overline{M}_i$, temperature dependence $\beta_i$ and extinction vector $\ri$. Section~\ref{sec:individual} examines the individual parameters of the Cepheids.

\subsection{Residuals of the Fit}
\label{sec:residuals}

\begin{figure*}
\begin{center}
\includegraphics[width=0.47\textwidth]{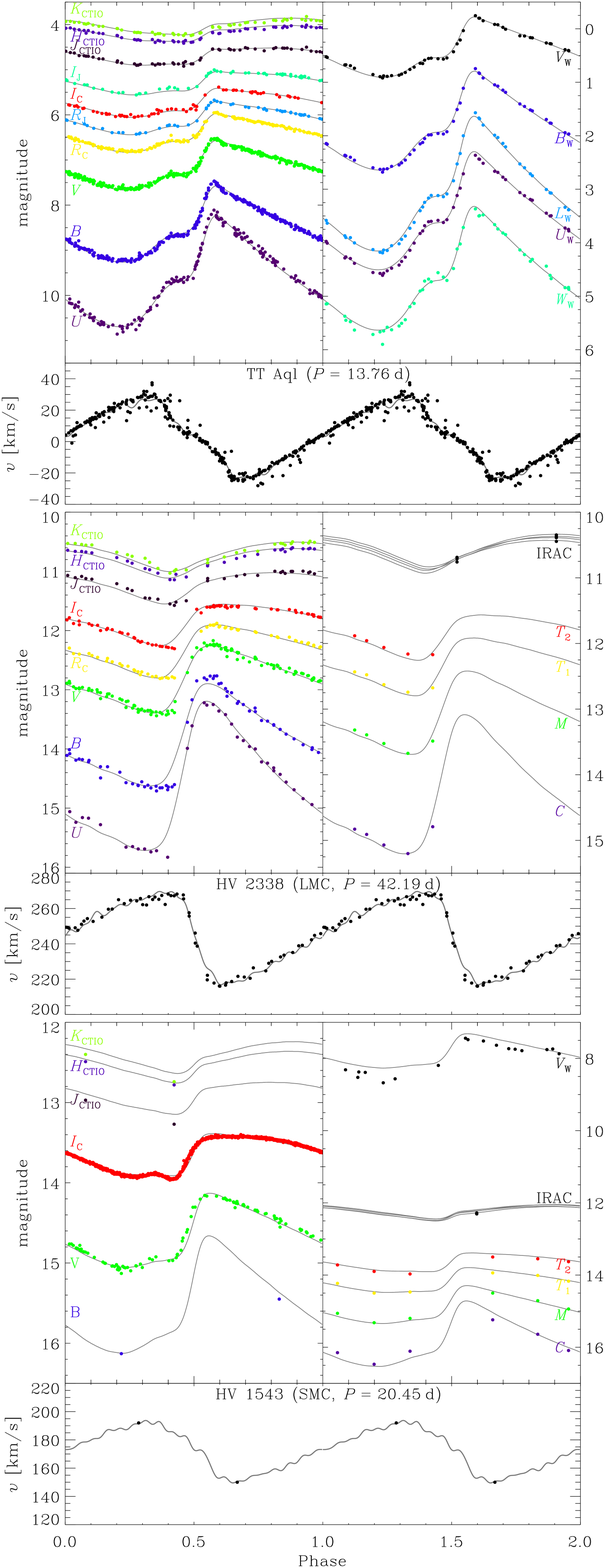}
\includegraphics[width=0.47\textwidth]{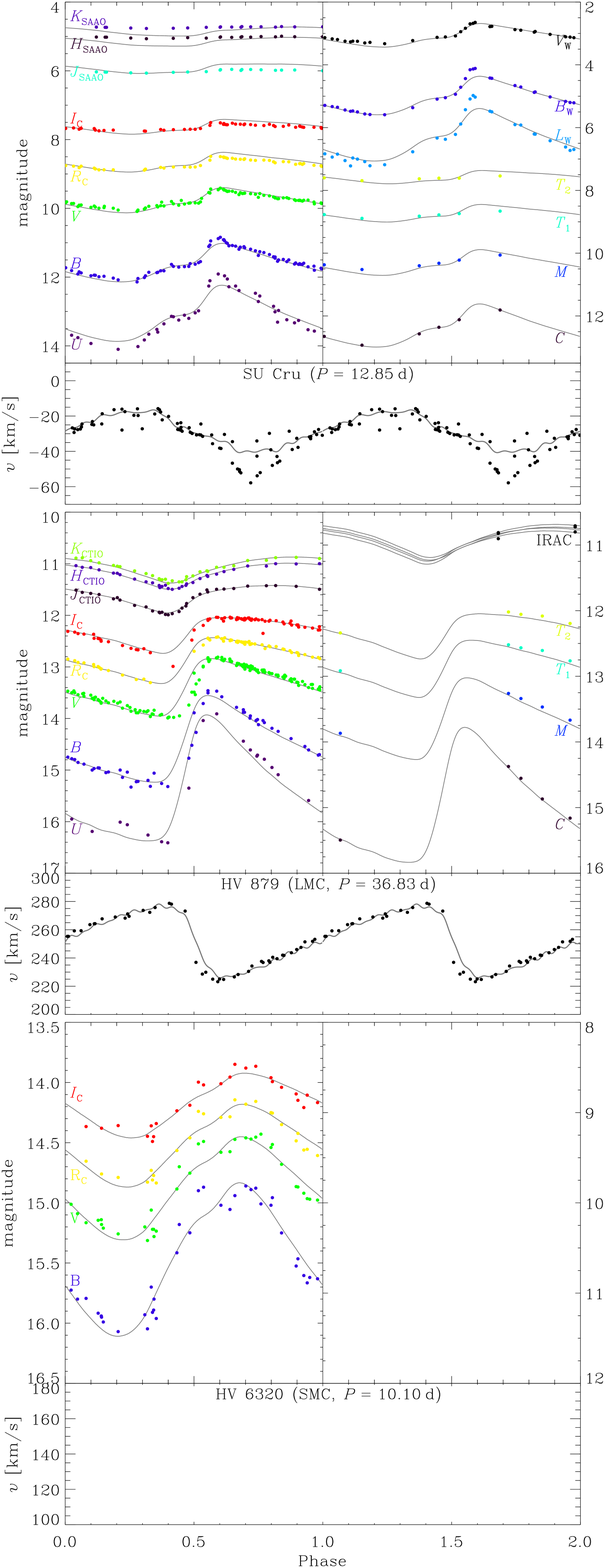}
\end{center}
\caption{Examples of light curves and radial velocity curves for Cepheids in the three galaxies (Galaxy at top, LMC at center, SMC at bottom). The fits are shown with grey solid lines. The left column shows Cepheids with data in many photometric bands and median-quality fits, while the right column shows the Cepheids with poor fits to their photometric data.}
\label{fig:lc}
\end{figure*}

Figure~\ref{fig:lc} shows the data and the best-fit models for six Cepheids in the Galaxy, LMC and SMC. For the left column, we have chosen stars with a large number of datapoints in many filters and a typical fit quality, while the right column shows Cepheids with fits among the worst $5\%$ for the given galaxy based on the photometric $\chi^2$ per degree of freedom. We see that the model fits the data very well, and that even sparse data can strongly constrain the fits given the light curve structures required by the better sampled bands (e.g.\ HV 1543, in the lower left panel of Figure~\ref{fig:lc}). The common problems are noisy data (HV~6320, in the lower right panel), inadequacy of the template (SU~Cru, upper right panel, see the discussion in Section~\ref{sec:disc_residuals}) and a small phase shift between data from different sources (HV 879, middle right panel). The remarkable point, however, is that this simple physical model of a radius plus a temperature template combined with a single amplitude produces good fits for objects that have photometry in many filters across a broad wavelength range as well as their radial velocity measurements. 

\begin{figure*}
\center{\includegraphics[width=0.9\textwidth]{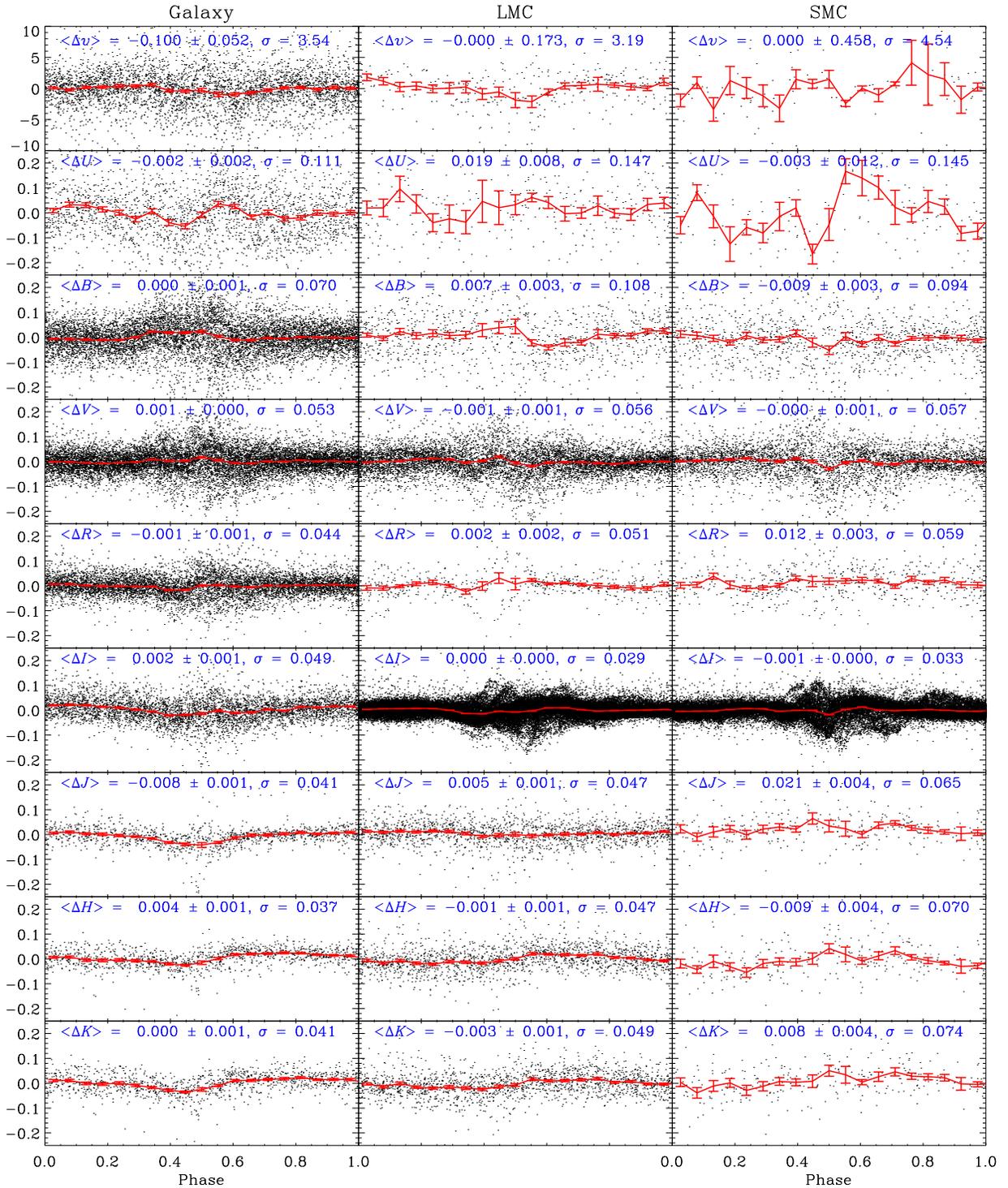}}
\caption{Residuals with respect to the model fits for radial velocities (top row; in km\,s$^{-1}$) and the most common filters (remaining panels). The red lines in each panel are averages in bins of $\Delta\phi= 0.05$ and the error bars are the uncertainties in these averages. Each panel gives the mean value of the residuals, its uncertainty, and the dispersion of the data around the mean. This dispersion is a combination of measurement errors and systematic residuals.}
\label{fig:residuals}
\end{figure*}

\begin{figure*}
\center{\includegraphics[width=0.9\textwidth]{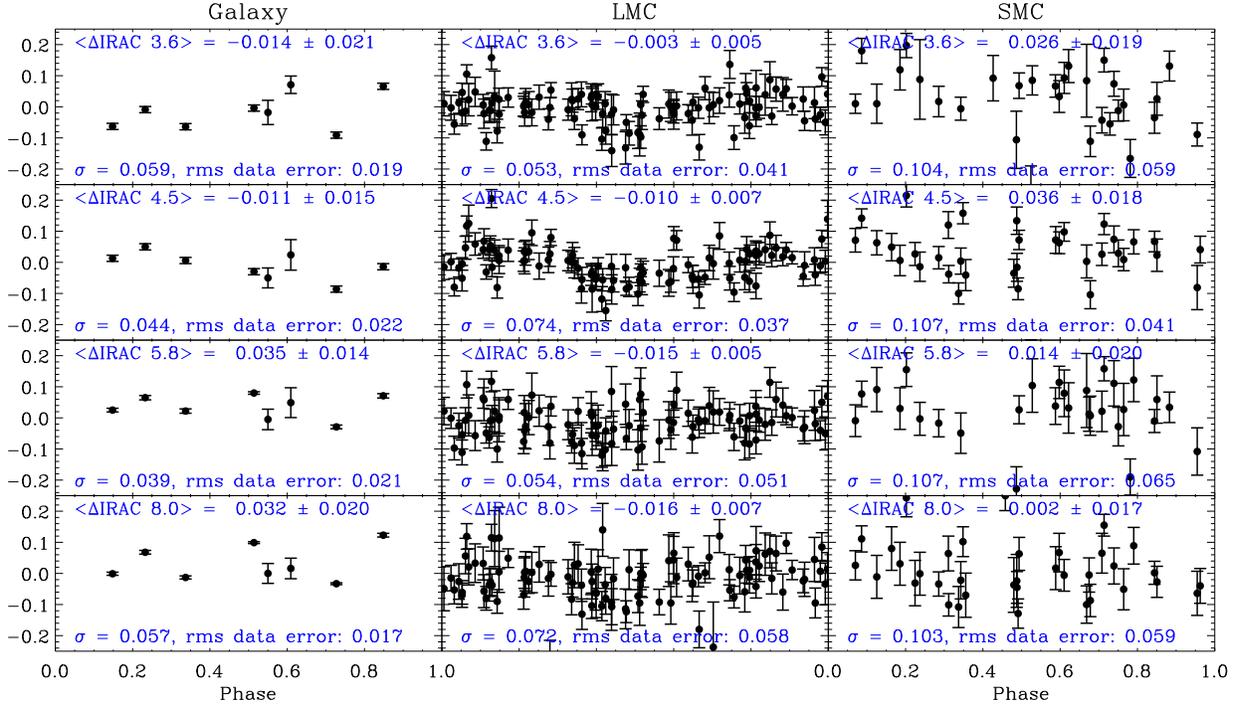}}
\caption{Residuals for the Spitzer IRAC bands. The format is the same as in Figure~\ref{fig:residuals}, but we do not show the phase-binned averages due to the small number of data points. For comparison to the dispersion of the residuals about the mean, we also report the average photometric uncertainty of the data.}
\label{fig:residuals_irac}
\end{figure*}

Figure~\ref{fig:residuals} shows the residuals of the fits as a function of phase for the radial velocities and the most common optical and near-IR bands. Figure~\ref{fig:residuals_irac} shows the residuals for the Spitzer IRAC bands. There are two separate ways to evaluate these residuals. First, for the mean properties of the Cepheids, we want the mean residuals to be consistent with zero given their uncertainties. This is generally true to high accuracy and it is reassuring that the average residuals are essentially zero for all filters with few systematic trends in the residuals as a function of phase. They are not exactly zero, however, which will be an important point in Section~\ref{sec:disc_residuals}. Second, we would like the dispersion of the residuals to be consistent with the measurement uncertainties. This is more difficult to evaluate, because not all the data sources include error estimates and we used as a weight the maximum of the actual measurement error and the rms residual for each band from each data source. However, all measurements in the Spitzer IRAC bands have associated errors and therefore we can compare them to the fit residuals in Figure~\ref{fig:residuals_irac}. We see that in the LMC the mean data error bars are similar to the scatter around the fit, suggesting that our model fits the data very well. In the SMC and especially in the Galaxy, there is a noticeable overall offset between the data and the model in some Spitzer bands, which causes formally a larger scatter around the model. The origin of the offsets will be discussed further in Section~\ref{sec:disc_residuals}. Furthermore, we see in Figure~\ref{fig:residuals} that the residuals are not entirely free of structure. The most obvious structure is that seen in the $I_{\rm C}$-band LMC and SMC residuals at phases between $\phi=0.4$ and $0.7$. These phases correspond to the fast rise from minimum to maximum light. We defer the discussion of these patterns in the residuals to Section~\ref{sec:discussion}.

\subsection{Global Quantities}
\label{sec:global}

\begin{figure}
\plotone{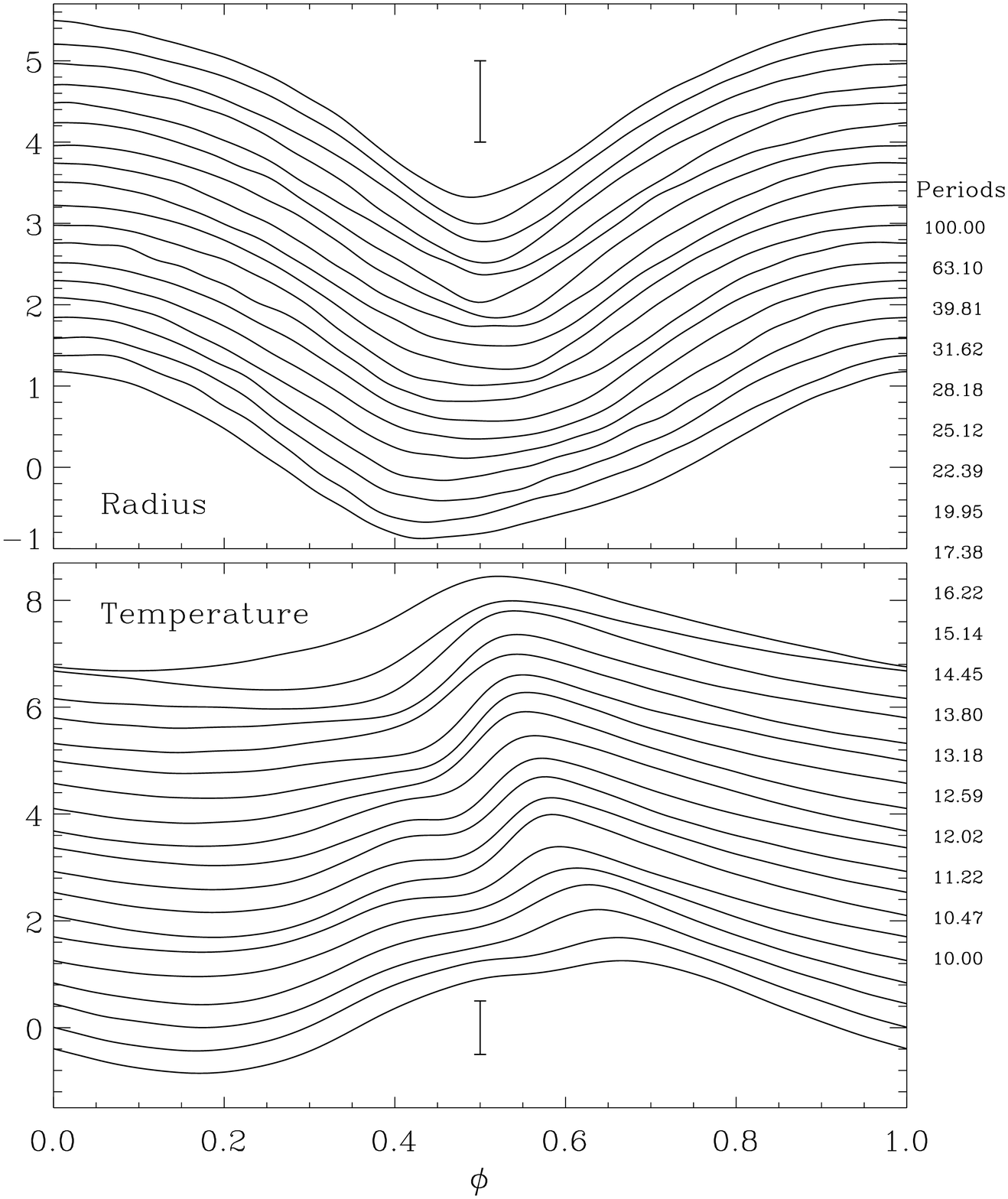}
\caption{The temperature and radius templates at the period grid points as a function of phase $\phi = (t-t_0)/P$. The templates are shifted with respect to each other for the sake of clarity. The period increases from the bottom to the top, and the actual period values are given to the right. The vertical line segments in both panels have unit length.}
\label{fig:templ_radtem}
\end{figure}

Figure~\ref{fig:templ_radtem} shows the radius and temperature templates for the anchor points of our period grid. The values of the coefficients are given in Table~\ref{tab:templates}. We make the templates publicly available\footnote{\url{http://www.astronomy.ohio-state.edu/~pejcha/cepheids}} including a code to calculate the average template light curve and radial velocity curve for any period within the period range and any filter along with the radius and temperature changes for that band. We clearly see the Hertzsprung progression in the temperature templates by following the shifts in the position of maximum and the bump on the rising branch to earlier phases at longer periods. The changes in the radial templates are more subtle, and the main feature is that the minimum becomes more sharply peaked as the period increases.

\begin{figure*}
\plotone{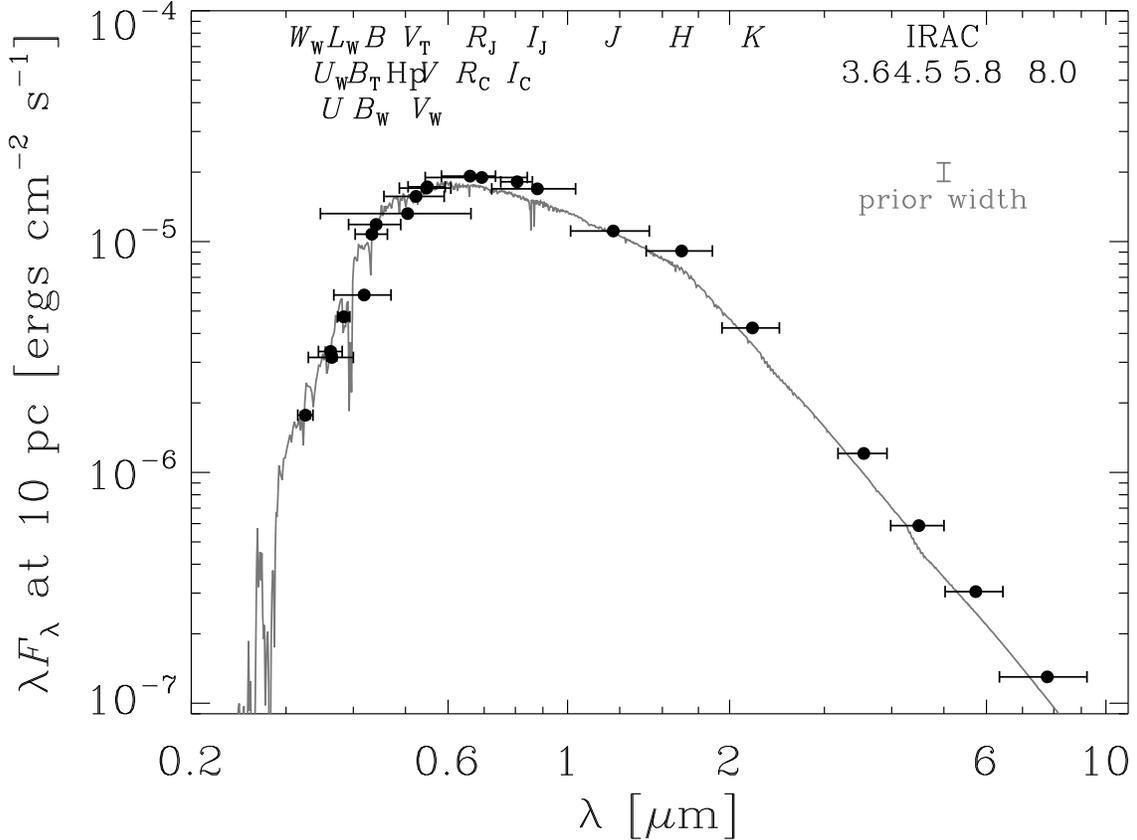}
\caption{The luminosity of the ``mean'' Cepheid with $R_0 = 10\,R_\sun$ and $T_0 = 5400$\,K located at $10$\,pc constructed from the values of $\overline{M}_i$ and converted to $\lambda F_\lambda$ using the LMC calibration distance of $18.50$\,mag. The horizontal error bars show the width of each filter and the filter names are given at the top. The uncertainties on $\overline{M}_i$ are shown, but they are generally smaller than the size of the symbol. For comparison, we also show an error bar corresponding to the width of the prior on $\overline{M}_i$. The grey line shows the flux ($\lambda F_\lambda$ at $10$\,pc) of a supergiant model atmosphere with $T_0=5400$\,K, $R_0 = 10\,R_\sun$, and $\log g = 1.5$ interpolated from the \citet{castelli04} grid of atmosphere models.}
\label{fig:mbar}
\end{figure*}

The magnitude zero-points $\overline{M}_i$ in Equation~(\ref{eq:phys_model_m}) represent the flux of the ``mean'' Cepheid with radius $R_0$ and temperature $T_0$. Comparing the fitted values to a theoretical model of a Cepheid atmosphere allows us to judge the results. In Figure~\ref{fig:mbar}, we convert the values of $\overline{M}_i$ to fluxes and compare them to a supergiant model atmosphere \citep{kurucz79,castelli04} with our reference $T_0=5400$\,K and $R_0=10\,R_\sun$ at $10$\,pc assuming a fixed LMC distance modulus of $18.50$ mag. We see that our fit gives fluxes consistent with the \citet{castelli04} model. There appears to be a small systematic offset in the sense that our data are about $0.08$\,mag too bright. Such a small shift can be caused by a number of reasons such as slightly incorrect flux calibrations, a small change of the LMC distance or a slightly different mean LMC extinction. The errors on $\overline{M}_i$ presented in Table~\ref{tab:filters} are much smaller than the assumed width of the prior, indicating that $\overline{M}_i$ is well constrained by our data. We note that the temperature scale of our model is controlled by the prior on $\overline{M}_i$ so the mean period--temperature relation (Eq.~\ref{eq:taubar_prior}) adjusts to match this temperature scale. This calibration leads to Cepheid temperatures that are systematically higher than the period--temperature relation from the \citet{bono98,bono05} pulsational models. If we removed the $\overline{M}_i$ prior and instead fixed the coefficients in Equation~(\ref{eq:rho_per})--(\ref{eq:tau_per}), we would get perfect agreement of the mean temperatures with their prior values (Eq.~[\ref{eq:tau_coefs}]) at the cost of poor match to the spectral energy distributions ($\overline{M}_i$) of the \citet{castelli04} model for that temperature. Clearly, having a theoretical stellar atmosphere model that is known to be correct would allow us to use a stronger prior to constrain our model and remove most of these ambiguities.

\begin{figure*}
\plotone{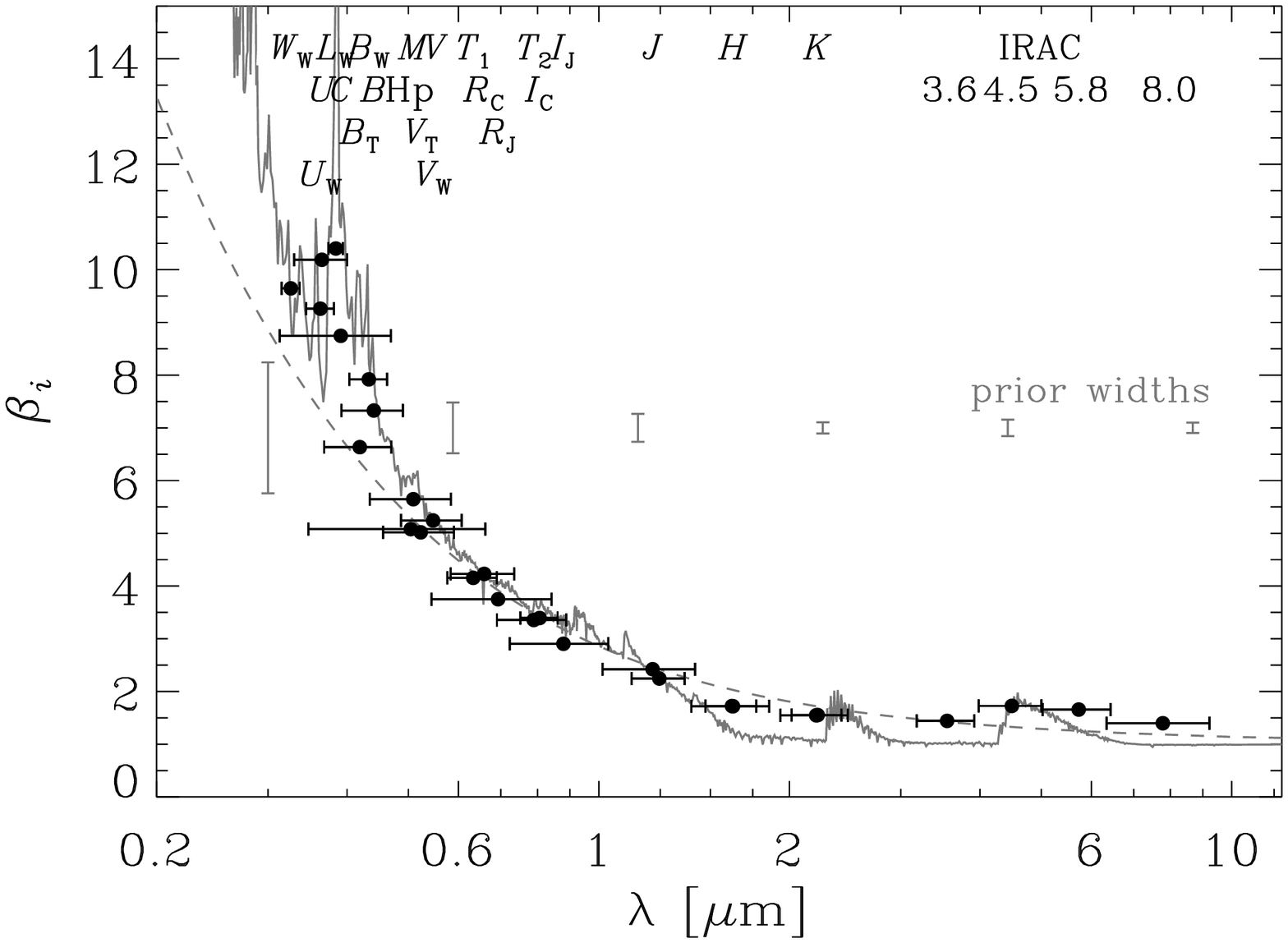}
\caption{The temperature coefficients $\beta_i$ (Eq.~[\ref{eq:beta_general}]) as a function of wavelength. The horizontal error bars show the widths of the filter passbands and the uncertainties in the $\beta_i$ are smaller than the size of symbols. The solid grey line is the prior on $\beta_i$ based on the \citet{castelli04} atmosphere models with $T_0=5400$\,K and $\log g = 1.5$. The widths of the prior are illustrated by the error bars. The dashed grey line is the estimate of $\beta_i$ for a black body with $T_0=5400$\,K. The biggest deviations from the black-body agree well with the \citet{castelli04} model. There are, however, systematic disagreements with the \citet{castelli04} model in the infrared that are probably caused by problems in the molecular opacities.}
\label{fig:beta}
\end{figure*}

As discussed in Section~\ref{sec:priors} (Eq.~[\ref{eq:beta_general}]), the temperature coefficients $\beta_i$ are the logarithmic derivatives of the flux in any band with respect to temperature at temperature $T_0$. Figure~\ref{fig:beta} compares the fitted values of $\beta_i$ to those for a $5400$\,K black body and to the prior based on the \citet{castelli04} atmosphere models with $\log g = 1.5$ and at $T_0=5400$\,K. Recall that $\beta_V\equiv 5.14$ is fixed to its prior value based on the model atmospheres. Our results clearly deviate from the black body in the sense predicted by the model atmospheres. At short wavelengths, the $\beta_i$ are larger than the black-body model because of the effects of metal opacities, and the model may even resolve some of the expected spectral features. In the infrared, we clearly see the CO band head at $\sim\!5\,\mu$m. We obtained essentially the same result with a black body prior on $\beta_i$, which means that the weakly imposed prior on $\beta_i$ is simply overwhelmed by the statistical power of the data -- the values of $\beta_i$ are robustly determined by the color changes of the Cepheids during the pulsational cycle. There are clear systematic differences from the \citet{castelli04} model, especially in the infrared, which are likely due to real problems in the molecular opacities of the \citet{castelli04} model \citep{fremaux06}.

\begin{figure*}
\plotone{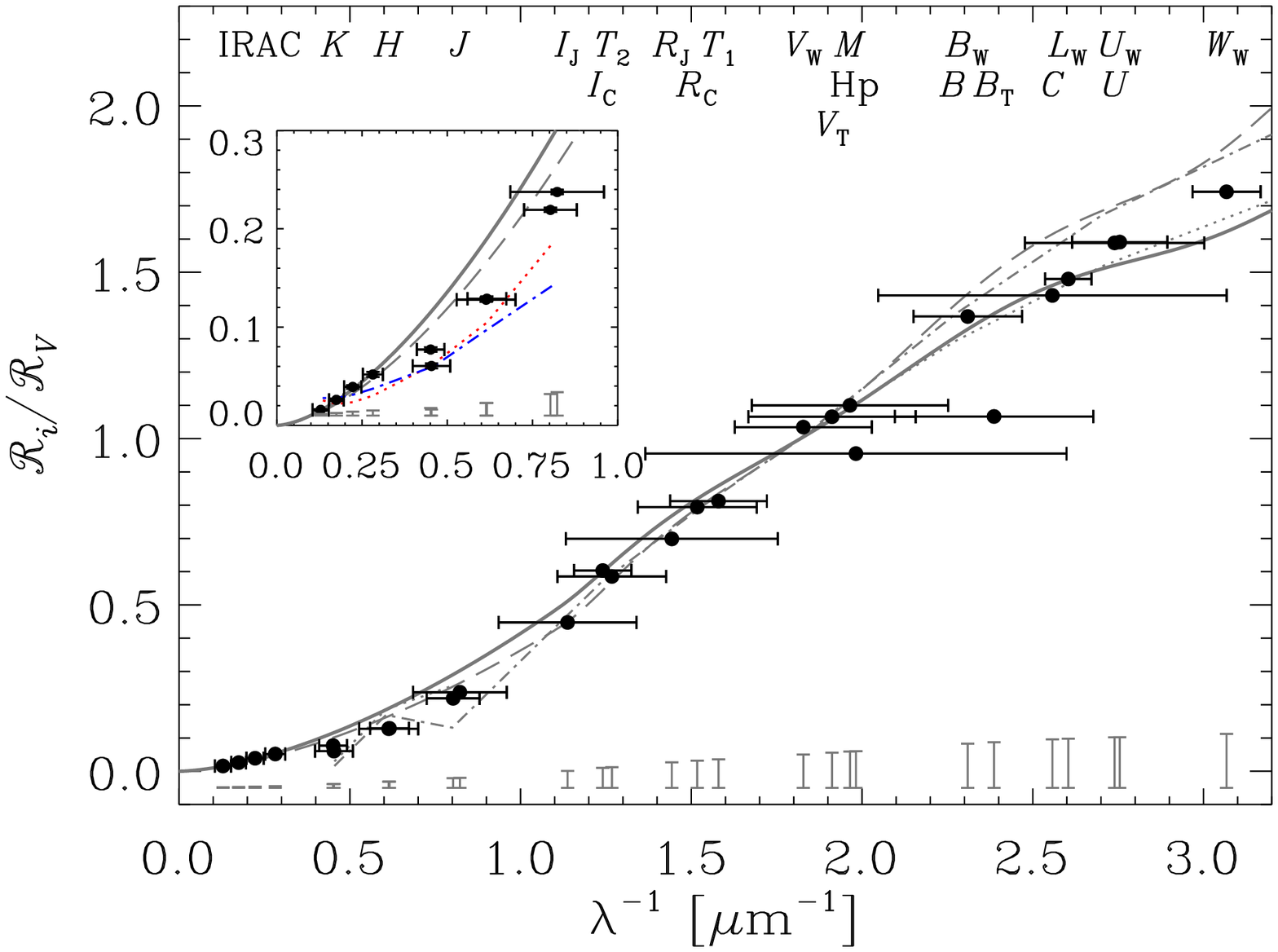}
\caption{The ratio $\ri/\rv$ of the total to the selective extinction $\ri$ in filter $i$ relative to $\rv$ as a function of inverse wavelength. The horizontal error bars show the widths of the filter passbands, and the errors in the estimates of $\mathscr{R}_i$ are smaller than the size of the symbol. The $\mathscr{R}_V \equiv 3.3$ \citet{cardelli89} prior is shown by the solid grey line. The widths of the prior at the filter wavelengths are indicated by the error bars at the bottom of the plot. The dashed gray line shows an $\rv=2.5$ \citet{cardelli89} reddening curve, and the gray dotted (dash-dotted) lines show the LMC (SMC) reddening curve of \citet{gordon03}. The inset shows the infrared region in more detail. The meaning of the grey solid and dashed lines is the same as in the bigger plot. The red dotted and blue dash-dotted lines show the infrared extinction curves of \citet{nishiyama09} and \citet{roman07}, respectively. The curves in the inset were normalized to our fitted value for $\mathscr{R}_{K_{\rm SAAO}}$. The IRAC values of $\ri$ are dominated by prior, while the $JHK$ values are not.}
\label{fig:extinction}
\end{figure*}

In Figure~\ref{fig:extinction} we compare the final extinction curve to the \citet{cardelli89} $\mathscr{R}_V=3.3$ extinction curve used as a prior. We focus on the shape of the extinction curve $\ri/\rv$ because we fixed $\rv\equiv 3.3$ in the models. In Figure~\ref{fig:extinction} one should also focus on the narrower band pass filters because we have not convolved the extinction law with the average Cepheid spectrum and filter bandpass. For example, the very broad Hp filter has anomalously low $\ri$, presumably because it is effectively a much redder band pass in the presence of significant extinction. An advantage of the model $\ri$ is that they ``correctly'' include all these band pass averages. Figure~\ref{fig:extinction} shows that between $0.5$ and $1.5\,\mu$m$^{-1}$, the results fall below the prior, while for $\lambda^{-1} \gtrsim 2\,\mu$m$^{-1}$ they are above. This suggests that the ``mean'' extinction law of our photometric dataset is different from the canonical $\rv \equiv 3.3$ \citet{cardelli89} law, falling between the $\rv\equiv 3.3$ \citet{cardelli89} law and  the empirical LMC and SMC curves of \citet{gordon03}. The $\ri$ values are controlled by the data except for the Spitzer bands, where the values and uncertainties closely follow the \citet{cardelli89} prior. That the best fit extinction curve is not simply a \citet{cardelli89} $\rv \equiv 3.3$ law should not be a surprise. There is no unique extinction law even in the Galaxy \citep{berry11}, and the LMC and SMC extinction laws show further differences \citep[e.g.][]{misselt99,gordon03}. Here we have determined the best average extinction law for Cepheids in these three galaxies. We can attempt to parameterize our results within the framework of the \citet{cardelli89} extinction models. Because we keep $\rv$ fixed, we perform a fit of the \citet{cardelli89} extinction law evaluated for $\lambda_{\rm center}$ of filter $i$ to $\ri/\rv$ with $\rv$ as a free parameter. We find that the best fit value is $\rv = 3.127 \pm 0.002$, but most of the $\ri/\rv$ values still show significant offsets of several percent from the best fit value. Interestingly, the largest differences are for the near-IR bands, where $K_{\rm SAAO}$ and $K_{\rm CTIO}$ are shifted by about $47\%$ and $31\%$ from their respective best-fit \citet{cardelli89} values, as can be seen also in the inset of Figure~\ref{fig:extinction}. We also fit the Cepheids data using $\rv \equiv 3.1$. The change in $\chi^2$ is negligible and the $\ri/\rv$ curve tilts to a slightly steeper slope to accommodate $\mathscr{R}_B \equiv \rv+1$, but it is still discernibly different from the $\ri/\rv$ ratios of the $\rv=3.1$ \citet{cardelli89} reddening law.

\begin{figure}
\plotone{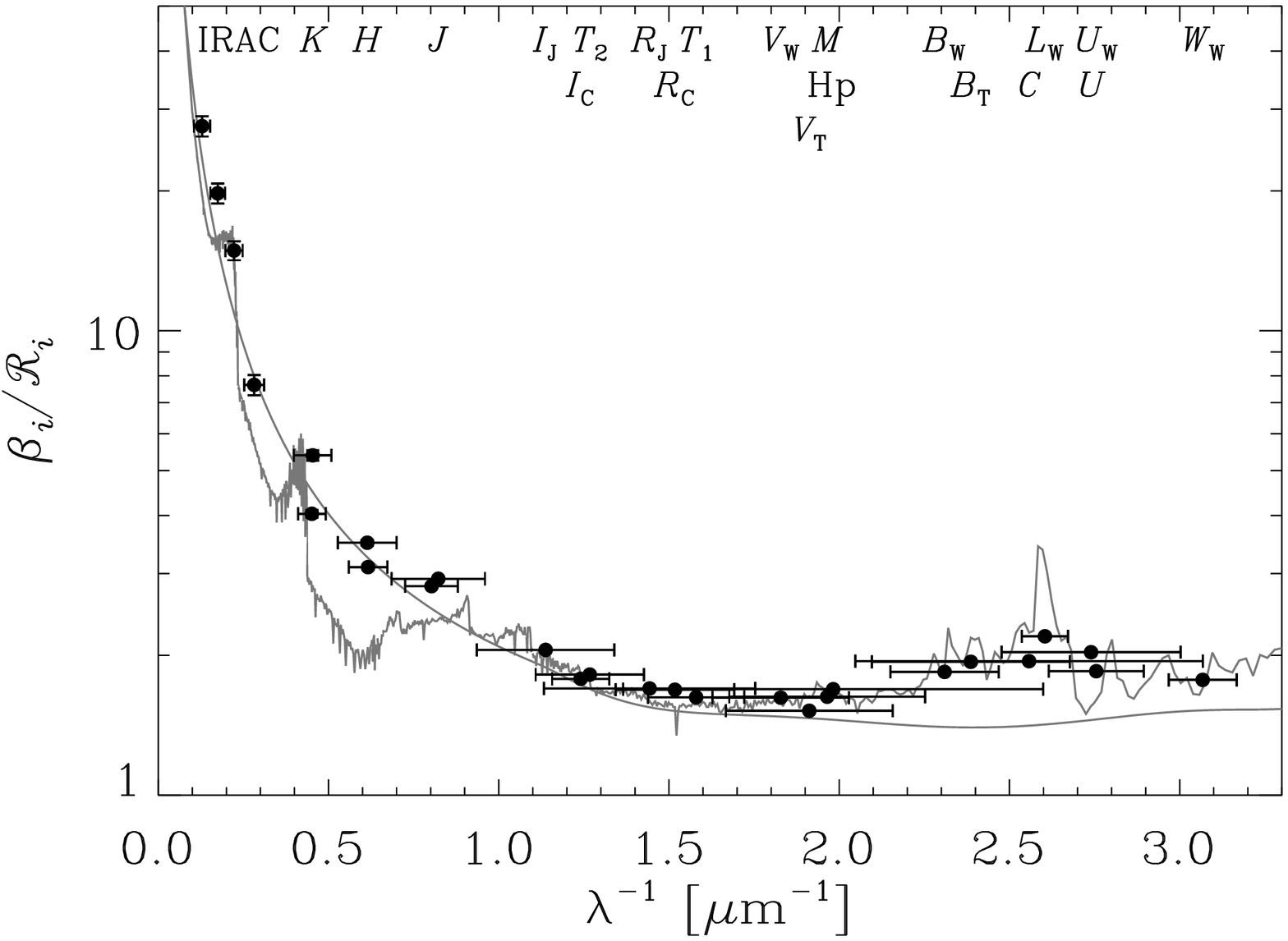}
\caption{The ratio $\beta_i/\ri$ of the temperature coefficient $\beta_i$ to the reddening factor $\ri$. Horizontal error bars show the widths of the filter passbands and the uncertainties in the ratios are generally smaller than the size of the symbol. The gray solid lines are the predictions for the \citet{castelli04} and black-body expressions for $\beta_i$ combined with an $\mathscr{R}_V\equiv 3.3$ \citet{cardelli89} law for $\mathscr{R}_i$. The parabolic shape indicates that our span of wavelengths is large enough to ensure that the temperature and extinction estimates are not mutually degenerate.}
\label{fig:beta_r}
\end{figure}

After establishing that $\beta_i$ and $\mathscr{R}_i$ represent the physical quantities intended in Equation~(\ref{eq:phys_model_m}), we can assess whether temperature and extinction can be determined independently. In Figure~\ref{fig:beta_r} we show the ratio of $\beta_i$ to $\mathscr{R}_i$ as a function of inverse wavelength. Because changes in temperature and reddening can have similar effects on the observed colors for a limited wavelength range, a change in $\bar{\tau}$ can be mimicked by a change in $E(B-V)$ if $\beta_i/\mathscr{R}_i$ is constant over that range. We see from Figure~\ref{fig:beta_r} that the ratio $\beta_i/\mathscr{R}_i$ has an approximately parabolic shape in $\lambda^{-1}$, and therefore we should be able to separate the effects of reddening and temperature quite robustly if there is enough wavelength range. Determining both temperature and extinction requires a minimum of three filters (two colors) and the efficacy of any choice can be assessed by connecting the longest and shortest wavelength bands by a line in Figure~\ref{fig:beta_r}, and then examining the distance of the middle band from the line. Temperature and extinction degeneracies are minimized by maximizing that distance. We generalize this procedure in Appendix~\ref{app:opti}. In assessing Figure~\ref{fig:beta_r}, it is important to recognize the large vertical scale. The curvature between filter wavelengths need only be large compared to the photometric errors ($<0.1$\,mag). For the same errors, however, a larger wavelength baseline is almost always better.

\begin{figure}
\plotone{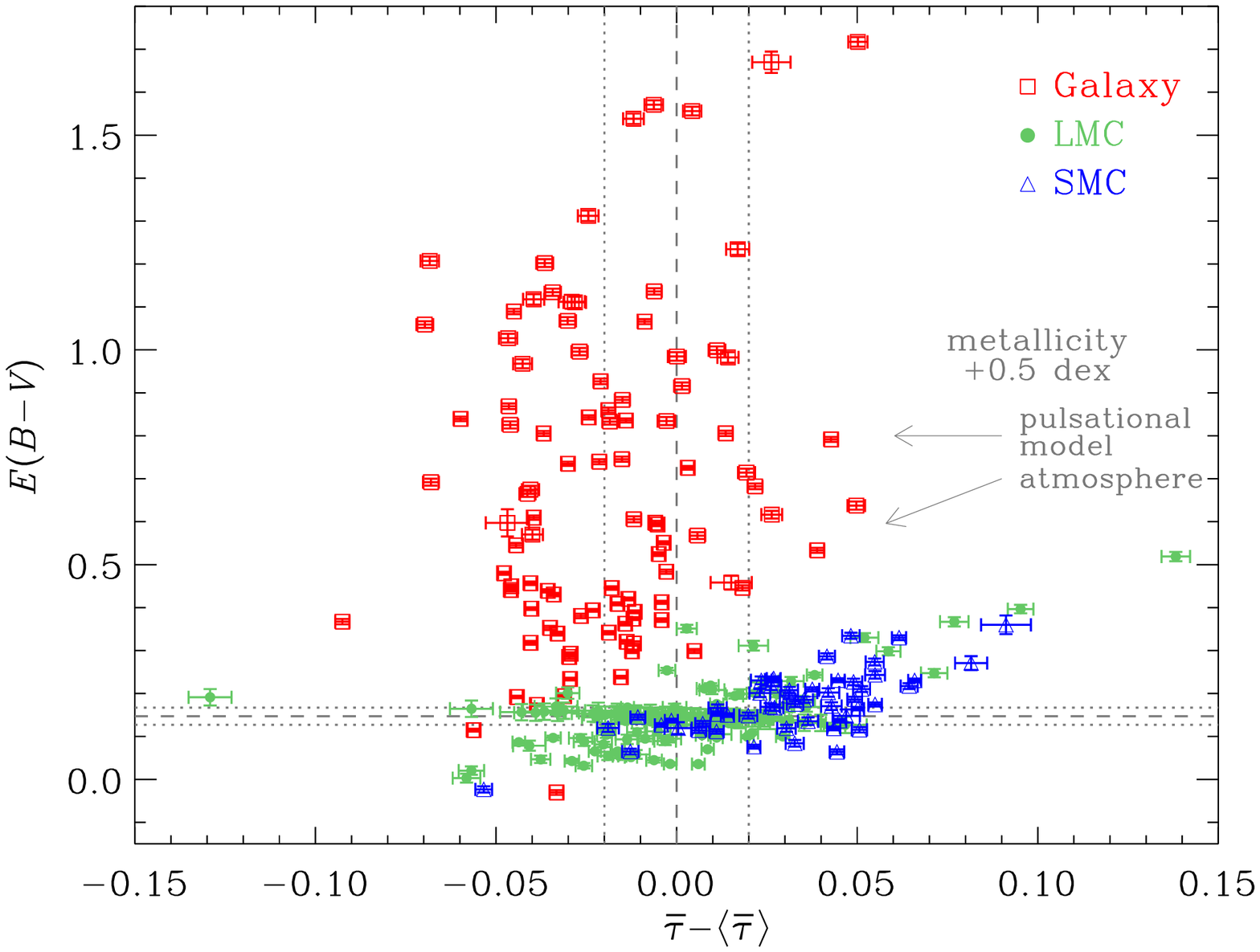}
\caption{Reddenings $E(B-V)$ as a function of temperature $\bar{\tau}$ relative to the mean period--temperature relation $\langle\bar{\tau}\rangle$ for the Galaxy (red open squares), LMC (green filled circles), and SMC (blue open triangles). The horizontal grey dashed line is the LMC reddening prior $\langle E(B-V)\rangle \equiv 0.147$\,mag with a width of $0.02$\,mag and the vertical dotted lines show the width of the $\langle\bar{\tau}\rangle$ prior. The grey arrows show the expected change of $\bar{\tau}$ and $E(B-V)$ as a result of a metallicity increase of $0.5$\,dex due to changes in the atmospheric metallicity at fixed temperature (lower arrow) and changes in the physical properties of the star at fixed period based on pulsational models (upper arrow). See Section~\ref{sec:metal_pred} for details.}
\label{fig:tem_ext}
\end{figure}

We can check for possible temperature/extinction correlations by looking at the individual temperatures and extinctions shown in Figure~\ref{fig:tem_ext}. In the LMC, there is a group of stars with only $V$ and $I_{\rm C}$ photometry, which does not allow for independent extinction and temperature determination. These stars concentrate along the prior $\langle E(B-V)\rangle \equiv 0.147$\,mag. There is also a tail of LMC and SMC stars with both high $\bar{\tau}$ and $E(B-V)$. These stars lack  photometry in bands bluer than $V$, so part of the problem may lie in degeneracies created by small systematic effects in the photometry calibration and the extinction law. For the most extreme cases the model predicts $U$-band magnitudes at maximum brighter than in the $B$ band, which cannot be true. We consider the temperatures and extinctions well separated for most of the stars, because Cepheids in the Galaxy have much higher extinctions than those in the LMC and SMC but noticeably lower temperatures. Metallicity differences also introduce apparent shifts in temperature and extinction. Arrows in Figure~\ref{fig:tem_ext}, based on the results of Section~\ref{sec:metal_pred}, show the effect of increasing the metallicity by $0.5$\,dex, which is approximately the spread in LMC Cepheid metallicities and the typical difference between LMC and SMC Cepheids \citep{romaniello08}. While the amplitude and sense of the shifts are consistent with metallicity effects, none of the outliers has a measured metallicity to verify this origin.

Some choices for our priors are unimportant as the data so strongly constrain the problem that the same final results will be obtained even with very different choices for the priors. For example, using a black body prior for $\beta_i$ instead of the estimate based on the \citet{castelli04} model atmospheres changes the results by a negligible amount. The departures from a black body due to metal line blanketing and molecular opacities are a robust result, as can be seen by comparing the fit errors on the global parameters with the prior widths in Table~\ref{tab:filters} (prior widths are $5\%$ for $\ri$, $10\%$ for $\beta_i$, and $0.1$ or $0.2$\,mag for $\overline{M}_i$, while the typical fit uncertainties are typically $1\%$ for $\ri$, $0.7\%$ for $\beta_i$ and $0.01$ to $0.02$ mag for $\overline{M}_i$, respectively). Our choice of priors is also overdetermined. For example, the prior on $\overline{M}_i$ essentially drives our temperature scale to be the same as in \citet{castelli04}, but the prior on $\bar{\tau}$ (Eq.~\ref{eq:bono_prior}) drives it (weakly) to the temperature scale of the \citet{bono05} pulsational models. Similarly, radial velocities and multi-filter photometry implicitly define the distance to the LMC through the Baade-Wesselink method and so in theory no prior on the LMC distance would be necessary. We discuss these issues further in the Conclusions (Section~\ref{sec:conclusions}).

\begin{figure}
\plotone{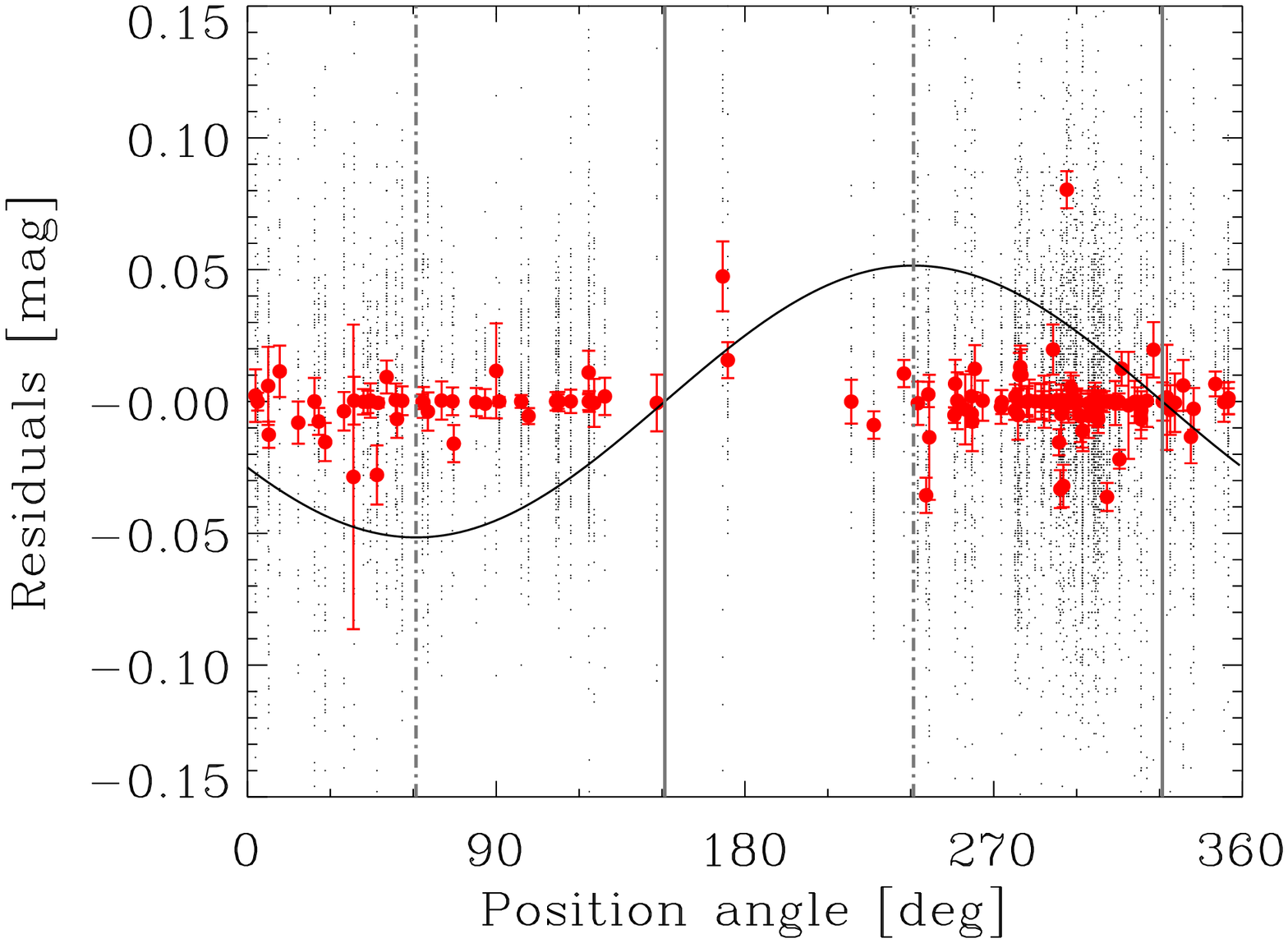}
\caption{The $I_{\rm C}$-band residuals of the LMC Cepheids (dots) as a function of the position angle of the Cepheid relative to the center of the LMC. The red circles show the mean residuals for each Cepheid where the error bar is the uncertainty in the mean. The solid black line shows the expected signal in the residuals for Cepheids at the observed mean distance from the LMC center of $1^{\circ}$ if we did not correct for the LMC tilt. Vertical solid and dashed lines mark the  position angles of the line of nodes and the perpendicular direction, respectively.}
\label{fig:theta_res}
\end{figure}

Finally, we check our assumption that the LMC Cepheids are distributed in a thin disk with the inclination and position angle given by \citet{nikolaev04} at an LMC center distance of $\mu=18.50$\,mag. In Figure~\ref{fig:theta_res} shows the $I_{\rm C}$-band residuals for the LMC Cepheids as a function of their position angle relative to the center of the LMC. An incorrect inclination or position angle of the disk would produce excess residuals at some position angles, but we see very little variation in the scatter of the Cepheid mean residuals as a function of the position angle. We note, however, that modest inconsistencies in distance can be partially absorbed by changes in the mean radius, especially for Cepheids without radial velocity measurements.

\subsection{Individual Properties of Cepheids}
\label{sec:individual}

In this Section we discuss our results on the individual properties of Cepheids, which are given in Tables~\ref{tab:galaxy}, \ref{tab:lmc}, and \ref{tab:smc}. First, we compare our distances for the Galactic Cepheids with measured parallaxes to check our assumption that the LMC distance modulus is $18.50$ mag. The overlap between our sample and the sample of \citet{benedict07}, who measured Cepheid parallaxes using the Fine Guidance Sensor on the Hubble Space Telescope, is only 2 stars, $\ell$~Car and $\zeta$~Gem, because of our restriction to $P \ge 10$\,days. The parallax measurements of \citet{benedict07} give $\mu = 8.48 \pm 0.22$ and $7.78 \pm 0.14$ for these two stars, respectively. We obtain $\mu=8.29 \pm 0.01$ and $7.89 \pm 0.03$ for $\ell$~Car and $\zeta$~Gem, respectively, given our assumed LMC distance of $\mu=18.50$\,mag. We note that the uncertainties in our distances include {\em all} the uncertainties in the model (ie.\ including $\ri$, $\beta_i$, etc.) and are not simply the uncertainties in the distance for fixed model parameters. For these two stars, the distances are in good agreement with their parallaxes.

\citet{storm11a,storm11b} determine Baade-Wesselink/IR distances and extinctions (some from \citealp{ref167}) for a sample of Galactic and LMC Cepheids. For the $21$ LMC stars we have in common, our distance moduli are higher by $0.054\pm 0.043$\,mag, where the error is the uncertainty in the mean, consistent with their lower mean LMC distance modulus of $\mu_{\rm LMC} = 18.45\pm 0.04$\,mag compared to our fixed mean value of $\mu_{\rm LMC} \equiv 18.50$\,mag. Our LMC extinction estimates are also systematically higher by $\Delta E(B-V) = 0.036\pm 0.020$\,mag. In the Galaxy, we have $30$ stars in common, and in this case our distance moduli are shorter by an average of $0.140\pm 0.058$\,mag. Given that our LMC distance is about $0.05$\,mag longer, our Galactic distance scale is about $0.19$\,mag shorter than that of \citet{storm11a}. For the Galactic Cepheids, we find slightly lower average extinction with $\Delta E(B-V) =-0.015\pm 0.009$\,mag. The LMC extinction offset could be reconciled by setting the mean LMC extinction to $\langle E(B-V)\rangle \simeq 0.10$ rather than the $\langle E(B-V)\rangle = 0.147$ we adopted from \citet{udalski99}. However, lowering the mean LMC $\langle E(B-V)\rangle$ to $0.10$\,mag would create a noticeable overall shift of the $PL$ relation.

\begin{figure}
\plotone{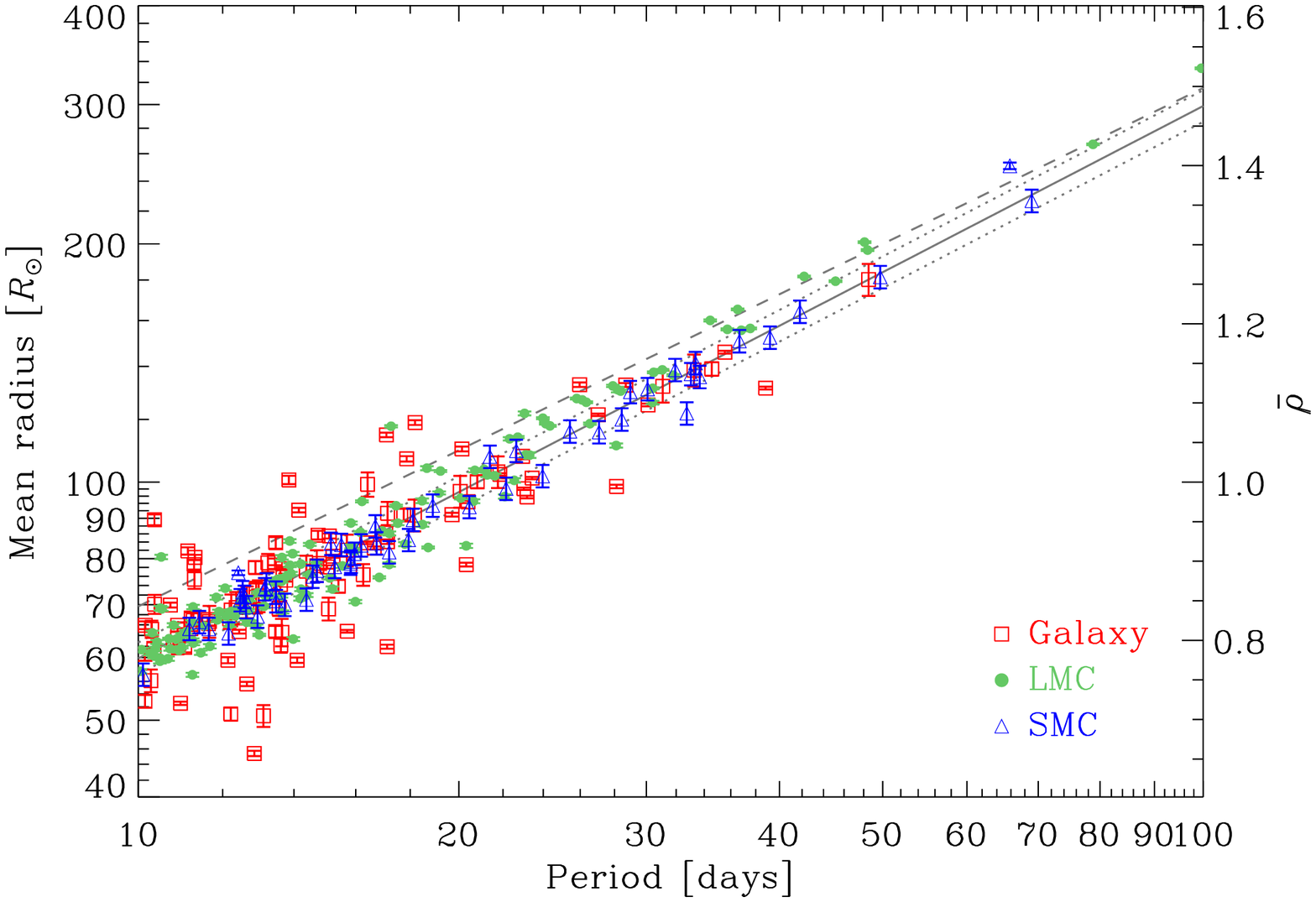}
\caption{Radii of individual Cepheids as a function of period for the Galaxy (red open squares), LMC (green filled circles), and SMC (blue open triangles). The grey solid line is the final mean period--radius relation $\langle\bar{\rho}(P)\rangle$ of our dataset (Eq.~[\ref{eq:rho_per_mes}]) with the width indicated by the grey dotted lines. The grey dashed line is the prior on $\langle\bar{\rho}(P)\rangle$ based on the \citet{bono98,bono05} pulsational models (Eq.~[\ref{eq:rho_coefs}]).}
\label{fig:radius}
\end{figure}

Figure~\ref{fig:radius} shows the distribution of the Cepheids in radius (Eq.~[\ref{eq:rhobar_prior}]). The overall distribution of the Cepheids tracks the prior well, but most Cepheids have errors in their radii that are smaller than the width of the prior, which indicates that the radius is constrained by the data (or other priors) more than by the prior on the radius. The larger scatter in the Galactic Cepheids is caused by a much greater spread in distances and reddenings compared to the LMC and SMC. A potential origin for the large positive outliers is blending, where additional light from an unresolved companion causes the amplitude to be smaller and the Cepheid to be brighter. During the process of cleaning our dataset we removed one obvious outlier (HV 2326) created by blending \citep{ref166}. Our fit also provides the updated values of coefficients $a_{\bar{\rho}}$ and $b_{\bar{\rho}}$, which describe how the mean radii change with the period. We find
\begin{subequations}
\label{eqs:rhotau_per_mes}
\beq
\langle\bar{\rho}\rangle = (0.777 \pm 0.002)+(0.698 \pm 0.006)\, \log\!\left(\! \frac{P}{10\,{\rm d}}\!\right).\label{eq:rho_per_mes}
\eeq
Compared to the prior, we find that the zero point is lower by about $0.07$\,dex, about three times the width of the prior on $a_{\bar{\rho}}$, and the output error in the zero point is much smaller than the width of the prior ($0.02$), which suggests that the data constrain it well. On the other hand, the value and uncertainty in the slope are more similar to the prior, which suggests that the prior is important for the slope estimates. Our limited period range and the small number of long-period Cepheids are not particularly well-suited for the determination of slopes.

\begin{figure}
\plotone{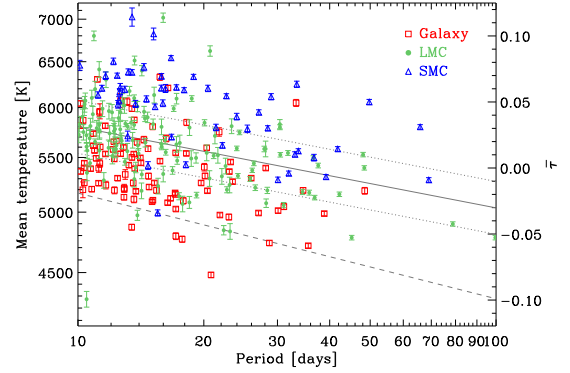}
\caption{Same as Figure~\ref{fig:radius}, but for temperature. The grey solid line shows the final mean period--temperature relation $\langle\bar{\tau}(P)\rangle$ of our dataset (Eq.~[\ref{eq:tau_per_mes}]) with its width denoted by the gray dotted lines. The grey dashed line is the prior on $\langle\bar{\tau}(P)\rangle$ based on the \citet{bono98,bono05} pulsational models (Eq.~[\ref{eq:tau_coefs}]).}
\label{fig:temperature}
\end{figure}

Figure~\ref{fig:temperature} shows the distribution in mean temperature. The final zero point and slope of the period--temperature relation are very different from the theoretical prior (Eq.~[\ref{eq:tau_coefs}]) based on the \citet{bono98,bono05} pulsational models. The zero point offset generally stems from the fact that our temperature scale is constrained by the prior on $\overline{M}_i$, which does not exactly match the theoretical pulsational model of $\langle\bar{\tau}(P)\rangle$, as was discussed in Section~\ref{sec:global}. We again see that the temperature error bars are frequently far smaller than the width of the prior and so must be tightly constrained by the data. Figure~\ref{fig:temperature} also shows that the SMC Cepheids seem to be systematically hotter than their LMC and Galactic counterparts. We will consider the differences between galaxies in more detail in Section~\ref{sec:discussion}. Our fit also provides the updated values of coefficients $a_{\bar{\tau}}$ and $b_{\bar{\tau}}$, which describe how the mean temperatures change with the period. We find a final period--temperature relation of
\beq
\langle\bar{\tau}\rangle = (0.031 \pm 0.002)- (0.061 \pm 0.005)\,  \log\!\left(\! \frac{P}{10\,{\rm d}}\!\right).\label{eq:tau_per_mes}
\eeq
\end{subequations}
These values for $a_{\bar{\tau}}$ and $b_{\bar{\tau}}$ differ noticeably from the prior based on the \citet{bono98,bono05} pulsational models given in Equation~(\ref{eqs:rhotau_coefs}) in the sense that we find a higher zero point and a shallower slope. The uncertainties in both quantities are smaller than the width of the prior, which suggests that the data constrain these parameters well.

\begin{figure*}
\plotone{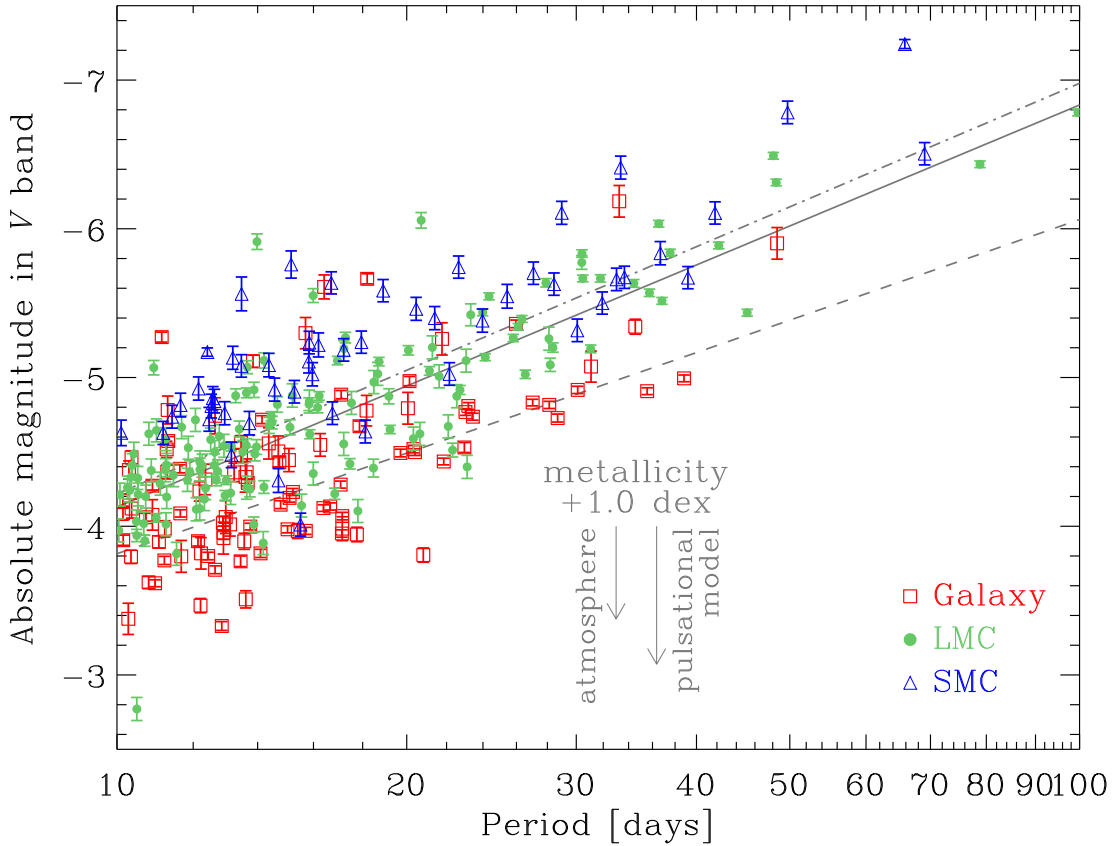}
\caption{The $V$-band period--luminosity relation for the Cepheids in our sample. The meaning of the symbols is the same as in Figure~\ref{fig:radius}. The dot-dashed gray line is the LMC $PL$ relation of \citet{udalski99} (Eq.~[\ref{eq:pl_udalski}]) assuming $\mu_{\rm LMC}= 18.50$\,mag and corrected for extinction based on red clump stars. The dashed line is the $PL$ relation defined by Equation~(\ref{eq:pl_theo}) constructed from the priors on the mean period--radius and period-temperature relations (Eq.~[\ref{eq:rho_coefs}]--[\ref{eq:tau_coefs}]). The solid grey line shows the ``mean'' $PL$ relation constructed from the final period--radius and period--temperature relations in Equations~(\ref{eq:rho_per_mes})--(\ref{eq:tau_per_mes}). Our output $PL$ is consistent with \citet{udalski99} even though the priors are not, which suggests that the \citet{bono98,bono05} priors are inconsistent, but are overwhelmed by the statistical power of the data. The arrows show the expected change of $M_V$ as a result of a metallicity increase of $1.0$\,dex due to changes in the atmospheric metallicity at fixed temperature (left arrow) and changes in the physical properties of the star at fixed period based on pulsational models (right arrow).}
\label{fig:pl}
\end{figure*}

Equations~(\ref{eq:rho_per_mes})--(\ref{eq:tau_per_mes}) provide priors for the mean radii and temperatures for stars which do not have enough data to constrain these parameters independently and also define the mean $PL$ and template light curves as outlined in Section~\ref{sec:connect}. By combining the radii and temperatures of individual Cepheids as shown in Section~\ref{sec:connect} we can construct a $PL$ relation as defined in Equation~(\ref{eq:pl_theo}). We show the $V$-band $PL$ relation in Figure~\ref{fig:pl}, along with the theoretical $PL$ relation implied by the \citet{bono98,bono05} pulsational models (eqs.~[\ref{eq:rhobar_prior}--\ref{eq:taubar_prior}]), which has slope of $-2.25$. We also show the $V$-band $PL$ relation of \citet{udalski99} with a slope of $-2.76$ that was obtained by dereddening the LMC Cepheids using nearby red clump stars. We see that despite having the priors from \citet{bono98,bono05} theoretical models, the data control the fit in the sense that the final $PL$ relation of the stars in our sample is now more similar to the empirical $PL$ of \citet{udalski99}. The theoretical period--radius and period--temperature relations from Equation~(\ref{eqs:rhotau_coefs}) predict a $PL$ relation that is incompatible with both our data and the earlier results of \citet{udalski99}. It is important to understand that the actual mean magnitudes of each Cepheid in any band are essentially equal to the mean of the observations, as seen in Figure~\ref{fig:residuals}, and that any differences in Figure~\ref{fig:pl} arise only from different assumptions about the mean extinctions and distances. To put this on more quantitative grounds, Equations~(\ref{eq:rho_per_mes})--(\ref{eq:tau_per_mes}) together with Equation~(\ref{eq:pl_theo}) imply a $PL$ relation
\beq
M_V = (-4.130\pm 0.024)- (2.703 \pm 0.070)\,\log\!\left(\! \frac{P}{10\,{\rm d}}\!\right),\label{eq:pl_mes_radtem}
\eeq
where $M_V$ is the absolute magnitude in the $V$ band. The errors of the slope and zero point consistently include the uncertainties in $\overline{M}_V$, $a_{\bar{\rho}}$, $b_{\bar{\rho}}$, $a_{\bar{\tau}}$, and $b_{\bar{\tau}}$ and all their mutual covariances. The standard deviation of the Cepheids about this relation is $0.46$\,mag. Longer wavelengths show less scatter, with a standard deviation of $0.33$\,mag at $I_{\rm C}$ band and only $0.22$\,mag for the IRAC [3.6] band. For comparison, \citet{udalski99} give a standard deviation for the $V$-band $PL$ relations of $0.16$ and $0.26$\,mag for the LMC and SMC, respectively, and $0.11$ and $0.21$\,mag for the $I_{\rm C}$ band, respectively. \citet{scowcroft11} gives a scatter of $0.14$\,mag in the Spitzer [3.6] band for the $P\ge 10$\,days LMC Cepheids based on phase-resolved Spitzer photometry. Thus, the scatters about our $PL$ relations are larger, but we are far more tightly constrained due to the large number of bands and we expect some additional contributions to the scatter from missing physics because our sample also includes stars from the LMC, SMC, and the Galaxy, as we will discuss in Section~\ref{sec:discussion}. We point out that given the limited period range of our sample and the lack of long-period Cepheids, the slopes of relations are poorly constrained. Adding short-period Cepheids or long-period stars in external galaxies would better constrain slopes of our period--radius, period--temperature and consequently period--luminosity relations.

The absolute $V$-band magnitude of a Cepheid with $P=10$\,days (Eq.~[\ref{eq:pl_mes_radtem}]) is approximately $-4.13\pm 0.02$\,mag, which is somewhat fainter than the $-4.22\pm 0.04$ found from the \citet{udalski99} $V$-band PL relation of
\beq
M_V = (-1.458\pm 0.021 ) - (2.760\pm 0.031 )\, \log\! \left(\!\frac{P}{1\,{\rm d}}\!\right).
\label{eq:pl_udalski}
\eeq
As we experimented with our models, we would find small offsets between the zero point of our $PL$ relation and the OGLE $PL$. These offsets are very sensitive to the definition of the $E(B-V)$ extinction zero point in the LMC. For example, shifting the LMC extinction zero point by $\Delta E(B-V) = 0.03$\,mag from our default $\langle E(B-V)\rangle =0.147$\,mag offsets the $PL$ zero point by $\rv\Delta E(B-V) = 0.10$\,mag. We also see in Figure~\ref{fig:pl} that there are potential differences between the individual galaxies such that a $PL$ fit only to the LMC Cepheids will be in better agreement with Equation~(\ref{eq:pl_udalski}). We discuss this in more detail in Section~\ref{sec:metal_global}.

\begin{figure}
\plotone{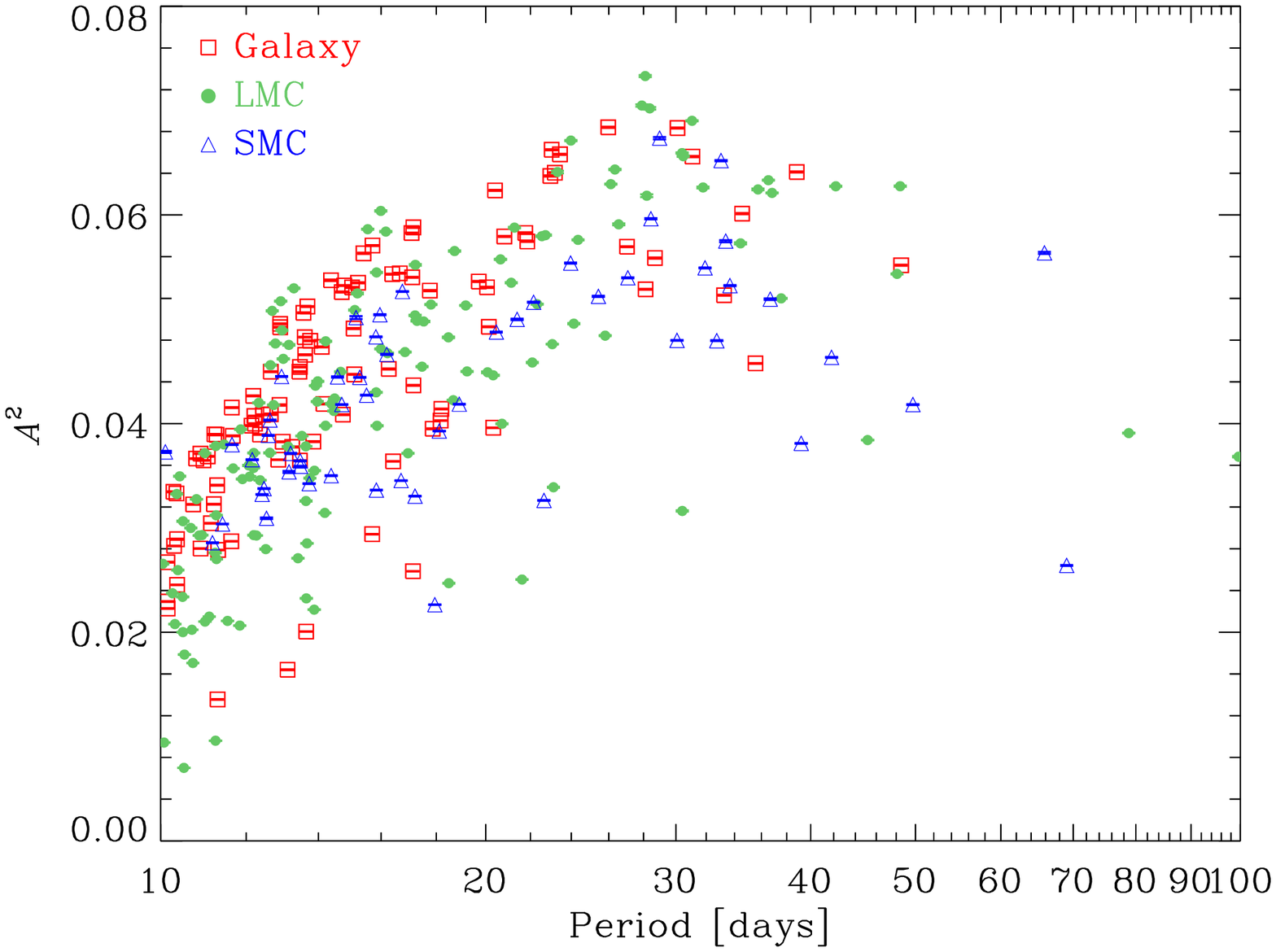}
\caption{Amplitude $A^2$ as a function of period. The meaning of the symbols is the same as in Figure~\ref{fig:radius}.}
\label{fig:amplitude}
\end{figure}

Figure~\ref{fig:amplitude} shows the amplitude $A^2$ (eqs.~[\ref{eq:rho}--\ref{eq:tau}]) as a function of period. There is a trend with the period: amplitude increases with period up to $P\sim 30$\,d and then either declines or becomes constant at higher periods. Interestingly, for a fixed $P$, the distribution of $A^2$ appears to be skewed towards lower values of $A^2$. In the case of LMC and SMC Cepheids, blending can decrease the amplitude somewhat, but the same behavior is seen is seen also in the Galactic Cepheids. We also do not see any correlation between $A^2$ and $\bar{\rho}-\langle\bar{\rho}\rangle$. This suggests that a fraction of Cepheids have intrinsically smaller amplitude, presumably due to their position in the instability strip \citep{szabo07}. There are small differences between the individual galaxies, which we discuss in Section~\ref{sec:metal_individual}. Our findings are in agreement with previous results \citep[e.g.][]{sandage71,vangenderen78,berdnikov86,paczynski00,sandage04,klagyivik09,szabados11}.

\section{Discussion}
\label{sec:discussion}

Our physical model, which includes only radius and temperature, does not include all relevant Cepheid physics -- in particular it does not include composition. In this Section we discuss the evidence for additional physics that was not included in Equations~(\ref{eq:phys_model_m}), (\ref{eq:phys_model_v}) and (\ref{eqs:rhotau_per}). We are interested in differences between the Cepheids in the three galaxies because of their different metallicities. \citet{romaniello08} find mean spectral abundances of $\feh \simeq 0.00$, $-0.33$ and $-0.75$\,dex for the Galactic, LMC and SMC Cepheids with typical ranges of $-0.18 \lesssim \feh \lesssim 0.25$, $-0.62 \lesssim \feh \lesssim -0.10$, and $-0.87 \lesssim \feh \lesssim -0.63$, respectively. It is important to remember that any effect of composition that can be mimicked by a parameter in the model has been! For example, a zero point difference $\overline{M}_i$ between the galaxies due to metallicity effects is automatically compensated for, because we fit for individual distances $\mu$ for the SMC and Galactic Cepheids. Similarly, differences in color are absorbed into reddening and temperature before leaving any trace in the residuals. We start by investigating these issues using theoretical model atmospheres in Section~\ref{sec:metal_pred}. Bearing these issues in mind, we can search for extra physics on three levels. First, in the model residuals (Section~\ref{sec:disc_residuals}), second, in global statistical properties like distances and $PL$ relations (Section~\ref{sec:metal_global}), and, third, in the distributions of the properties of the individual Cepheids (Section~\ref{sec:metal_individual}). Finally, in Section~\ref{sec:second_order} we evaluate the significance of second-order terms and the first-order metallicity correction relative to the first-order temperature term $2.5\beta\bar{\tau}$ in Equation~(\ref{eq:phys_model_m}).

\subsection{Theoretical Predictions for the Effects of Metal Content}
\label{sec:metal_pred}

\begin{figure}
\plotone{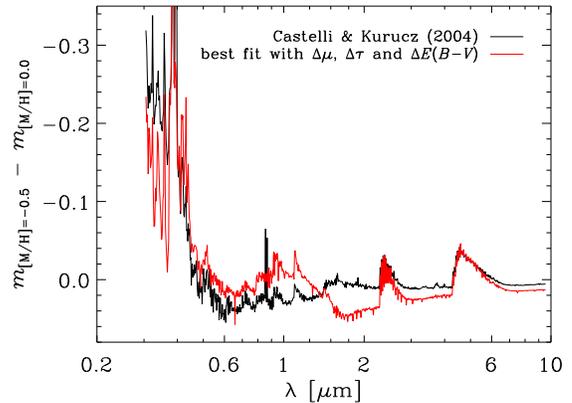}
\caption{Magnitude differences as a function of wavelength between \citet{castelli04} stellar atmospheres with $\mh =-0.5$ and $0.0$ at fixed temperature $T_0 = 5400$\,K, radius $R_0 = 10R_\sun$, and surface gravity $\log g = 1.5$ (black line). The red line is the best fit our model of Eq.~(\ref{eq:metal_model}) can make to the effects of metallicity. }
\label{fig:msmc-mgal}
\end{figure}

Composition affects both the stellar atmospheres (changing Eq.~[\ref{eq:phys_model_m}]) and the mean properties at fixed period (changing Eq.~[\ref{eqs:rhotau_per}]). The observed properties will be a combination of both effects. We can get a sense of metallicity effects on the mean Cepheid properties by examining model period--radius and period--temperature relations at different metallicities. In the \citet{bono98} models the radius at fixed period increases as the metallicity decreases, by $6\%$ for Cepheids with $P=10$\,days when the metal content is decreased from Galactic to SMC metallicity. Combining the period--radius relation of \citet{bono98} with the temperature--luminosity relation of \citet{bono05}, we find that the temperature increases by $5\%$ for Cepheids with $P=10$\,days when the metal content is decreased from the Galactic metallicity to that of the SMC. In combination, the luminosity increases by $\sim 35\%$ at the lower metallicity. \citet{fiorentino02} and \citet{marconi05} also investigated the effects of metallicity on the mean properties using pulsational models, finding that increasing the metal content moves the instability strip to lower temperatures. These effects will be absorbed into shifts in $\bar{\tau}$ and $\mu$/$\bar{\rho}$.

Metallicity also modifies the SED of the stellar atmosphere at fixed $T$, $L$ and $R$ by changing the strengths of spectral lines and the amount of line blanketing. Figure~\ref{fig:msmc-mgal} compares the flux of a \citet{castelli04} model atmosphere with $\mh = -0.5$ to one with solar metallicity at fixed temperature and radius ($T_0=5400$\,K and $R=10\,R_\sun$). As expected, the lower metallicity star is brighter blueward of $\sim 0.45\,\mu$m, due to the reduced effects of metal-line blanketing, and in the molecular band heads at $\sim 2.5\,\mu$m and $\sim 5\,\mu$m. Now let us assume that we are fitting photometric measurements of stars with different metallicities using the model in Equation~(\ref{eq:phys_model_m}), which does not explicitly take metallicity into account. This leads to shifts in distance, reddening, and temperature, because these quantities will try to absorb as much of the variance in the appearance of the stellar atmosphere as possible. To quantify this effect, we fit the magnitude difference $\Delta m$ between model atmospheres with different metallicities (Figure~\ref{fig:msmc-mgal}) with a model based on Equation~(\ref{eq:phys_model_m}),
\beq
\Delta m = \Delta\mu + \mathscr{R}\Delta E(B-V) - 2.5\beta\Delta \bar{\tau}.
\label{eq:metal_model}
\eeq
For the purposes of this discussion, changes in $\mu$ are degenerate with changes in $\bar{\rho}$, and we thus include only $\Delta\mu$. We also assume the $\mathscr{R}_V=3.3$ \citet{cardelli89} reddening curve for $\mathscr{R}$, and the  $\beta$ computed from the same \citet{castelli04} atmosphere that was shown in Figure~\ref{fig:beta}. As we see in Figure~\ref{fig:msmc-mgal}, the model provides a reasonable match to many of the effects of metallicity on the atmosphere by changing the distance/radius, temperature and reddening of the star. In particular, the metal-line blanketing effects blueward of $\sim 0.45\,\mu$m and the molecular bands in the IR are fit reasonably well. The introduced shifts in the model parameters are $(\Delta \mu, \Delta E(B-V), \Delta \bar{\tau}) = (-0.152, -0.166, -0.053)$ per dex in the sense that at fixed $L$ and $T$ a metal-poor atmosphere can be mimicked by making the Cepheid fainter (more distant or smaller), hotter, and more reddened than the same Cepheid with higher metallicity. These changes are then superposed on any shifts in the period--radius and period--temperature relations. A formalism related to Equation~(\ref{eq:metal_model}) can be extended to determine the optimal set of filters for determinations of the distance, reddening, temperature and metallicity, as outlined in Appendix~\ref{app:opti}. 

\subsection{Metallicity Effects in the Fit Residuals}
\label{sec:disc_residuals}

\begin{figure}
\plotone{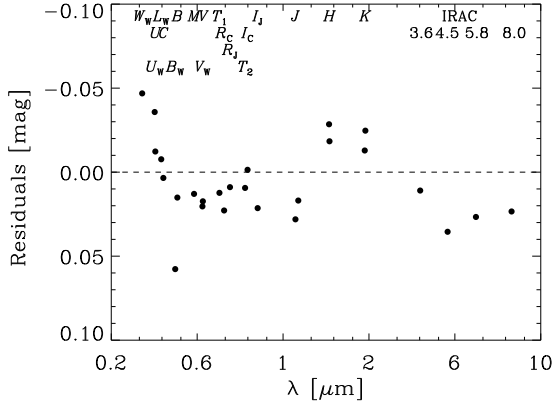}
\caption{Theoretical residuals from a fit to the magnitude difference due to metallicity between model atmospheres with $\Delta \mh = 0.5$\,dex ($\Delta m = m_{\mh=-0.5}-m_{\mh=0.0}$) in each of our filters with respect to a model that allows only for changes in distance, reddening and temperature of the Cepheid (Eq.~[\ref{eq:metal_model}]). Here we convolved $\Delta m$, $\mathscr{R}$, and $\beta$ with a tophat function with the central wavelength and width of each filter, and performed the fit weighting each filter by the number of measurements in our dataset.}
\label{fig:metal_residuals}
\end{figure}

\begin{figure}
\plotone{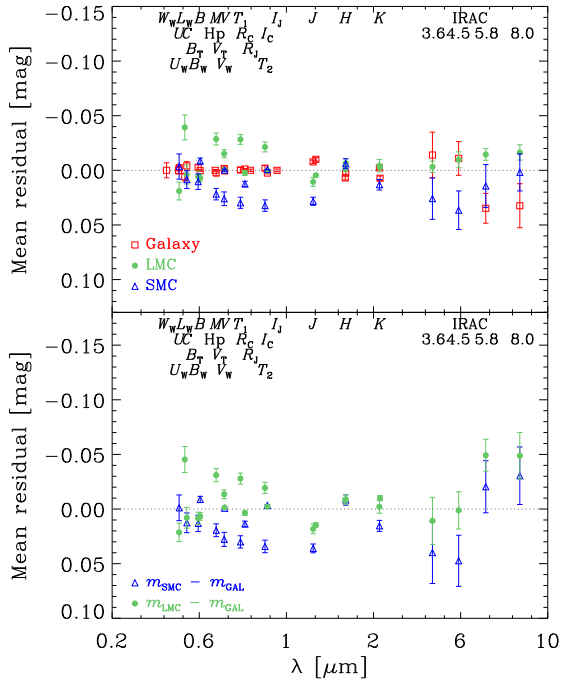}
\caption{{\em Top}: Actual mean fit residuals as a function of wavelength for the Galaxy (red open squares), LMC (green filled circles), and SMC (blue open triangles). The vertical error bars are the uncertainties in the mean. {\em Bottom}: Relative differences of the residuals of the LMC and SMC with respect to the Galaxy.}
\label{fig:residuals_lambda}
\end{figure}

We start looking for signs of composition effects in the fit residuals, but it is again important to remember that most of the effects of metallicity are automatically projected onto the individual parameters of the Cepheids ($\bar{\tau}$, $\bar{\rho}$, $E(B-V)$, $\mu$). We discuss metallicity-dependent signatures in these parameters in Section~\ref{sec:metal_individual}. The fit residuals contain only the manifestations of metallicity that are orthogonal to changes in distance, radius, reddening and temperature. They can be identified by examining the residuals of the fits as a function of wavelength and metallicity. In order to get an idea of the expected signal, we repeated the procedure from Section~\ref{sec:metal_pred}, but instead of integrating over the model atmosphere, we convolved the $m_{\mh=-0.5}-m_{\mh=0.0}$ ``spectra'' and $\beta$ constructed from the \citet{castelli04} model atmospheres, and the \citet{cardelli89} $\rv=3.3$ extinction law with tophat functions having the central wavelength and width of each of our filters, and performed the fit in Equation~(\ref{eq:metal_model}) for these synthetic bandpasses. Individual filters were weighted by the number of measurements for each filter in our dataset (see Table~\ref{tab:phot}). We did not include the Hipparcos and Tycho filters as there are no LMC or SMC data in these passbands. The residuals of this fit as a function wavelength are shown in Figure~\ref{fig:metal_residuals}. With these changes in the weight at any given wavelength, the shifts in stellar parameters are $(\Delta \mu, \Delta E(B-V), \Delta \bar{\tau}) = (-0.199, -0.208, -0.064)$\,dex$^{-1}$, broadly consistent with the uniformly weighted theoretical comparison in Section~\ref{sec:metal_pred}. We see that changing the metallicity of the stellar atmosphere by $0.5$\,dex is largely compensated for by changes in the distance, reddening and temperature, leading to residuals of $\lesssim 0.05$\,mag in the majority of the filters. That the residuals are so small despite fitting $29$ bands from $0.3\,\mu$m to $8\,\mu$m helps to explain why pinning down the effects of composition is so difficult!

In Figure~\ref{fig:residuals_lambda} we show the actual fit residuals as a function of wavelength $\lambda$ for each of the three galaxies. In order to facilitate comparison to our theoretical model in Figure~\ref{fig:metal_residuals}, the bottom panel shows differences in the residuals of the LMC and SMC from the Galaxy. We see that our results are qualitatively similar to Figure~\ref{fig:metal_residuals}. In particular, for $m_{\rm SMC}-m_{\rm Galaxy}$ we see that for $\lambda \lesssim 0.5\,\mu$m the residuals are negative and decreasing with decreasing $\lambda$, are approximately constant and positive between $0.6\,\mu$m and $1.0\,\mu$m, followed by negative residuals in the $H$ and $K$ bands. The data for the LMC do not show such a clear trend, but the LMC metallicity is closer to the Galaxy and should show smaller differences than the SMC. Although the residuals here differ in detail from Figure~\ref{fig:metal_residuals} (because of the mismatch between theoretical and actual profiles of $\mathscr{R}$ and $\beta$ (Figures~\ref{fig:beta} and \ref{fig:extinction}) and weighting by the data), we see similar patterns. In particular, when comparing metal-poor to metal-rich we see negative residuals bluewards of $\sim 0.45\,\mu$m, slightly negative and flat residuals between $0.45\,\mu$m and $1.0\,\mu$m, and negative residuals in the near-IR.

\citet{scowcroft11} suggest that differences between the Spitzer [3.6] and [4.5] bands are a potential metallicity indicator because the [4.5] band includes the CO bandhead near $5\,\mu$m. In Figure~\ref{fig:residuals_lambda} we see that there are indeed substantial differences between the Galaxy and the LMC and SMC. The differences are smallest for the [3.6] band and increase for the [4.5], [5.8], [8.0] bands. The pattern of these differences does not agree with what we predicted in Figure~\ref{fig:metal_residuals}---not only is the offset larger but the sign has reversed. This could be another symptom of the problems in the \citet{castelli04} model molecular opacities we discussed in Section~\ref{sec:global}, but it could also be created by different calibrations of the IRAC data, as the data for each galaxy are from different groups of authors. Nonetheless, we conclude that the pattern of the residuals is likely evidence for composition effects. Quantitatively interpreting Figure~\ref{fig:residuals_lambda} is non-trivial, not because of any uncertainties in the model, but because of the effects created by the different weightings of the filters in the Galactic, LMC, and SMC data sets.

\begin{figure}
\plotone{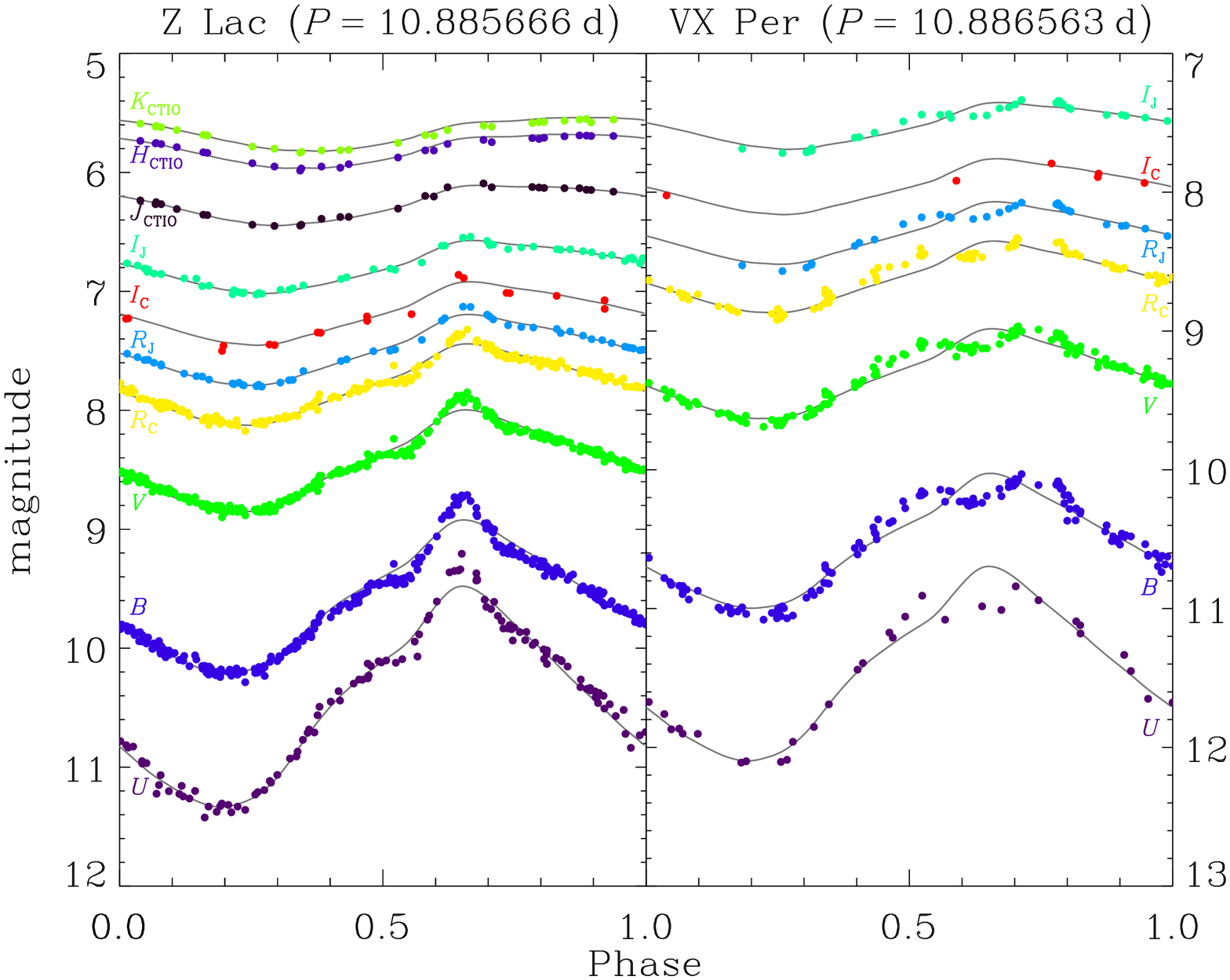}
\caption{Light curves of two Galactic Cepheids Z~Lac and VX~Per with essentially identical periods, $\Delta P/P \simeq 8\times 10^{-5}$ and metallicities \citep[$\feh=0.01$ and $-0.05$ for Z~Lac and VX~Per, respectively,][]{andrievsky02}. The light curve shapes differ due to the well-known resonance between the fundamental mode and the second overtone.}
\label{fig:lc_compare}
\end{figure}

Finally, we return to the phase dependent structures visible in Figure~\ref{fig:residuals} that we noted in Section~\ref{sec:individual}. Some of the residuals are due to limitations in the underlying ansatz that the light curve structure is solely determined by $P$, $A^2$, $\bar{\tau}$ and $\bar{\rho}$ (see Section~\ref{sec:phys_model}), but some of the residuals are caused by physical differences between otherwise almost identical Cepheids. Figure~\ref{fig:lc_compare} shows the light curves of two Galactic Cepheids whose relative period difference is only $\Delta P/P = 8\times 10^{-5}$. There is a significant difference between phases $0.5$ and $0.7$, where Z~Lac ($\feh=0.01$, \citealp{andrievsky02}) shows a bump, while VX~Per ($\feh=-0.05$) exhibits a dip. This is a demonstration of the well-known strong resonance between the fundamental mode and second overtone for $10\le P \lesssim 13$\,days \citep[e.g.][]{simonlee81,antonello96}. This resonance is responsible for most of the structure seen in the residuals in Figure~\ref{fig:residuals}, and for the ``jump'' in the upper envelope of amplitudes at $\sim 13$\,days in Figure~\ref{fig:amplitude}. Similar (but smaller) discrepancies, especially in the depth of the dip preceding the rise to the maximum are also seen for longer periods, but we could not find any discernible pattern. While these resonance effects introduce noise, they should not significantly affect the mean properties of the Cepheids or the global variables as the fit residuals are quite small. We thus do not discuss this issue in more detail.

\subsection{Metallicity Effects on the Global Parameters}
\label{sec:metal_global}

Additional signs of the effects of composition can be searched for in the global parameters of our fit. Metallicity effects can modify the $PL$ relations of the individual galaxies, where the zero point comparison is degenerate with distance uncertainties, but the slopes are not. The $PL$ relations found by linear least squares fits to the mean extinction- and distance-corrected $V$-band magnitudes are
\begin{subequations}
\label{eqs:pl}
\begin{eqnarray}
M_V^{\rm SMC}& =& (-4.53 \pm 0.11) - (2.76 \pm 0.30)\,\log\!\left(\!\frac{P}{10\,{\rm d}}\!\right),\\
M_V^{\rm LMC}& =& (-4.16 \pm 0.05) - (2.70 \pm 0.15)\,\log\!\left(\!\frac{P}{10\,{\rm d}}\!\right),\\
M_V^{\rm Galaxy}&=& (-3.93 \pm 0.07) - (2.35 \pm 0.29)\,\log\!\left(\!\frac{P}{10\,{\rm d}}\!\right),
\end{eqnarray}
\end{subequations}
where $M_V$ is the absolute magnitude in the $V$ band. All stars within each galaxy had a uniform weight in the fit and the error estimates from the fits are very close to those obtained by bootstrap resampling. The uncertainties do not include covariances in our model, unlike the $PL$ relation in Equation~(\ref{eq:pl_mes_radtem}). The rms scatter of the Cepheids about these mean relations are $\sigma^{\rm SMC} = 0.43$, $\sigma^{\rm LMC} = 0.35$, and $\sigma^{\rm Galaxy} = 0.43$\,mag. However, we find that the exact values of the slopes and to a lesser extent the zero points depend sensitively on the statistical methods used for performing the linear regression.

There is a significant difference between the zero points in the sense that SMC Cepheids are brighter than LMC Cepheids which are in turn brighter than Cepheids in the Galaxy. This is in line with theoretical inferences from atmosphere and pulsational models, which predict that metal poor Cepheids are brighter. Arrows in Figure~\ref{fig:pl} indicate the direction of zero-point shifts in the $V$-band $PL$ relation due to metallicity as predicted in Section~\ref{sec:metal_pred}. We see that the implied shifts are compatible with what is observed. This issue is discussed in greater detail in Section~\ref{sec:metal_individual} and in Figure~\ref{fig:deltar_deltat}. Another issue to consider for differences between the galaxies are our various distance priors. We find that the mean distance modulus of SMC Cepheids is $\mu_{\rm SMC} = 18.95 \pm 0.02$\,mag with a scatter of $0.11$ mag, which is very close to our prior of $\mu_{\rm SMC} = 18.90$\,mag with a width of $0.10$\,mag. Given the LMC distance modulus of $\mu_{\rm LMC} = 18.50$\,mag, the difference is $\mu_{\rm SMC} - \mu_{\rm LMC}  = 0.45 \pm 0.02$\,mag. This is in close agreement with $0.44 \pm 0.05$\,mag determined by \citet{cioni00} from a large sample of TRGB stars. \citet{storm11b} found $\mu_{\rm SMC} - \mu_{\rm LMC}  = 0.47 \pm 0.15$\,mag using the infrared surface brightness method. For the Galactic Cepheids, where we did not use any distance priors, the distances are constrained by the implicit Baade-Wesselink aspects of our model, so it is interesting that we see only a marginal zero point shift between the Galaxy and the LMC as we also saw in our earlier comparison to \citet{storm11a,storm11b}. Our sample contains two Galactic Cepheids with parallax measurement and we found in Section~\ref{sec:individual} that our distances are in good agreement with parallax measurements. We see in Equations~(\ref{eqs:pl}) that the slopes agree within their uncertainties with some evidence that the slope of Galactic Cepheids is somewhat shallower. However, the worry here is that recovering the $PL$ from Galactic Cepheids with widely varying distances and reddenings is quite difficult, and our sample has no Galactic Cepheids with $P>50$\,days. Adding both short and long period stars would help us to better characterize the differences in slopes. In summary, there are hints of differences in the $PL$ relation slopes, but this is not the best probe for composition effects given the nature of the data.

\subsection{Metallicity Effects on the Individual Cepheid Parameters}
\label{sec:metal_individual}

\begin{figure*}
\plotone{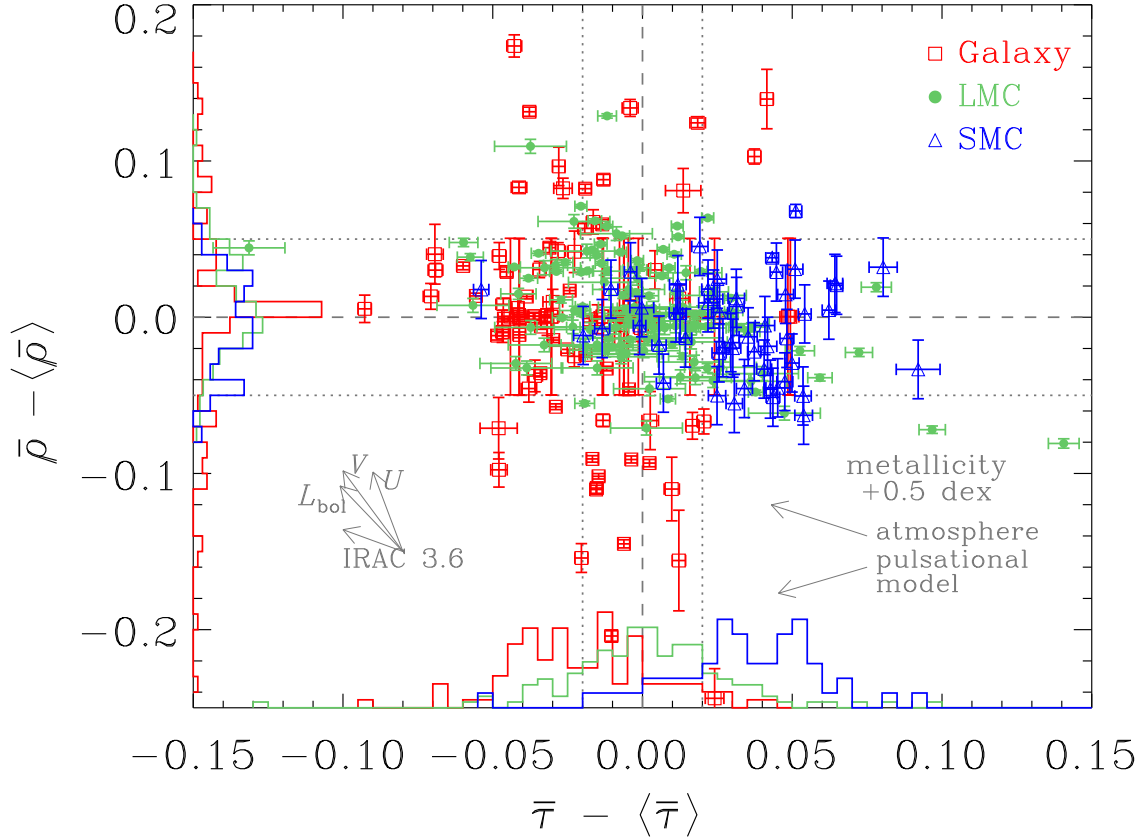}
\caption{Residuals with respect to the radius and temperature priors. The meaning of the symbols is the same as in Figure~\ref{fig:radius}. The individual priors and their widths are shown with grey dashed lines and grey dotted lines, respectively. The histograms on the bottom and left sides of the plot show distributions of $\bar{\rho}-\langle\bar{\rho}\rangle$ and $\bar{\tau}-\langle\bar{\tau}\rangle$ for each of the galaxies normalized by the number of Cepheids in each galaxy. The grey arrows in the lower left corner indicate the directions in which changes in $\bar{\rho}$ and $\bar{\tau}$ exactly cancel each other and there is no net change of $m_i$ for the $U$, $V$, and $[3.6]$ bands as well as for the bolometric luminosity. The arrows above the temperature histogram show the expected change of $\bar{\tau}$ and $\bar{\rho}$ as a result of a metallicity increase of $0.5$\,dex due to changes in the atmospheric metallicity at fixed temperature (upper arrow) and changes in the physical properties of the star at fixed period based on pulsational models (lower arrow).}
\label{fig:deltar_deltat}
\end{figure*}

As was shown in Section~\ref{sec:disc_residuals}, most metallicity effects on the stellar atmospheres will be absorbed into changes in distance/radius, temperature and extinction. There are also the direct changes in the mean $T$ and $R$ at fixed $P$, which also cause shifts in $E(B-V)$ and $\mu$ when a single $PL$ relation ($\langle \bar{\rho}(P)\rangle$ and $\langle \bar{\tau}(P)\rangle$) is used. Without independent extinction scales, we have no means of confirming composition effects on extinction inferences given the very different intrinsic extinctions of the three galaxies seen in Figure~\ref{fig:tem_ext}. In Figure~\ref{fig:deltar_deltat} we show the residuals of the Cepheid radii and temperatures from their respective updated priors, $\bar{\rho}-\langle\bar{\rho}\rangle$ and $\bar{\tau}-\langle\bar{\tau}\rangle$ (Eq.~\ref{eqs:rhotau_per_mes}). In this way, most of the period dependence of radius and temperature are removed and we can search for systematic differences between the individual galaxies. As a reminder, based on the discussions in Sections~\ref{sec:metal_pred} and \ref{sec:disc_residuals} we expect shifts of order $(\Delta \bar{\rho}, \Delta \bar{\tau}) \simeq (0.04, -0.06)$\,dex$^{-1}$ from the changes in the stellar atmospheres and $(\Delta \bar{\rho}, \Delta \bar{\tau}) \simeq (-0.03, -0.02)$\,dex$^{-1}$ from the shifts in the pulsational models.

Looking at the histogram of the radii, we see that SMC Cepheid radii have large errors, while the LMC and Galactic Cepheids generally have much smaller errors. There is no noticeable shift in mean radii between the individual galaxies. Unfortunately, the three galaxies have very different distance priors and available numbers of radial velocities. Since radius estimates will be strongly correlated with distance, the differences in the radius residuals are dominated by these systematic issues. This is particularly visible for the SMC, where the lack of radial velocity measurements means that the radius determinations are dominated by the prior on $\bar{\rho}$ and the Gaussian prior on the SMC distance. This explains the relatively large errors on the SMC Cepheid radii. In the LMC, the distances are fixed and any variations in the mean luminosity are thus absorbed by the radius. Hence we get small errors in the radius as the mean luminosity is well determined and any variations in luminosity are absorbed into a tightly constrained radius. In essence, these three galaxies are not ideal for examining metallicity effects on radius. Moreover, we see that the combined effects of shifts in the period--radius relation and the changes in the radius estimate created by changes in the stellar atmospheres tend to cancel.

The apparent temperature distributions are a more promising area to look for differences because the broad wavelength baselines and well constrained extinctions lead to well-constrained temperature estimates up to the shifts created by changes in the stellar atmospheres (Fig.~\ref{fig:tem_ext}). While the distributions are broad, we see a clear trend that Galactic Cepheids are generally cooler than the LMC Cepheids and the SMC Cepheids are generally hotter. Errors on the temperatures are comparable for all galaxies and are smaller than the width of the temperature prior. These relative shifts in temperature are robust and do not depend on the choice of the priors that fix the temperature scale. The simplest explanation of this pattern is the effect of composition as outlined in Section~\ref{sec:metal_pred}.

Figure~\ref{fig:deltar_deltat} also shows directions in the radius--temperature space along which there are no changes in luminosity for several filters as well as for the bolometric luminosity. There is considerably more freedom for the fit along these directions because the luminosities are very well determined. We also show the shifts in temperature due to changes in the stellar temperature at fixed period and from the changes in the model atmosphere at fixed temperature. These two effects have the same sign, leading to a net effect of order $\Delta \bar{\tau} \simeq 0.12$\,dex$^{-1}$. This broadly agrees with the observed shifts, given that the metallicity difference between the SMC and the LMC is $\Delta\feh \sim 0.4$ and between the LMC and the Galaxy is $\Delta \feh \sim 0.3$ \citep{romaniello08}.

Finally, it is known that the light curve amplitudes of SMC Cepheids are smaller than those of their LMC counterparts and that this effect might be related to the metallicity differences \citep{vangenderen78,paczynski00,klagyivik09,szabados11}. This is apparent in Figure~\ref{fig:amplitude}, which confirms that SMC Cepheids with periods between $13$ and $30$ days do not reach amplitudes as high as in the LMC and Galaxy. There does not seem to be any discernible difference in amplitudes between the LMC and the Galaxy. Unfortunately, the scatter is such that the statistical pattern is not useful for individual Cepheids.

\subsection{The Importance of Higher-Order Terms}
\label{sec:second_order}

\begin{figure}
\plotone{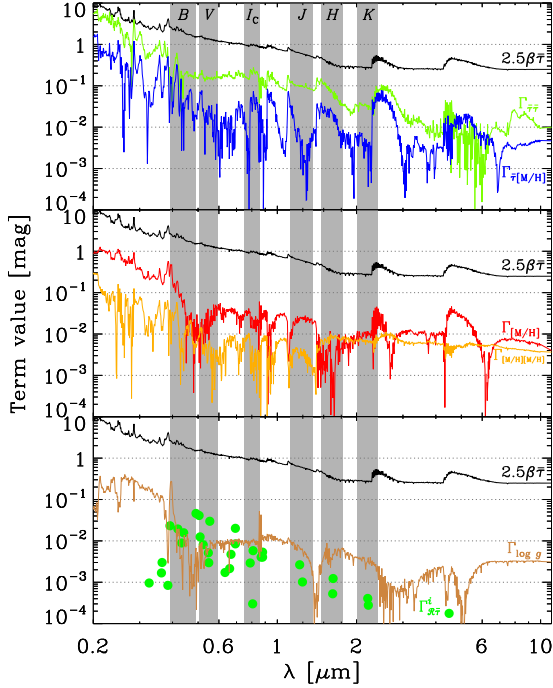}
\caption{Absolute values of the higher-order terms compared to the leading temperature term $2.5\beta\bar{\tau}$. The top panel compares it to the typical scale of the second temperature derivative $\Gamma_{\bar{\tau}\bar{\tau}}$ and the temperature/metallicity cross-derivative $\Gamma_{\bar{\tau}\mh}$. The middle panel compares it to the typical scale of the first ($\Gamma_{\mh}$) and second ($\Gamma_{\mh\mh}$) derivatives of $\overline{M}$ with respect to metallicity $\mh$. The bottom panel compares $2.5\beta\bar{\tau}$ to the derivative $\Gamma_{\log g}$ with respect to surface gravity $\log g$ and to the change of the reddening coefficients in filter $i$ with temperature $\Gamma_{\mathscr{R}\bar{\tau}}^i$. The grey vertical bands show the pass bands of several common filters.}
\label{fig:second_order}
\end{figure}

The results of Sections~\ref{sec:metal_pred}--\ref{sec:metal_individual} strongly suggest that adding a metallicity term to our model is a next logical step. We should, however, consider whether any other higher-order terms are equally important. These can come in the form of additional terms in Equation~(\ref{eq:phys_model_m}) or in more complex period--radius and period--temperature priors (Eq.~[\ref{eqs:rhotau_per}]). Here we focus on the first question. The typical value of the first-order metallicity correction, essentially $\partial \overline{M} / \partial \mh$, is
\beq
\Gamma_{\mh} = -2.5\frac{\partial \log F }{\partial \mh}\Delta\mh = \gamma\Delta\mh,
\label{eq:gamma_mh}
\eeq
where $\gamma = -2.5\partial \log F/\partial \mh$ is the ``standard'' metallicity correction vector and $\Delta\mh = 0.5$\,dex is the typical spread in metallicity of Cepheid samples. Another missing first-order term is the surface gravity correction
\beq
\Gamma_{\log g} = -2.5 \frac{\partial \log F}{\partial \log g} \Delta \log g,
\eeq
where $\Delta\log g = 0.5$ is the typical spread in Cepheid surface gravities. Possible second-order terms in temperature and metallicity are the second derivative of the spectrum with respect to temperature (i.e.\ the term proportional to $\partial \beta/\partial \bar{\tau}$)
\beq
\Gamma_{\bar{\tau}\bar{\tau}} = -2.5\frac{1}{2}\frac{\partial^2 \log F}{\partial \bar{\tau}^2}\bar{\tau}^2,
\eeq
the metallicity dependence of the temperature vector (the term proportional to $\partial \beta/\partial \mh$)
\beq
\Gamma_{\bar{\tau}\mh} = -2.5\frac{\partial^2\log F}{\partial\bar{\tau}\partial\mh}\bar{\tau}\Delta\mh,
\eeq
and the second derivative of the spectrum with respect to metallicity
\beq
\Gamma_{\mh\mh} = -2.5 \frac{1}{2}\frac{\partial^2 \log F}{\partial \mh^2}(\Delta\mh)^2.
\eeq
Here, $\bar{\tau} = 0.1$ is the typical spread in Cepheid temperatures for $10 \le P \le 100$\,days (Fig.~\ref{fig:temperature}).

Finally, the reddening coefficients will also change with the temperature of the star, because the flux and the corresponding wavelength-weighted reddening will shift within the filter pass band. We thus consider a reddening-temperature cross-term
\beq
\Gamma_{\mathscr{R}\bar{\tau}}^i = \frac{\partial \ri(\bar{\tau})}{\partial \bar{\tau}}\bar{\tau}\Delta E(B-V),
\label{eq:gamma_rt}
\eeq
where $\Delta E(B-V)=0.2$\,mag is the assumed spread in reddenings, and $\ri(\bar{\tau})$ is defined as
\beq
\ri(\bar{\tau}) = \frac{-2.5}{\Delta E(B-V)}\log \left(\frac{\int_0^\infty \lambda^{-1}\Pi_i(\lambda)F_\nu(\lambda,\bar{\tau})e^{-\mathscr{T}_\lambda}\intd \lambda     }{\int_0^\infty \lambda^{-1}\Pi_i(\lambda)F_\nu(\lambda, \bar{\tau})\intd \lambda}\right).
\eeq
Here, $\Pi_i(\lambda)$ is the transmission function of filter $i$, which we assume to be a top hat for simplicity, $F_\nu$ is the energy flux per unit frequency, and $\mathscr{T}_\lambda$ is the optical depth to dust at wavelength $\lambda$ defined as $\mathscr{R}_\lambda \Delta E(B-V) = -2.5\log e^{-\mathscr{T}_\lambda}$.

In Figure~\ref{fig:second_order} we compare these terms to the first-order temperature term $-2.5\beta\bar{\tau}$. Here, we obtained our results by evaluating Equations~(\ref{eq:gamma_mh})--(\ref{eq:gamma_rt}) using the theoretical model atmospheres of \citet{castelli04} with $T=5250$\,K\footnote{Here we choose a different value for the temperature than in the rest of the paper ($T_0 = 5400$\,K) in order to avoid interpolating in the grid of \citet{castelli04} model atmospheres. We would get essentially the same results with calculation centered on $T_0=5400$\,K.}  and $\log g=1.5$. The first point to note is that the first order temperature term we use for our models is clearly the dominant term except at very short $\lambda$, which is why our simple ansatz works so well. We see that the dominant higher-order term is the second derivative with respect to temperature $\Gamma_{\bar{\tau}\bar{\tau}}$, which is a $10$--$20\%$ correction over most of the wavelength range considered. Its importance grows in the near-UV, where it can be almost as important as the first-order temperature term. The metallicity dependence of the zero point $\Gamma_{\mh}$ and the metallicity-temperature cross-term $\Gamma_{\bar{\tau}\mh}$ are modestly smaller, typically being $\sim\! 3\%$ corrections except in the UV. The second metallicity derivative $\Gamma_{\mh\mh}$, the reddening-temperature cross-term $\Gamma_{\mathscr{R}\bar{\tau}}$ and the zero point dependence on surface gravity $\Gamma_{\log g}$ are generally small but represent $\sim\! 1\%$ corrections in the optical bands. It is interesting to note that some of these terms may also contribute to variance in the light curve shapes, since any higher order terms involving the temperature or first order terms in $\log g$ must also modify the light curves.

\section{Conclusions}
\label{sec:conclusions}

In this work, we present a method for self-consistently determining distances, reddenings, radii, and temperatures of individual Cepheids along with the mean reddening law and phase-dependent variations of radius and temperature as a function of period. Our approach is in some senses a global version of the Baade-Wesselink method and it is in some senses an implementation of the ideas of \citet{freedman10b}. Our method provides statistically consistent error estimates for all parameters of the model (Section~\ref{sec:model_data}). We fit our physical model to $\sim\! 5,000$ radial velocity and $\sim\! 177,000$ magnitude measurements in 29 bandpasses covering wavelength range from $0.3\,\mu$m to $8\,\mu$m for $287$ Cepheids from the Galaxy, LMC, and SMC. With only four to six variables per Cepheid, depending on the availability of velocity data and the amount of freedom in the distance estimates, we fit the complete phase dependent data set with a magnitude scatter of $0.051$\,mag and radial velocity scatter of $3.5$\,km\,s$^{-1}$, as shown in Figures~\ref{fig:residuals} and \ref{fig:residuals_irac}. 

Our templates are built to span the period range $10 \le P \le 100$\,days and can be used to produce model light curves or $PL$ relations for any of the $29$ bands. Predicting the light curve for an uncalibrated band is straightforward provided that the value of a single parameter $\beta_i$, the logarithmic flux derivative with respect to the temperature (Eq.~[\ref{eq:beta_general}], Figure~\ref{fig:beta}), can be calculated or estimated for the new filter. The templates do not provide perfect fits, particularly for periods between $10$ and $13$\,days where strong resonance effects are present. The mean phase radius and temperature variations we obtain match the Hertzsprung progression in the longer period stars.

We obtain a spectral energy distribution (Figure~\ref{fig:mbar}) and the logarithmic flux derivative with respect to the temperature $\beta_i$ (Figure~\ref{fig:beta}) for a ``mean'' Cepheid. When we compare our results with estimates based on the black body law or theoretical stellar atmosphere models, we completely rule out black body models. We find better agreement with the \citet{castelli04} model atmospheres, but also strong evidence for problems in the atmosphere models \citep[see][]{fremaux06,hauschildt99a,hauschildt99b}. The coefficients $\beta_i$, which correspond to the logarithmic derivative of the spectrum with respect to temperature, deviate from the black body law due to metal line blanketing and molecular opacities again in general agreement with theoretical atmosphere models. We are even able to resolve several spectral features. However, the theoretical atmosphere models are inadequate in describing the infrared spectrum, probably indicating problems with the molecular opacities in the models. There are no difficulties separating temperature and reddening in our models (Figures~\ref{fig:beta_r} and \ref{fig:tem_ext}) up to degeneracies created by the effect of composition.

We obtain a very high precision measurement of the mean extinction law for $29$ filters from $0.3\,\mu$m to $8\,\mu$m. The mean reddening law of Cepheids in our sample (Figure~\ref{fig:extinction}) shows departures from the $\mathscr{R}_V = 3.3$ \citet{cardelli89} law usually assumed for Cepheids. In part this is due to averaging over the three galaxies, but it clearly illustrates that extinction laws need to be considered in efforts to increase the precision of Cepheid distances. This is true despite the trend towards infrared wavelengths, because the near-IR $\mathscr{R}$ values have larger fractional uncertainties. Our results for the extinction curve also let us directly evaluate the Wessenheit factor $\mathcal{R}_{VI_{\rm C}} = \mathscr{R}_V/(\mathscr{R}_V-\mathscr{R}_{I_{\rm C}})$ commonly used in Cepheid studies \citep[e.g.][]{madore82,madore91,madore_freedman09,ngeow05}. For $\rv\equiv 3.3$ $(3.1)$ we find that $\mathcal{R}_{VI_{\rm C}}= 2.521 \pm 0.005$ ($2.475\pm 0.005$), where the error comes only from the determination of $\mathscr{R}_{I_{\rm C}}$, because we hold $\mathscr{R}_V$ fixed. The result for $\rv\equiv 3.3$ is significantly different from the usually assumed $2.45$ \citep{wozniak96,freedman01} and introduces a distance shift of about $\Delta\mu \simeq 0.01$\,mag for a Cepheid with $E(V-I_{\rm C})=0.10$\,mag. More generally, we can calculate the Wessenheit factors for any combination of filters and we can correctly include the mutual covariances between the filters in the error estimate. For example, \citet{riess11a} assumed the optical/near-IR Wessenheit factor of  $\mathcal{R}_{HVI_{\rm C}} = \mathscr{R}_H/(\rv-\mathscr{R}_{I_{\rm C}}) = 0.410$ to reduce the uncertainty on the measurement of the Hubble constant. From our analysis, we find $\mathcal{R}_{HVI_{\rm C}} = 0.322\pm 0.006$ ($0.290 \pm 0.006$) again at fixed $\rv\equiv 3.3$ ($3.1$), inconsistent with the assumptions by \citet{riess11a} in either case. This illustrates that extinction laws need to be better understood to correctly measure precise distances in the universe, particularly since these estimates still all assume $\rv\equiv 3.3$ (or $3.1$) and so underestimate the uncertainties.

We find weak evidence of metallicity effects in the fit residuals. They are weak because much of the metallicity effect can be mimicked by shifts in the distance, reddening, and temperature of a Cepheid (Fig.~\ref{fig:msmc-mgal}). The effect of metallicity on stellar atmospheres  that is orthogonal to these parameters only leads to residuals $\lesssim 0.05$\,mag for a metallicity difference of $0.5$\,dex (Fig.~\ref{fig:metal_residuals}) which is approaching the regime where we may also need to consider absolute calibration differences arising from the heterogeneous data sets. The clearest effect of metallicity is seen in the temperature distributions of the Cepheids, because with $29$ bands the temperature is well constrained independently of the extinction and distance/radius. The SMC Cepheid temperatures are typically higher than those of the LMC, which are in turn higher than for the Galaxy, although the distributions overlap. This can be explained as a combination of projecting the metallicity effects on stellar atmospheres into changes in temperature as well as shifts in the period--temperature relation. The shifts we observe are roughly consistent with expectations. There may also be small shifts in radius (Fig.~\ref{fig:deltar_deltat}), but for our sample this is difficult to disentangle from the effects of distance priors and would be better addressed given samples of Cepheids truly at a common distance and with a range of metallicities. Similarly, we find small differences in the zero-points of our $PL$ relations, which can also be attributed to the projection of metallicity effects, but this has many of the same limitations as the search for shifts in radius. We do find shifts in the $PL$ relation slopes, but our limited period range ($P \ge 10$\,days) is not well suited for a robust investigation of slopes.

In Appendix~\ref{app:opti} we outline a procedure to select optimal filter sets for disentangling distance, temperature, reddening and metallicity or to maximize or minimize metallicity effects when one is concerned solely with distance determination. We find that sets of filters spanning the longest wavelength range possible, for example $UV$[4.5], yield the best fit results for parameter estimation. Such filter combinations produce parameter uncertainties $0.5$ to $1$ order of magnitude smaller than the $BVI_{\rm C}$ combination originally planned for the HST Key Project. We find that the maximum effect on distance determination due to metallicity can be expected for filter combinations like $UBV$ and $BVR_{\rm C}$. On the other hand, the slope of the dependence of distance on metallicity can be reduced to less than $10^{-3}$\,mag\,dex$^{-1}$ for filter combinations like $VHK$ or $I_{\rm C}$[3.6][4.5]. However, some filter combinations with infrared filters are quite sensitive to metallicity (e.g.\ $UVJ$). For these metrics, there are no particular benefits to the mid-IR over the near-IR, so Cepheid studies in the JWST era might better focus on the near-IR where Cepheids are brighter and the PSF will be more compact (to reduce systematic errors from crowding and blending). Also, mid-IR Cepheid measurements might be affected by circumstellar dust \citep[see, e.g.][]{kervella06,gallenne11}. These statements assume the validity of the \citet{castelli04} model atmospheres, which we did find are partly inconsistent with the Cepheid data, so filter choices should be evaluated for a broader range of atmosphere models.

Our goal in this paper was simply to carry out a complete analysis with the simplest possible model. That the model does so well helps to explain why efforts to identify higher-order corrections such as metallicity are so challenging. At least based on the \citet{castelli04} atmosphere models, the next most important terms are the second derivative of the zero point with respect to the temperature followed by the first-order metallicity correction and the metallicity correction to the logarithmic flux derivative with respect to the temperature. The second derivative of the zero point with respect to metallicity, the first order surface gravity derivative  and the dependence of the extinction correction $\ri$ on temperature are relatively unimportant unless working in the UV. There is also a clear metallicity dependence to the mean temperature, which also indicates the need for higher-order terms in the period-radius and period-temperature relations. These estimates of higher-order effects are again based on the theoretical atmosphere models of \citet{castelli04}.

There is clearly a broad range of possible future extensions. Adding additional data or galaxies is trivial, as is extending to shorter periods. It is also possible to fit for signatures of a companion to the Cepheid which will distort the SED. The virtue of our approach, particularly in a cosmological setting, is that it forces a correct use of prior assumptions in the parameter estimation and uncertainties, allows the data to overrule those assumptions, if necessary, and produces error estimates incorporating the full uncertainties of the entire model. Improved stellar atmosphere models would lead to better priors, although this is less of a concern given the demonstrated ability of the data to constrain the SED of Cepheids and their dependence on temperature. Particularly with better atmosphere models it should be possible to relax the assumptions about extinction ($\rv\equiv 3.3$ and mean $\langle E(B-V)\rangle =0.147$\,mag for the LMC based on \citealt{udalski99}) and put the Cepheids on their own absolute extinction scale. This would particularly help in finding the expected environmental dependencies of the extinction law and examining their effects on distance uncertainties.

The final important lesson from this study, as also previously stated by \citet{freedman10b}, is that there is no need to obtain a light curve in order to add almost all the information available from observing at a new wavelength. Given one band with enough data to determine the period and phase of the Cepheid at the time of any observation at another band, a single observation in this new band suffices because the model correctly includes all the phase correlations between the bands. In practice, it would be wise to obtain at least two observations so as to have an internal check on the results. This removes any major barrier to systematically using the short wavelength filters that provide the greatest leverage for the control and measurement of systematic errors. Much of the drive towards longer wavelengths has been that with lower amplitudes the light curves can be more poorly sampled to still yield an accurate mean magnitude.  With our models, {\it any} band can be poorly sampled provided its phase is well-determined. In some sense, this is a multi-band version of the method used by \citet{gerke11}, where Cepheids in M81 were identified and phased with ground based images, and then calibrated using single epochs of HST data combined with the \citet{stetson96} template models. Furthermore, when fitting an individual Cepheid, all the bands can be simultaneously co-phased by using our models and measuring the goodness of fit to all the data as a function of period.

\acknowledgements

We thank Vicky Scowcroft, Chow-Choong Ngeow and Nicolas Nardetto for assistence in using their datasets. We thank Marc Pinsonneault and Todd Thompson for fruitful discussions, Lucas Macri and Kris Stanek for critical reading of the manuscript, and the referee, Barry Madore, for his comments. O.P.\ is supported in part by an Alfred P.\ Sloan Foundation Fellowship to T.A.\ Thompson. C.S.K.\ is supported by NSF grant AST-0908816.

\appendix

\section{Filter Optimization for Cepheids}
\label{app:opti}

We can use our approach for estimating the effect of metallicity on individual filters presented in Section~\ref{sec:metal_pred} to design a method to evaluate and optimise filter choices for different metrics such as minimizing the overall error in the physical parameters or to yield the smallest or largest possible signal of metallicity. The method is general and related to the ``dark energy figure of merit'' \citep{albrecht06,wang08}.

Assume that we obtain a set of mean magnitudes of a Cepheid  $\mathbf{m}$, where each component of $\mathbf{m}=(m_1,\ldots,m_{\nf})$ is a mean magnitude in one of the $\nf$ distinct filters. We need to determine the distance $\mu$, reddening $E(B-V)$, temperature $\tau$, and metallicity $\mh$ by fitting
\beq
\mathbf{m} = \mu \mathbf{1} + \mathscr{R}E(B-V) -2.5\bm{\beta}\tau + \mh\bm{\gamma}.
\label{eq:app_func}
\eeq
Here, $\mathbf{1}$ is a vector with all components equal to unity, $\mathscr{R}$ is the reddening vector, $\bm{\beta}$ is the logarithmic change of flux with temperature, and $\bm{\gamma}$ is the metallicity correction vector. Without radial velocities, the radius is degenerate with distance and we merge these two quantities into $\mu$. In order to keep the analysis simple, we use the $\mathscr{R}_V=3.3$ \citet{cardelli89} reddening law for $\mathscr{R}$ and we calculate $\bm{\beta}$ from the \citet{castelli04} model atmospheres for $T_0 = 5400$\,K, $\log g = 1.5$, and $\mh=0.0$. The metallicity vector $\bm{\gamma}$ is defined in Equation~(\ref{eq:gamma_mh}) from Section~\ref{sec:second_order}. We then convolved the models with a simple top hat model for the filters to compute the values of $\mathscr{R}$, $\bm{\beta}$, and $\bm{\gamma}$ for each filter.

If we optimize the fit implied by Equation~(\ref{eq:app_func}) to any data set, the best fit parameters defined by minimizing the $\chi^2$ fit statistics are
\beq
   \left( \begin{array}{c} \mu \\ E(B-V)\\ \tau\\ \mh  \end{array} \right) = \frac{\mathbf{C}^{-1}}{\sigma^2}\left(\begin{array}{c}\mathbf{m}\cdot\mathbf{1} \\ \mathbf{m} \cdot \mathscr{R} \\ -2.5\mathbf{m}\cdot\bm{\beta} \\ \mathbf{m}\cdot\bm{\gamma}   \end{array}\right),
\eeq
where $\sigma$ represents the measurement errors. $\mathbf{C}$ is the covariance matrix of the basis functions $\mathbf{1}$, $\mathscr{R}$, $-2.5\bm{\beta}$, and $\bm{\gamma}$
\beq
\mathbf{C} = \frac{1}{\sigma^2}\left( \begin{array}{cccc} 
\mathbf{1}^2 & \mathbf{1}\cdot \bm{\mathscr{R}} & -2.5\mathbf{1}\cdot\bm{\beta} & \mathbf{1}\cdot\bm{\gamma} \\
\mathbf{1}\cdot\mathscr{R} & \mathscr{R}^2 & -2.5\mathscr{R}\cdot \bm{\beta} & \mathscr{R}\cdot\bm{\gamma} \\
-2.5\mathbf{1}\cdot\bm{\beta} & -2.5\bm{\mathscr{R}}\cdot\bm{\beta} &(2.5\bm{\beta})^2 & -2.5\bm{\beta}\cdot\bm{\gamma} \\
\mathbf{1}\cdot\bm{\gamma} & \bm{\mathscr{R}}\cdot\bm{\gamma} &-2.5\bm{\beta}\cdot\bm{\gamma} & \bm{\gamma}^2
  \end{array} \right).
\eeq
The covariance matrix $\mathbf{C}$ has rank $4$ and we thus require measurements in at least four filters to uniquely determine all four parameters. However, we have a prior knowledge on the temperature of the Cepheid (essentially the width of the instability strip), which we can use to reduce the number of necessary bands to three. We assume that the instability strip has a width of $\sigma_\tau = 0.02$\,mag (see Section~\ref{sec:priors}), and we add term $1/\sigma_\tau^2$ to the diagonal entry of $\mathbf{C}$ that corresponds to the temperature, which then reads $(2.5\bm{\beta})^2+1/\sigma_\tau^2$. We assume measurement errors of $\sigma=0.05$\,mag.

The best choice of filters then depends on the desired metric for evaluating success. One possible metric is the overall size the error ellipse
\beq
\mathcal{E} =\log \left|\frac{1}{\det \mathbf{C}}\right|^{1/2}.
\eeq
This metric minimizes the generalized area of the error ellipse and hence tries to obtain the best joint estimate of all four parameters $(\mu,E(B-V),\tau,\mh)$. A second possible metric is 
\beq
\mathcal{E}' = \log |C^{-1}_{\mu\mu}|^{1/2},
\eeq
which tries to obtain the smallest error in distance given that all four parameters need to be estimated. An alternate set of metrics is to examine the sensitivity of distance measurements to the metallicity correction $\bm{\gamma}$. We rewrite Equation~(\ref{eq:app_func}) as
\beq
\mathbf{m}-\mh\bm{\gamma} = \mu \mathbf{1} + \mathscr{R}E(B-V) -2.5\bm{\beta}\tau,
\eeq
and interpret this equation as fitting a ``true'' metallicity-independent magnitude $\mathbf{m}$ plus a metallicity correction proportional to $\bm{\gamma}$. The solution is
\beq
\label{eq:app_dist}
\left( \begin{array}{c} \mu \\ E(B-V) \\ \tau \end{array} \right) = \frac{\mathbf{\tilde{C}}^{-1}}{\sigma^2}
\left(\begin{array}{c}\mathbf{m}\cdot \mathbf{1} \\ \mathbf{m}\cdot \mathscr{R} \\ \mathbf{m}\cdot\bm{\beta}  \end{array}   \right)
-\mh\ \frac{\mathbf{\tilde{C}}^{-1}}{\sigma^2} \left( \begin{array}{c}\bm{\gamma}\cdot\mathbf{1}\\ \bm{\gamma}\cdot \mathscr{R} \\ -2.5\bm{\gamma}\cdot\bm{\beta}\end{array}\right),
\eeq
where $\mathbf{\tilde{C}}$ is the $3\times 3$ upper-left submatrix of $\mathbf{C}$, including the prior contribution $1/\sigma_\tau^2$ to the diagonal temperature term. The second term in Equation~(\ref{eq:app_dist}),
\beq
\bm{\delta} = \frac{\mathbf{\tilde{C}}^{-1}}{\sigma^2} \left( \begin{array}{c}\bm{\gamma}\cdot\mathbf{1}\\ \bm{\gamma}\cdot \mathscr{R} \\ -2.5\bm{\gamma}\cdot\bm{\beta}\end{array}\right),
\eeq
is an estimate of the sensitivity of the parameter vector $(\mu,E(B-V),\tau)$ to metallicity effects. In particular, the change in distance modulus due to the change in metallicity is proportional to $\delta_\mu$, which has units of mag\,dex$^{-1}$. Thus, the metric $\mathcal{E}''$
\beq
\mathcal{E}'' = \log |\delta_\mu|
\label{eq:app_e2prime}
\eeq
expresses the sensitivity of the distance estimate to metallicity for any given filter set.

In order to calculate these metrics, we investigated three groups of filters. The first group has the Johnson-Cousins filters $UBV(RI)_{\rm C}$, the second one has Sloan filters $u'g'r'i'z'$, and the third includes all these filters plus the Walraven and Washington photometric systems. In addition, all three groups include the $JHK$ and Spitzer IRAC bands. We first evaluated $\mathcal{E}$ for all possible triplets of filters within each group. Table~\ref{tab:app_flts} shows the nine filter triplets with the smallest $\mathcal{E}$ in each group normalized as $\mathcal{E}-\mathcal{E}_{BVI_{\rm C}}$ with respect to the originally planned choice of the Hubble Key project \citep{freedman01}, $BVI_{\rm C}$, which has $\mathcal{E}_{BVI_{\rm C}} = -2.379$. The Key Project as realized used only two bands, and thus has $\mathcal{E}_{\rm KP} = \infty$ and no significant control over systematic uncertainties when examining individual Cepheids. We see that in all three groups, the best choices improve over the $BVI_{\rm C}$ filter set by about $0.9$\,dex, or about a factor of $8$. The common feature of all the optimal solutions is broad wavelength coverage from the UV through the IRAC bands. Near-IR bands are only slightly worse than mid-IR bands. In the third group, the best filters are clearly dominated by the Walraven filter set. These filters were designed to be especially sensitive to temperature and metallicity of early-type stars and have relatively narrow widths that make them very sensitive to temperature and metallicity variations. The filters of the Washington system, which were designed for $G$ and $K$ giants, do not stand out as clearly, probably because the width of these filters is larger. However, the improvement from these ``exotic'' filter sets is at most $0.14$\,dex, or only a factor of $1.4$.

The fluxes of Cepheids in the near-UV bands are generally low (see Fig.~\ref{fig:mbar}) and measurements in these filters are not easily obtained from the ground. We repeated the calculation without filters having central wavelengths shorter than $400$\,nm. The results are given in the middle section of Table~\ref{tab:app_flts}. The best combinations simply swap $U/u'$ for $B/g'$ and we also see a decrease in the overall precision of about $0.4$ to $0.6$\,dex, which corresponds to an increase in the parameter errors by a factor of $2.5$ to $4.0$ compared to using the $U/u'$ bands. Still, the total gain in precision with respect to $BVI_{\rm C}$ is still $0.5$\,dex, which is approximately a factor of $3.2$.

Minimizing $\mathcal{E}$ corresponds to trying to obtain the smallest joint uncertainty in all four parameters, and the filter combinations that perform worst are those using only near/mid-IR filters such as [4.5][5.8][8.0], $JK$[5.8], and $K$[3.6][8.0]. These filter sets generally give $\mathcal{E}-\mathcal{E}_{BVI_{\rm C}} \approx 2$, which is two orders of magnitude worse than $BVI_{\rm C}$. The reason is that with wavelength coverage limited to the infrared, the extinction and temperature are not well constrained. We repeated the analysis allowing for four filters in order to see which additional filter is the most beneficial. We see that for both Johnson-Cousins and Sloan groups, adding the $H$ band or a second IRAC band gives the best results. It also shows that the best complement to the Walraven and Washington filters is one of the Spitzer IRAC bands. Adding a fourth filter decreases $\mathcal{E}$ by about $0.2$ to $0.4$\,dex, which corresponds to a decrease in the parameter errors by a factor of $1.6$ to $2.5$. 

If we are uninterested in any quantity other than distance, then $\mathcal{E}'$ may be a better metric. In Table~\ref{tab:app_dist_alt} we present the filter sets leading to the smallest errors in distance given that the four physical parameters must be determined independently, again for the same three filter groups and relative to the original Key Project choice of $BVI_{\rm C}$. Again, the final Key Project filter set of only $VI_{\rm C}$ gives $\mathcal{E}' =\infty$. The best results are obtained with one optical, one near-IR and one mid-IR filter, with an overall error decrease by about a factor of $10$ relative to $BVI_{\rm C}$. Unlike the total error, the near-UV filters are not required to get the lowest error in distance given three filters, but they are a good addition if four filters are available. We also see that Johnson-Cousins and Sloan groups do not differ significantly in their errors. Adding ``exotic'' filter sets like the Walraven and Washington does not noticeably improve the distance determination.

Finally, in Table~\ref{tab:app_dist} we present the filter sets with the smallest and largest sensitivities of the distance determination to metallicity $\mathcal{E}''$ (Eq.~[\ref{eq:app_e2prime}]). There are filter combinations that are very insensitive to metallicity with distance shifts less than $10^{-4}$\,mag\,dex$^{-1}$. The least-metal sensitive filter combinations usually involve one optical ($R_{\rm C}$, $I_{\rm C}$, or $V$) and two infrared filters, which usually include at least one of the Spitzer IRAC bands. Infrared-only filter combinations like $JH$[8.0] also yield small sensitivity to metallicity at the price of never being able to probe additional physics if it becomes necessary because they produce such large $\mathcal{E}$ in Table~\ref{tab:app_flts}. Interestingly, filter combinations that include the [4.5] Spitzer band, which is positioned on the CO bandhead feature and which is claimed to have potential for measuring metallicity effect in Cepheids \citep{scowcroft11}, can still have negligible metallicity sensitivity. With only three bands, it is relatively easy to construct filter combinations that are orthogonal to $\bm{\gamma}$ after determining $\mu$ and $\bar{\tau}$.

Conversely, filter combinations that yield largest sensitivity to metallicity involve bandpasses with small $\lambda_{\rm center}$. For example, $UBV$ has metallicity dependence of about $1.1$\,mag\,dex$^{-1}$ and commonly used set $BVI_{\rm C}$ has sensitivity of about $0.23$\,mag\,dex$^{-1}$. The $VI_{\rm C}H$ filter combination used by \citet{riess09a,riess11a,riess11b} has sensitivity of about $0.006$\,mag\,dex$^{-1}$. As expected, including ``exotic'' filters like the Walraven system greatly increases metallicity sensitivity as these filters were designed to do exactly that. The original Hubble Key Project choice of $BVI_{\rm C}$ has a moderate sensitivity of $0.23$\,mag\,dex$^{-1}$, while using only $VI_{\rm C}$ has $0.066$\,mag\,dex$^{-1}$.

This exploration does depend on the choices for $\bm{\beta}$, $\mathscr{R}$, and $\bm{\gamma}$. Here, we used the \citet{castelli04} theoretical model atmospheres to produce $\bm{\beta}$ and $\bm{\gamma}$, and the \citet{cardelli89} $\rv=3.3$ model of $\mathscr{R}$. We know these models have problems as we discussed in the main part of our study. If we use our empirical estimates of $\bm{\beta}$ and $\mathscr{R}$ we do not see significant changes in the general trends outlined in this Appendix. However, since the primary question is the effects of $\bm{\gamma}$, a well-designed survey should consider a range of models for $\bm{\gamma}$ rather than the single case we considered here.

\begin{deluxetable}{lcr}
\tabletypesize{\scriptsize}
\tablecolumns{3}
\tablewidth{0pc}
\tablecaption{References For Photometric Measurements.}
\tablehead{
\colhead{Reference} & \colhead{Filters} & \colhead{Datapoints}}
\startdata
\citet{ref162} & $1,2,3$ &  351\\
\citet{ref001} & $23,24,25,26,27$ &  975\\
\citet{ref002} & $1,2,3$ &  2430\\   
\citet{ref004} & $23,24,25,26,27$ &  5340\\
\citet{ref123} & $1,2,3$ &  426\\
\citet{ref086} & $1,2,3,5$ &  1666\\  
\citet{ref009,ref121} & $3,14,15,16,17$ &  3679\\  
\citet{ref088} & $2$ &  7\\
\citet{ref014} & $1,2,3$ &  1015\\
\citet{ref089} & $23,24,27$ &  1382\\	
\citet{ref118} & $2,3,4,5$ &  2489\\
\citet{ref018} & $2,3,9,10$ &  3894\\
\citet{ref017} & $11,12,13$ &  446\\
\citet{ref022} & $1,2,3,4,5$ &  3904\\
\citet{ref019} & $1,2,3,4,5$ &  726\\
\citet{ref091} & $2,3,4,5$ &  301\\
\citet{ref092} & $2,3,4,5$ &  65\\
\citet{ref093,ref036} & $6,7,8$ &  3464\\   
\citet{ref095,ref100} & $2,3,11,12,13$ &  795\\ 
\citet{ref149} & $2,3$ &  74\\
\citet{ref043} & $11,12,13$ &  189\\
\citet{ref065} & $1,2,3$ &  129\\
\citet{ref064} & $3,4$ &  44\\
\citet{ref102} & $3,5$ &  214\\
\citet{ref154} & $2,3,9,10,11,12,13$ &  641\\
\citet{ref163} & $18,19,20$ &  5954\\  
\citet{ref161} & $1,2,3$ &  270\\
\citet{ref145} & $2,3,4,5$ &  1879\\  
\citet{ref079} & $1,2,3,4,5$ & 24281 \\   
\citet{ref167} & $11,12,13$ &  4300\\
\citet{ref176} & $33,34,35,36$ &  416\\
\citet{ref166,ref175} & $3,5$ &  105402\\  
\citet{ref177} & $33,34,35,36$ &  32\\
\citet{ref178} & $33,34,35,36$ &  134\\
\tableline
total & & 177,326
\enddata

\label{tab:phot}
\end{deluxetable}

\begin{deluxetable}{lr}
\tabletypesize{\scriptsize}
\tablecolumns{2}
\tablewidth{0pc}
\tablecaption{References For Radial Velocity Measurements.}
\tablehead{
\colhead{Reference} & \colhead{Datapoints}}
\startdata
\citet{ref129} & 25\\
\citet{ref126} & 62\\
\citet{ref125} & 58\\
\citet{ref128,ref010} & 72\\  
\citet{ref123} & 53\\
\citet{ref022} & 794\\
\citet{ref019} & 232\\
\citet{ref090} & 258\\
\citet{ref159} & 21\\
\citet{ref092} & 62\\	
\citet{ref024,ref029} & 370\\  
\citet{ref149} & 36\\
\citet{ref097} & 81\\	
\citet{ref032} & 63\\
\citet{ref034,ref042} & 145\\ 
\citet{ref119a,ref119b,ref119c} & 882\\
\citet{ref056} & 681\\  
\citet{ref055,ref156} & 189\\  
\citet{ref157} & 33\\
\citet{ref079} & 382\\   
\citet{ref171} & 266\\
\citet{ref173} & 33\\
\citet{ref172} & 163\\
\citet{ref174} & 13\\
\citet{ref169} & 57\\
\tableline
total & 5,031
\enddata

\label{tab:rv}
\end{deluxetable}

\begin{deluxetable}{rcrrrccc}
\tabletypesize{\scriptsize}
\tablecolumns{8} 
\tablewidth{0pc} 
\tablecaption{Filters and Filter Properties.}
\tablehead{
\colhead{No.} & \colhead{Name} & \colhead{ $ \lambda_{\rm center}\,[{\rm nm}]$ }& \colhead {$\Delta\lambda\,[{\rm nm}]$ } & \colhead{ Datapoints } & \colhead {$\overline{M}_i$ } & \colhead {$\beta_i$} & \colhead{ $\mathscr{R}_i$} }
\startdata
 1 &  $U$                  &  365 &   35 & 2891 & \phs$1.6600 \pm 0.0368$ & $10.188 \pm  0.006 $ & $ 5.246 \pm  0.003$ \\
 2 &  $B$                  &  441 &   49 & 11574 & \phs$1.0306 \pm 0.0264$ &\phn $7.329 \pm  0.003 $ & $ \equiv  4.303 $\\
 3 &  $V$                  &  549 &   43 & 25893 & \phs$0.1557 \pm 0.0186$ &$ \equiv 5.1404 $ & $ \equiv  3.303 $\\
 4 &  $R_{\rm C}$          &  659 &   75 & 7833 & $-0.2792 \pm 0.0154$ &\phn $4.229 \pm  0.002 $ & $ 2.623 \pm  0.002$ \\
 5 &  $I_{\rm C}$          &  806 &   54 & 100369 & $-0.6877 \pm 0.0125$ &\phn $3.393 \pm  0.002 $ & $ 1.993 \pm  0.003$ \\
 6 &  $J_{\rm SAAO}$       & 1216 &  202 & 1156 & $-1.1488 \pm 0.0095$ &\phn $2.422 \pm  0.006 $ & $ 0.784 \pm  0.007$ \\
 7 &  $H_{\rm SAAO}$       & 1629 &  229 & 1156 & $-1.5679 \pm 0.0075$ &\phn $1.719 \pm  0.006 $ & $ 0.422 \pm  0.007$ \\
 8 &  $K_{\rm SAAO}$       & 2205 &  269 & 1152 & $-1.6356 \pm 0.0071$ &\phn $1.548 \pm  0.006 $ & $ 0.200 \pm  0.008$ \\
 9 &  $R_{\rm J}$          &  693 &  149 & 1042 & $-0.4787 \pm 0.0138$ &\phn $3.750 \pm  0.007 $ & $ 2.308 \pm  0.003$ \\
10 &  $I_{\rm J}$          &  879 &  156 & 1040 & $-0.9137 \pm 0.0110$ &\phn $2.904 \pm  0.007 $ & $ 1.478 \pm  0.005$ \\
11 &  $J_{\rm CTIO}$       & 1246 &  120 & 2013 & $-1.1898 \pm 0.0089$ &\phn $2.246 \pm  0.005 $ & $ 0.724 \pm  0.007$ \\
12 &  $H_{\rm CTIO}$       & 1623 &  150 & 2013 & $-1.5571 \pm 0.0074$ &\phn $1.724 \pm  0.005 $ & $ 0.426 \pm  0.007$ \\
13 &  $K_{\rm CTIO}$       & 2217 &  200 & 2005 & $-1.6435 \pm 0.0070$ &\phn $1.553 \pm  0.005 $ & $ 0.255 \pm  0.008$ \\
14 &  $C$                  &  391 &   78 & 736 & \phs$1.3852 \pm 0.0316$ &\phn $8.746 \pm  0.008 $ & $ 4.725 \pm  0.003$ \\
15 &  $M$                  &  509 &   74 & 736 & \phs$0.3660 \pm 0.0205$ &\phn $5.647 \pm  0.007 $ & $ 3.633 \pm  0.002$ \\
16 &  $T_1$                &  633 &   56 & 736 & $-0.2774 \pm 0.0152$ &\phn $4.155 \pm  0.007 $ & $ 2.682 \pm  0.002$ \\
17 &  $T_2$                &  789 &   99 & 736 & $-0.6781 \pm 0.0124$ &\phn $3.356 \pm  0.006 $ & $ 1.934 \pm  0.003$ \\
18 &  Hp                   &  504 &  157 & 5816 & \phs$0.3534 \pm 0.0184$ &\phn $5.080 \pm  0.003 $ & $ 3.154 \pm  0.001$ \\
19 &  $B_{\rm T}$          &  419 &   51 & 69 & \phs$1.7390 \pm 0.0249$ &\phn $6.634 \pm  0.034 $ & $ 3.523 \pm  0.018$ \\
20 &  $V_{\rm T}$          &  523 &   67 & 69 & \phs$0.2065 \pm 0.0196$ &\phn $5.018 \pm  0.034 $ & $ 3.520 \pm  0.018$ \\
23 &  $V_{\rm W}$          &  547 &   60 & 2268 & $-6.6645 \pm 0.0190$ &\phn $5.243 \pm  0.004 $ & $ 3.417 \pm  0.001$ \\
24 &  $B_{\rm W}$          &  433 &   29 & 2199 & $-5.7640 \pm 0.0286$ &\phn $7.920 \pm  0.005 $ & $ 4.515 \pm  0.002$ \\
25 &  $U_{\rm W}$          &  363 &   18 & 1244 & $-4.3928 \pm 0.0335$ &\phn $9.261 \pm  0.007 $ & $ 5.254 \pm  0.003$ \\
26 &  $W_{\rm W}$          &  326 &   10 & 524 & $-3.5206 \pm 0.0350$ &\phn $9.646 \pm  0.011 $ & $ 5.753 \pm  0.008$ \\
27 &  $L_{\rm W}$          &  384 &   10 & 1462 & $-4.7697 \pm 0.0377$ & $10.402 \pm  0.008 $ & $ 4.887 \pm  0.006$ \\
33 &  IRAC 3.6\,$\mu$m     & 3550 &  370 & 142 & $-1.7672 \pm 0.0075$ &\phn $1.442 \pm  0.023 $ & $ 0.171 \pm  0.008$ \\
34 &  IRAC 4.5\,$\mu$m     & 4493 &  510 & 149 & $-1.7263 \pm 0.0081$ &\phn $1.726 \pm  0.020 $ & $ 0.129 \pm  0.006$ \\
35 &  IRAC 5.8\,$\mu$m     & 5731 &  705 & 142 & $-1.7603 \pm 0.0082$ &\phn $1.657 \pm  0.027 $ & $ 0.086 \pm  0.004$ \\
36 &  IRAC 8.0\,$\mu$m     & 7782 & 1440 & 149 & $-1.7901 \pm 0.0071$ &\phn $1.397 \pm  0.018 $ & $ 0.052 \pm  0.002$
\enddata

\tablecomments{\ List of filters along with their central wavelength $\lambda_{\rm center}$, zero point $\overline{M}_i$, temperature coefficient $\beta_i$, and extinction coefficient $\mathscr{R}_i$.}
\label{tab:filters}
\end{deluxetable}

\begin{deluxetable}{clllll}
\tabletypesize{\scriptsize}
\tablecolumns{6}
\tablewidth{0pc}
\tablecaption{Template Coefficients.}
\tablehead{
\colhead{$\log\left(P/10\,{\rm d}\right)$} & \colhead{$j$}& \colhead{$c_{\tau,j}$} & \colhead{$s_{\tau,j}$} & \colhead{$c_{\rho,j}$} & \colhead{$s_{\rho,j}$} }
\startdata
0.00 &  1 &   $-$0.655942      &   $-$0.799722      & \phs 1.00000       & \phs 0.000000     \\
     &  2 & \phs 0.432002E-01  &   $-$0.940700E-02  & \phs 0.424302E-01  & \phs 0.816138E-01 \\
     &  3 & \phs 0.166285E-01  & \phs 0.658372E-01  &   $-$0.105897E-01  &   $-$0.390460E-01 \\
     &  4 & \phs 0.395908E-02  &   $-$0.288813E-01  & \phs 0.822020E-02  & \phs 0.118774E-01 \\
     &  5 &   $-$0.588958E-02  & \phs 0.805280E-02  &   $-$0.352807E-03  &   $-$0.958669E-03 \\
     &  6 & \phs 0.735751E-02  & \phs 0.151437E-02  & \phs 0.511735E-02  &   $-$0.142921E-02 \\
     &  7 &   $-$0.615025E-02  &   $-$0.394491E-02  &   $-$0.172599E-05  & \phs 0.270471E-02 \\
     &  8 & \phs 0.418875E-02  & \phs 0.347581E-02  & \phs 0.957207E-03  &   $-$0.494847E-02 \\
     &  9 & \phs 0.190958E-03  & \phs 0.254600E-04  & \phs 0.328564E-02  &   $-$0.339728E-04 \\
     & 10 & \phs 0.704976E-03  & \phs 0.614612E-03  & \phs 0.373839E-03  &   $-$0.345807E-02 \\
     & 11 &   $-$0.463388E-03  &   $-$0.341031E-03  & \phs 0.287978E-02  &   $-$0.331189E-03 \\
     & 12 &   $-$0.286117E-03  & \phs 0.178882E-03  &   $-$0.168316E-03  & \phs 0.228903E-03 \\
     & 13 &   $-$0.190860E-03  &   $-$0.288726E-03  &   $-$0.493956E-03  & \phs 0.255491E-03 \\
     & 14 &   $-$0.113362E-03  &   $-$0.314758E-04  &   $-$0.389795E-03  &   $-$0.563042E-03 \\
     & 15 &   $-$0.254223E-03  &   $-$0.177749E-03  & \phs 0.117149E-02  & \phs 0.560073E-03 \\
     & 16 &   $-$0.560226E-04  & \phs 0.128664E-04  &   $-$0.120511E-02  & \phs 0.640936E-03 \\
     & 17 &   $-$0.186283E-03  &   $-$0.141655E-03  &   $-$0.483337E-03  &   $-$0.642305E-03 \\
     & 18 &   $-$0.158575E-03  &   $-$0.150494E-04  & \phs 0.965609E-03  & \phs 0.200574E-03 \\
     & 19 &   $-$0.768975E-04  &   $-$0.552935E-04  &   $-$0.662260E-04  & \phs 0.343053E-03 \\
     & 20 &   $-$0.230873E-03  &   $-$0.156252E-03  & \phs 0.510411E-03  &   $-$0.553118E-03 \\
0.02 &  1 &   $-$0.654341      &   $-$0.763829      & \phs 1.00000       & \phs 0.000000     \\
     &  2 & \phs 0.238339E-01  &   $-$0.128259E-01  & \phs 0.964536E-02  & \phs 0.115707     \\
     &  3 & \phs 0.490282E-01  & \phs 0.632282E-01  &   $-$0.141993E-01  &   $-$0.394291E-01 \\
     &  4 &   $-$0.146646E-01  &   $-$0.365181E-01  & \phs 0.126668E-01  & \phs 0.169339E-01 \\
     &  5 &   $-$0.527408E-02  & \phs 0.141425E-01  &   $-$0.464929E-02  &   $-$0.613523E-03
\enddata

\tablecomments{The table lists Fourier coefficients of the templates at $N_P\equiv 19$ anchor points from $\log(P/10\,{\rm d}) = 0.00$ to $1.00$. Each anchor point contains four columns, $c_{\tau,j}$, $s_{\tau,j}$, $c_{\rho,j}$, and $s_{\rho,j}$, where each line gives the $j$-th coefficient, $j=1\ldots\nf$, where $\nf\equiv 20$ (see Eqs.~[\ref{eq:fou_rho}]--[\ref{eq:fou_tau}]). The template at any period is obtained by linear interpolation of all coefficients in $\log P$ bracketing the desired period. The complete Table is provided in the electronic edition.}
\label{tab:templates}
\end{deluxetable}

\begin{deluxetable}{lccccccc}
\tabletypesize{\scriptsize}
\tablecolumns{8} 
\tablewidth{0pc} 
\tablecaption{Galactic Cepheids}
\tablehead{ 
\colhead{Name} & \colhead {$ P$ [d]} & \colhead{$\mu$ [mag]} & \colhead{$E(B-V)$ [mag]} & \colhead{$\bar{v}$ [km s$^{-1}$]} & \colhead{$ \bar{\rho} $} & \colhead{$\bar{\tau}$} & \colhead{$ A^2 \times 10^2$}}
\startdata
         BZ Cyg & 10.1420 & $11.376 \pm 0.040 $ & $0.916 \pm 0.008 $ & \phn $  -11.4 \pm 0.1 $ & $  0.782 \pm 0.008 $ & \phs $  0.032 \pm 0.002 $ & $ 2.2943 \pm 0.0038 $ \\
         SY Aur & 10.1446 & $11.708 \pm 0.040 $ & $0.446 \pm 0.006 $ & \phs \phn \phn $    0.0 \pm 0.1 $ & $  0.723 \pm 0.008 $ & \phs $  0.049 \pm 0.002 $ & $ 2.6711 \pm 0.0047 $ \\
    $\zeta$ Gem & 10.1505 & \phn $ 7.892 \pm 0.025 $ & $-0.031 \pm 0.006 $ & \phs \phn \phn $    6.1 \pm 0.0 $ & $  0.819 \pm 0.005 $ & $ -0.003 \pm 0.002 $ & $ 2.2276 \pm 0.0036 $ \\
         FQ Car & 10.2742 & $13.236 \pm 0.079 $ & $0.597 \pm 0.032 $ & \phn \phn $   -3.7 \pm 0.5 $ & $  0.749 \pm 0.014 $ & $ -0.016 \pm 0.006 $ & $ 3.3464 \pm 0.0044 $ \\
         AN Aur & 10.2893 & $12.981 \pm 0.055 $ & $0.568 \pm 0.008 $ & \phs \phn \phn $    1.9 \pm 0.2 $ & $  0.814 \pm 0.011 $ & \phs $  0.036 \pm 0.002 $ & $ 2.8273 \pm 0.0032 $ \\
          Y Sct & 10.3415 & $11.021 \pm 0.039 $ & $0.735 \pm 0.005 $ & \phs \phn $   14.2 \pm 0.2 $ & $  0.790 \pm 0.008 $ & \phs $  0.000 \pm 0.002 $ & $ 3.3299 \pm 0.0034 $ \\
         MZ Cen & 10.3538 & $13.842 \pm 0.035 $ & $0.665 \pm 0.007 $ & \phn $  -24.4 \pm 0.2 $ & $  0.952 \pm 0.007 $ & $ -0.011 \pm 0.002 $ & $ 2.8914 \pm 0.0047 $ \\
         FO Car & 10.3560 & $13.582 \pm 0.061 $ & $0.380 \pm 0.005 $ & \phn $  -12.1 \pm 0.4 $ & $  0.846 \pm 0.012 $ & \phs $  0.004 \pm 0.002 $ & $ 2.4539 \pm 0.0026 $ \\
         FR Car & 10.7170 & $12.799 \pm 0.016 $ & $0.285 \pm 0.005 $ & \phn \phn $   -6.9 \pm 0.1 $ & $  0.845 \pm 0.003 $ & $ -0.001 \pm 0.002 $ & $ 3.2252 \pm 0.0031 $ \\
         TW Nor & 10.7861 & $11.700 \pm 0.032 $ & $1.090 \pm 0.006 $ & \phn $  -56.1 \pm 0.2 $ & $  0.797 \pm 0.006 $ & $ -0.016 \pm 0.002 $ & $ 3.6645 \pm 0.0034 $ \\
          Z Lac & 10.8857 & $11.291 \pm 0.100 $ & $0.374 \pm 0.003 $ &  \nodata & $  0.803 \pm 0.020 $ & \phs $  0.017 \pm 0.002 $ & $ 3.7122 \pm 0.0028 $ \\
         VX Per & 10.8866 & $12.003 \pm 0.019 $ & $0.484 \pm 0.005 $ & \phn $  -36.3 \pm 0.1 $ & $  0.819 \pm 0.004 $ & \phs $  0.026 \pm 0.002 $ & $ 2.8020 \pm 0.0023 $ \\
         XX Cen & 10.9536 & $10.669 \pm 0.011 $ & $0.238 \pm 0.003 $ & \phn $  -18.1 \pm 0.0 $ & $  0.720 \pm 0.002 $ & \phs $  0.013 \pm 0.002 $ & $ 3.6481 \pm 0.0027 $ \\
         TY Sct & 11.0540 & $11.690 \pm 0.033 $ & $0.927 \pm 0.006 $ & \phs \phn $   25.0 \pm 0.2 $ & $  0.791 \pm 0.006 $ & \phs $  0.007 \pm 0.002 $ & $ 3.6861 \pm 0.0032 $ \\
         SV Per & 11.1292 & $12.520 \pm 0.023 $ & $0.533 \pm 0.006 $ & \phn \phn $   -5.8 \pm 0.1 $ & $  0.913 \pm 0.005 $ & \phs $  0.067 \pm 0.002 $ & $ 3.0448 \pm 0.0029 $ \\
         DR Vel & 11.2013 & $11.314 \pm 0.017 $ & $0.609 \pm 0.005 $ & \phs \phn $   20.5 \pm 0.1 $ & $  0.815 \pm 0.003 $ & $ -0.012 \pm 0.002 $ & $ 3.2269 \pm 0.0032 $ \\
       V438 Cyg & 11.2108 & $11.297 \pm 0.039 $ & $1.111 \pm 0.013 $ & \phn \phn $   -9.1 \pm 0.2 $ & $  0.828 \pm 0.008 $ & $ -0.000 \pm 0.003 $ & $ 3.8962 \pm 0.0048 $ \\
         AD Cam & 11.2622 & $13.878 \pm 0.042 $ & $0.983 \pm 0.012 $ & \phn $  -60.8 \pm 0.2 $ & $  0.826 \pm 0.008 $ & \phs $  0.042 \pm 0.003 $ & $ 3.8923 \pm 0.0047 $ \\
         HZ Per & 11.2796 & $13.844 \pm 0.033 $ & $1.312 \pm 0.012 $ & \phn $  -28.5 \pm 0.2 $ & $  0.896 \pm 0.006 $ & \phs $  0.003 \pm 0.003 $ & $ 3.4095 \pm 0.0044 $ \\
       V340 Nor & 11.2894 & $11.689 \pm 0.061 $ & $0.458 \pm 0.016 $ & \phn $  -40.7 \pm 0.1 $ & $  0.877 \pm 0.012 $ & \phs $  0.043 \pm 0.006 $ & $ 1.3556 \pm 0.0037 $
\enddata

\tablecomments{For each Cepheid, the table lists the period $P$, distance modulus $\mu$, reddening $E(B-V)$, mean radial velocity $\overline{v}$, mean logarithmic radius $\bar{\rho}$, mean logarithmic temperature $\bar{\tau}$ and the amplitude $A^2$. The complete Table is provided in the electronic edition.}
\label{tab:galaxy}
\end{deluxetable}

\begin{deluxetable}{lccccccc}
\tabletypesize{\scriptsize}
\tablecolumns{8} 
\tablewidth{0pc} 
\tablecaption{LMC Cepheids}
\tablehead{ 
\colhead{Name} & \colhead {$ P$ [d]} & \colhead{$\mu$ [mag]} & \colhead{$E(B-V)$ [mag]} & \colhead{$\bar{v}$ [km s$^{-1}$]} & \colhead{$ \bar{\rho} $} & \colhead{$\bar{\tau}$} & \colhead{$ A^2 \times 10^2$}}
\startdata
       HV 01016 & 10.0438 & 18.480 & $0.147 \pm 0.019 $ &  \nodata & $  0.761 \pm 0.003 $ & \phs $  0.025 \pm 0.006 $ & $ 2.6563 \pm 0.0067 $ \\
      OIII 2562 & 10.0716 & 18.479 & $0.147 \pm 0.019 $ &  \nodata & $  0.788 \pm 0.003 $ & \phs $  0.033 \pm 0.006 $ & $ 0.9424 \pm 0.0018 $ \\
       HV 00923 & 10.2547 & 18.511 & $0.144 \pm 0.019 $ &  \nodata & $  0.784 \pm 0.002 $ & \phs $  0.039 \pm 0.006 $ & $ 2.3734 \pm 0.0030 $ \\
      OIII 0057 & 10.3075 & 18.558 & $0.150 \pm 0.019 $ &  \nodata & $  0.809 \pm 0.003 $ & \phs $  0.027 \pm 0.006 $ & $ 2.0783 \pm 0.0033 $ \\
       HV 02371 & 10.3479 & 18.514 & $0.142 \pm 0.019 $ &  \nodata & $  0.780 \pm 0.003 $ & \phs $  0.046 \pm 0.006 $ & $ 3.3251 \pm 0.0042 $ \\
      OIII 0114 & 10.3727 & 18.552 & $0.142 \pm 0.019 $ &  \nodata & $  0.789 \pm 0.003 $ & \phs $  0.048 \pm 0.006 $ & $ 2.5954 \pm 0.0040 $ \\
       HV 02277 & 10.4135 & 18.533 & $0.142 \pm 0.019 $ &  \nodata & $  0.798 \pm 0.003 $ & \phs $  0.051 \pm 0.006 $ & $ 3.4945 \pm 0.0050 $ \\
      OIII 3284 & 10.4772 & 18.447 & $0.147 \pm 0.019 $ &  \nodata & $  0.774 \pm 0.003 $ & \phs $  0.025 \pm 0.006 $ & $ 2.3392 \pm 0.0043 $ \\
       HV 05551 & 10.4852 & 18.544 & $0.093 \pm 0.011 $ &  \nodata & $  0.777 \pm 0.001 $ & \phs $  0.016 \pm 0.003 $ & $ 2.0011 \pm 0.0033 $ \\
       HV 13064 & 10.4872 & 18.482 & $0.191 \pm 0.019 $ &  \nodata & $  0.841 \pm 0.003 $ & $ -0.099 \pm 0.006 $ & $ 3.0641 \pm 0.0056 $ \\
      OIII 0016 & 10.5070 & 18.554 & $0.169 \pm 0.019 $ &  \nodata & $  0.905 \pm 0.003 $ & $ -0.004 \pm 0.006 $ & $ 0.6998 \pm 0.0018 $ \\
      OIII 0978 & 10.5199 & 18.517 & $0.157 \pm 0.019 $ &  \nodata & $  0.840 \pm 0.003 $ & \phs $  0.013 \pm 0.006 $ & $ 1.7861 \pm 0.0027 $ \\
      OIII 1784 & 10.6628 & 18.492 & $0.147 \pm 0.019 $ &  \nodata & $  0.777 \pm 0.003 $ & \phs $  0.023 \pm 0.006 $ & $ 2.9994 \pm 0.0042 $ \\
       HV 12078 & 10.6876 & 18.427 & $0.088 \pm 0.010 $ &  \nodata & $  0.802 \pm 0.001 $ & \phs $  0.004 \pm 0.003 $ & $ 2.0235 \pm 0.0030 $ \\
      OIII 0798 & 10.7104 & 18.526 & $0.145 \pm 0.019 $ &  \nodata & $  0.789 \pm 0.003 $ & \phs $  0.032 \pm 0.006 $ & $ 1.7056 \pm 0.0023 $ \\
       HV 12537 & 10.7923 & 18.531 & $0.142 \pm 0.019 $ &  \nodata & $  0.819 \pm 0.003 $ & \phs $  0.053 \pm 0.006 $ & $ 3.2753 \pm 0.0037 $ \\
       HV 02280 & 10.8548 & 18.525 & $0.143 \pm 0.019 $ &  \nodata & $  0.801 \pm 0.003 $ & \phs $  0.041 \pm 0.006 $ & $ 2.9274 \pm 0.0032 $ \\
       HV 02432 & 10.9192 & 18.494 & $0.247 \pm 0.010 $ &  \nodata & $  0.787 \pm 0.002 $ & \phs $  0.100 \pm 0.004 $ & $ 2.9310 \pm 0.0037 $ \\
       HV 02864 & 10.9853 & 18.426 & $0.095 \pm 0.006 $ & \phs $  259.8 \pm 0.1 $ & $  0.791 \pm 0.001 $ & \phs $  0.020 \pm 0.002 $ & $ 3.7166 \pm 0.0037 $ \\
       HV 02598 & 10.9913 & 18.486 & $0.138 \pm 0.019 $ &  \nodata & $  0.806 \pm 0.003 $ & \phs $  0.060 \pm 0.006 $ & $ 2.1020 \pm 0.0020 $ \\
       HV 00921 & 11.0905 & 18.506 & $0.147 \pm 0.019 $ &  \nodata & $  0.811 \pm 0.003 $ & \phs $  0.030 \pm 0.006 $ & $ 2.1498 \pm 0.0024 $
\enddata

\tablecomments{Same as in Table~\ref{tab:galaxy}, but for the LMC Cepheids. The complete Table is provided in the electronic edition.}
\label{tab:lmc}
\end{deluxetable}

\begin{deluxetable}{lccccccc}
\tabletypesize{\scriptsize}
\tablecolumns{8} 
\tablewidth{0pc} 
\tablecaption{SMC Cepheids}
\tablehead{ 
\colhead{Name} & \colhead {$ P$ [d]} & \colhead{$\mu$ [mag]} & \colhead{$E(B-V)$ [mag]} & \colhead{$\bar{v}$ [km s$^{-1}$]} & \colhead{$ \bar{\rho} $} & \colhead{$\bar{\tau}$} & \colhead{$ A^2 \times 10^2$}}
\startdata
       HV 06320 & 10.0998 & $19.017 \pm 0.071 $ & $0.148 \pm 0.015 $ &  \nodata & $  0.757 \pm 0.014 $ & \phs $  0.078 \pm 0.004 $ & $ 3.7284 \pm 0.0095 $ \\
       HV 02063 & 11.1663 & $18.881 \pm 0.071 $ & $0.167 \pm 0.007 $ &  \nodata & $  0.815 \pm 0.014 $ & \phs $  0.055 \pm 0.002 $ & $ 2.8573 \pm 0.0033 $ \\
       HV 02017 & 11.4091 & $18.870 \pm 0.071 $ & $0.175 \pm 0.007 $ &  \nodata & $  0.823 \pm 0.014 $ & \phs $  0.060 \pm 0.002 $ & $ 3.0375 \pm 0.0030 $ \\
       HV 01610 & 11.6431 & $18.944 \pm 0.071 $ & $0.171 \pm 0.009 $ &  \nodata & $  0.814 \pm 0.014 $ & \phs $  0.070 \pm 0.003 $ & $ 3.7982 \pm 0.0038 $ \\
       HV 00856 & 12.1557 & $19.038 \pm 0.071 $ & $0.273 \pm 0.008 $ &  \nodata & $  0.809 \pm 0.014 $ & \phs $  0.081 \pm 0.003 $ & $ 3.6536 \pm 0.0043 $ \\
       HV 01365 & 12.4127 & $19.483 \pm 0.016 $ & $0.231 \pm 0.005 $ & \phs $  126.8 \pm 0.1 $ & $  0.885 \pm 0.003 $ & \phs $  0.070 \pm 0.002 $ & $ 3.3213 \pm 0.0026 $ \\
       HV 02227 & 12.4663 & $18.866 \pm 0.071 $ & $0.201 \pm 0.010 $ &  \nodata & $  0.851 \pm 0.014 $ & \phs $  0.048 \pm 0.003 $ & $ 3.3765 \pm 0.0036 $ \\
       HV 02230 & 12.5294 & $18.823 \pm 0.071 $ & $0.218 \pm 0.010 $ &  \nodata & $  0.861 \pm 0.014 $ & \phs $  0.051 \pm 0.003 $ & $ 3.0924 \pm 0.0067 $ \\
       HV 02052 & 12.5778 & $18.858 \pm 0.071 $ & $0.084 \pm 0.008 $ &  \nodata & $  0.855 \pm 0.014 $ & \phs $  0.058 \pm 0.003 $ & $ 3.8876 \pm 0.0045 $ \\
       HV 01744 & 12.6237 & $18.925 \pm 0.071 $ & $0.135 \pm 0.009 $ & \phs $  141.9 \pm 5.8 $ & $  0.842 \pm 0.014 $ & \phs $  0.061 \pm 0.003 $ & $ 4.0334 \pm 0.0046 $ \\
       HV 01873 & 12.9396 & $19.026 \pm 0.071 $ & $0.185 \pm 0.007 $ &  \nodata & $  0.830 \pm 0.014 $ & \phs $  0.060 \pm 0.002 $ & $ 4.4509 \pm 0.0041 $ \\
       HV 02225 & 13.1477 & $18.877 \pm 0.071 $ & $0.119 \pm 0.014 $ &  \nodata & $  0.865 \pm 0.014 $ & \phs $  0.024 \pm 0.004 $ & $ 3.5367 \pm 0.0095 $ \\
       HV 02202 & 13.1921 & $18.852 \pm 0.071 $ & $0.227 \pm 0.008 $ &  \nodata & $  0.871 \pm 0.014 $ & \phs $  0.073 \pm 0.002 $ & $ 3.7139 \pm 0.0051 $ \\
       HV 00827 & 13.4642 & $18.935 \pm 0.071 $ & $0.186 \pm 0.006 $ &  \nodata & $  0.861 \pm 0.014 $ & \phs $  0.072 \pm 0.002 $ & $ 3.6416 \pm 0.0031 $ \\
       HV 02189 & 13.4703 & $18.989 \pm 0.071 $ & $0.360 \pm 0.022 $ &  \nodata & $  0.850 \pm 0.014 $ & \phs $  0.114 \pm 0.007 $ & $ 3.5912 \pm 0.0039 $ \\
       HV 01326 & 13.7250 & $19.041 \pm 0.071 $ & $0.169 \pm 0.009 $ &  \nodata & $  0.845 \pm 0.014 $ & \phs $  0.049 \pm 0.003 $ & $ 3.4243 \pm 0.0033 $ \\
       HV 01335 & 14.3814 & $19.079 \pm 0.071 $ & $0.243 \pm 0.009 $ &  \nodata & $  0.852 \pm 0.014 $ & \phs $  0.076 \pm 0.003 $ & $ 3.5012 \pm 0.0039 $ \\
       HV 02088 & 14.5796 & $18.955 \pm 0.071 $ & $0.208 \pm 0.008 $ &  \nodata & $  0.881 \pm 0.014 $ & \phs $  0.052 \pm 0.002 $ & $ 4.4474 \pm 0.0053 $ \\
       HV 00843 & 14.7136 & $18.927 \pm 0.071 $ & $0.120 \pm 0.009 $ & \phs $  166.8 \pm 5.8 $ & $  0.888 \pm 0.014 $ & \phs $  0.002 \pm 0.003 $ & $ 4.1825 \pm 0.0037 $ \\
       HV 02233 & 15.1677 & $18.807 \pm 0.071 $ & $0.271 \pm 0.016 $ &  \nodata & $  0.922 \pm 0.014 $ & \phs $  0.102 \pm 0.004 $ & $ 5.0158 \pm 0.0141 $ \\
       HV 01442 & 15.2886 & $18.964 \pm 0.071 $ & $0.232 \pm 0.006 $ & \phs $  130.0 \pm 7.1 $ & $  0.893 \pm 0.014 $ & \phs $  0.046 \pm 0.002 $ & $ 4.4427 \pm 0.0041 $
\enddata

\tablecomments{Same as in Table~\ref{tab:galaxy}, but for the SMC Cepheids. The complete Table is provided in the electronic edition.}
\label{tab:smc}
\end{deluxetable}

\begin{deluxetable}{cccccc}
\tabletypesize{\scriptsize}
\enlargethispage{2cm}
\tablecolumns{6} 
\tablewidth{0pc} 
\tablecaption{Filter combinations that yield the smallest overall error $\mathcal{E}$.}
\tablehead{
\colhead{$UBV(RI)_{\rm C}JHK$+IRAC} & \colhead{$\mathcal{E}- \mathcal{E}_{BVI_{\rm C}}$} & \colhead{$u'g'r'i'z'JHK$+IRAC} & \colhead{$\mathcal{E}- \mathcal{E}_{BVI_{\rm C}}$} & \colhead{All} & \colhead{$\mathcal{E}- \mathcal{E}_{BVI_{\rm C}}$}
}
\startdata
$U V         $[4.5] &  $-0.892$        &        $u'r' $[4.5]  &  $-0.838$    &     $M         L_{\rm W} $[5.8]  &  $-0.979$             \\
$U V         $[5.8] &  $-0.891$        &        $u'r' $[5.8]  &  $-0.834$    &     $M         L_{\rm W} $[8.0]  &  $-0.979$              \\
$U V         $[8.0] &  $-0.884$        &        $u'r' $[8.0]  &  $-0.823$    &     $M         L_{\rm W} $[4.5]  &  $-0.977$             \\
$U V         $[3.6] &  $-0.864$        &        $u'g' $[4.5]  &  $-0.811$    &     $g'        L_{\rm W} $[8.0]  &  $-0.966$             \\
$U R_{\rm C} $[4.5] &  $-0.861$        &        $u'g' $[5.8]  &  $-0.811$    &     $g'        L_{\rm W} $[5.8]  &  $-0.966$           \\
$U R_{\rm C} $[5.8] &  $-0.858$        &        $u'g' $[8.0]  &  $-0.806$    &     $g'        L_{\rm W} $[4.5]  &  $-0.963$             \\
$U R_{\rm C} $[8.0] &  $-0.847$        &        $u'r' $[3.6]  &  $-0.801$    &     $M         L_{\rm W} $[3.6]  &  $-0.963$              \\
$U V         K$     &  $-0.842$        &        $u'g' $[3.6]  &  $-0.789$    &     $(VL)_{\rm W} $[5.8]  &  $-0.959$                    \\
$U R_{\rm C} $[3.6] &  $-0.823$        &        $u'r' K$      &  $-0.778$    &     $(VL)_{\rm W} $[4.5]  &  $-0.958$                    \\
\hline
$B R_{\rm C} $[4.5] &  $-0.487$        &        $g'r' $[4.5]  &  $-0.221$    &     $B         T_1       $[4.5]  &  $-0.512$             \\
$B R_{\rm C} $[5.8] &  $-0.478$        &        $g'r' $[5.8]  &  $-0.208$    &     $B         r'        $[4.5]  &  $-0.509$              \\
$B R_{\rm C} $[8.0] &  $-0.455$        &        $g'r' $[8.0]  &  $-0.179$    &     $B         T_1       $[5.8]  &  $-0.505$             \\
$B V         $[4.5] &  $-0.452$        &        $g'r' $[3.6]  &  $-0.155$    &     $B         r'        $[5.8]  &  $-0.502$             \\
$B V         $[5.8] &  $-0.447$        &        $g'i' $[4.5]  &  $-0.149$    &     $B         R_{\rm C} $[4.5]  &  $-0.487$           \\
\hline
$UVH$[4.5]      &  $-1.057$    &   $u'g'r'$[4.5] &  $-1.009$    &    $M (WL)_{\rm W} $[8.0]  &  $-1.357$          \\
$UVH$[5.8]      &  $-1.052$    &   $u'g'r'$[5.8] &  $-1.006$    &    $M (WL)_{\rm W} $[5.8]  &  $-1.357$           \\
$UV$[4.5][8.0]  &  $-1.043$    &   $u'r'H $[4.5] &  $-0.998$    &    $M (WL)_{\rm W} $[4.5]  &  $-1.354$          \\
$UV$[4.5][5.8]  &  $-1.042$    &   $u'g'H $[4.5] &  $-0.996$    &    $g'(WL)_{\rm W} $[8.0]  &  $-1.347$          \\
$UV$[3.6][4.5]  &  $-1.041$    &   $u'g'r'$[8.0] &  $-0.995$    &    $g'(WL)_{\rm W} $[5.8]  &  $-1.346$          
\enddata
\tablecomments{$\mathcal{E}-\mathcal{E}_{BVI_{\rm C}}$ represents the logarithmic reduction in the parameter uncertainties compared to using the $BVI_{\rm C}$ filters. The upper part of the Table shows results for all filters, the middle part shows results for filters with $\lambda_{\rm center} \geq 0.4\,\mu$m, and the lower part for four filters in the set.}
\label{tab:app_flts}
\end{deluxetable}

\begin{deluxetable}{cccccc}
\tabletypesize{\scriptsize}
\enlargethispage{2cm}
\tablecolumns{6} 
\tablewidth{0pc} 
\tablecaption{Filter combinations that yield the smallest error in distance $\mathcal{E}'$.}
\tablehead{
\colhead{$UBV(RI)_{\rm C}JHK$+IRAC} & \colhead{$\mathcal{E}'- \mathcal{E}'_{BVI_{\rm C}}$} & \colhead{$u'g'r'i'z'JHK$+IRAC} & \colhead{$\mathcal{E}'- \mathcal{E}'_{BVI_{\rm C}}$} & \colhead{All} & \colhead{$\mathcal{E}'- \mathcal{E}'_{BVI_{\rm C}}$}
}
\startdata
$V         H K$        &  $-1.145$    &    $r'H$[3.6]  &  $-1.129$  &  $V         HK$      &  $-1.145$         \\
$R_{\rm C} H $[3.6]    &  $-1.122$    &    $g'H$[4.5]  &  $-1.116$  &  $T_1       H$[3.6]  &  $-1.130$          \\
$R_{\rm C} H $[8.0]    &  $-1.104$    &    $g'H$[5.8]  &  $-1.115$  &  $V_{\rm W} HK$      &  $-1.130$         \\
$B         H $[4.5]    &  $-1.088$    &    $r'H$[8.0]  &  $-1.109$  &  $r'        H$[3.6]  &  $-1.129$         \\
$I_{\rm C} H $[8.0]    &  $-1.087$    &    $i'H$[3.6]  &  $-1.107$  &  $R_{\rm C} H$[3.6]  &  $-1.122$       \\
$I_{\rm C} H $[3.6]    &  $-1.084$    &    $i'H$[8.0]  &  $-1.096$  &  $g'        H$[4.5]  &  $-1.116$         \\
$B         K $[4.5]    &  $-1.031$    &    $z'H$[8.0]  &  $-1.078$  &  $g'        H$[5.8]  &  $-1.115$          \\
$R_{\rm C} K $[8.0]    &  $-1.013$    &    $z'H$[3.6]  &  $-1.047$  &  $r'        H$[8.0]  &  $-1.109$         \\
$R_{\rm C} K $[3.6]    &  $-1.011$    &    $g'K$[5.8]  &  $-1.036$  &  $B_{\rm W} H$[4.5]  &  $-1.108$         \\
\hline
  $R_{\rm C} HK     $[3.6]  &  $-1.211$  &  $r'H K$[3.6]  &  $-1.217$   &    $(BW)_{\rm W} HK$  &   $-1.248$     \\
  $U         BH     K$      &  $-1.204$  &  $g'H K$[4.5]  &  $-1.214$   &    $(BW)_{\rm W} H$[3.6] &$-1.245$      \\
  $I_{\rm C} HK     $[3.6]  &  $-1.199$  &  $i'H K$[3.6]  &  $-1.207$   &    $(UB)_{\rm W} HK$  &   $-1.233$     \\
  $U         BH     $[3.6]  &  $-1.191$  &  $z'H K$[3.6]  &  $-1.197$   &    $B         W_{\rm W} HK$  &   $-1.233$     \\
  $V         H$[3.6] [5.8]  &  $-1.179$  &  $u'g'HK$      &  $-1.194$   &    $u'        B_{\rm W} HK$  &   $-1.228$
\enddata
\tablecomments{$\mathcal{E}'-\mathcal{E}'_{BVI_{\rm C}}$ represents the logarithmic reduction in the distance uncertainty compared to using the $BVI_{\rm C}$ filters. The upper part of the Table shows results for all filters and the lower part for four filters in the set.}
\label{tab:app_dist_alt}
\end{deluxetable}

\begin{deluxetable}{cccccc}
\tabletypesize{\scriptsize}
\tablecolumns{6} 
\tablewidth{0pc} 
\tablecaption{Filter combinations that yield the smallest and largest change in distance due to metallicity.}
\tablehead{
\colhead{$UBV(RI)_{\rm C}JHK$+IRAC} & \colhead{$\mathcal{E}''$} & \colhead{$u'g'r'i'z'JHK$+IRAC} & \colhead{$\mathcal{E}''$} & \colhead{All} & \colhead{$\mathcal{E}''$}
}
\startdata
\cutinhead{Smallest change in distance}	
$I_{\rm C} $[3.6][4.5]        &   $    -4.344$                &         $i'H    $[8.0]  &   $-4.161$     &        $T_1       V_{\rm W} H    $   &   $-5.149$         \\
$V         H         K$            &   $    -3.677$           &         $i'$[3.6][4.5]  &   $-3.970$     &        $I_{\rm C} $[3.6]     [4.5]   &   $-4.344$         \\
$R_{\rm C} H$[8.0]           &   $    -3.662$                 &         $r'$H    [8.0]  &   $-3.738$     &        $i'        H         $[8.0]   &   $-4.161$         \\
$R_{\rm C} $[3.6][4.5]        &   $    -3.599$                &         $r'$[3.6][4.5]  &   $-3.409$     &        $C         W_{\rm W} $[8.0]   &   $-4.057$         \\
$V         J         $[4.5]        &   $    -3.371$           &         $z'$[3.6][4.5]  &   $-3.342$     &        $i'        $[3.6]     [4.5]   &   $-3.970$         \\
$R_{\rm C} $[5.8][8.0]        &   $    -3.269$                &         $i'$[5.8][8.0]  &   $-3.319$     &        $R_{\rm C} i'        T_1  $   &   $-3.926$         \\
$I_{\rm C} $[5.8][8.0]        &   $    -3.255$                &         $z'$[5.8][8.0]  &   $-3.315$     &        $T_2       $[3.6]     [4.5]   &   $-3.840$         \\
$I_{\rm C} J         H$           &   $    -3.215$            &         $r'$[5.8][8.0]  &   $-3.250$     &        $V         r'        H    $   &   $-3.836$         \\
$J         H         $[8.0]        &   $    -3.148$           &         $r'J    H$      &   $-3.209$     &        $r'        H         $[8.0]   &   $-3.738$         \\
$J         $[5.8][8.0]        &   $    -3.129$                &         $J H    $[8.0]  &   $-3.148$     &        $I_{\rm C} M         $[4.5]   &   $-3.731$         \\
\cutinhead{Largest change in distance}
$UBV         $   & $0.024$          &          $u'g'r'$   &   $-0.094$         &         $Uu'        U_{\rm W}$   &   $ 0.936$              \\
$UVR_{\rm C} $   &   $    -0.103$          &          $u'r'i'$   &   $-0.301$         &         $B(BL)_{\rm W}$   &   $ 0.678$                     \\
$UBR_{\rm C} $   &   $    -0.186$          &          $u'g'i'$   &   $-0.357$         &         $C(WL)_{\rm W}$   &   $ 0.632$                     \\
$B         V         R_{\rm C} $   &   $    -0.327$          &          $u'r'z'$   &   $-0.441$         &         $B(WL)_{\rm W}$   &   $ 0.508$                     \\
$U         R_{\rm C} I_{\rm C} $   &   $    -0.338$          &          $u'i'z'$   &   $-0.471$         &         $(BWL)_{\rm W}$   &   $ 0.453$                     \\
$U         V         I_{\rm C} $   &   $    -0.384$          &          $u'g'z'$   &   $-0.486$         &         $B (BW)_{\rm W}$   &   $ 0.409$                    \\
$U         B         I_{\rm C} $   &   $    -0.447$          &          $u'r'J $   &   $-0.570$         &         $g'M         L_{\rm W}$   &   $ 0.363$             \\
$U         R_{\rm C} J         $   &   $    -0.564$          &          $u'i'J $   &   $-0.593$         &         $U B         W_{\rm W}$   &   $ 0.360$             \\
$U         V         J         $   &   $    -0.575$          &          $u'g'J $   &   $-0.623$         &         $U B         C        $   &   $ 0.323$             \\
$U         I_{\rm C} J         $   &   $    -0.581$          &          $u'z'J $   &   $-0.686$         &         $B C         W_{\rm W}$   &   $ 0.314$             
\enddata
\tablecomments{$\mathcal{E}''$ represents the logarithm of the absolute change in distance modulus with metallicity. The upper part of the Table shows filter sets with the smallest changes, while the lower shows those with the largest changes.}
\label{tab:app_dist}
\end{deluxetable}


\begin{thebibliography}{}
\bibitem[Albrecht et al.(2006)]{albrecht06} Albrecht, A., Bernstein, G., Cahn, R., et al.\ 2006, arXiv:astro-ph/0609591 
\bibitem[Andrievsky et al.(2002)]{andrievsky02} Andrievsky, S.~M., Kovtyukh, V.~V., Luck, R.~E., et al.\ 2002, \aap, 381, 32 
\bibitem[Antonello \& Morelli(1996)]{antonello96} Antonello, E., \& Morelli, P.~L.\ 1996, \aap, 314, 541 
\bibitem[Baade(1926)]{baade26} Baade, W.\ 1926, Astronomische Nachrichten, 228, 359 
\bibitem[Barnes et al.(1987)]{ref024} Barnes, T.~G., III, Moffett, T.~J., \& Slovak, M.~H.\ 1987, \apjs, 65, 307 
\bibitem[Barnes et al.(1988)]{ref029} Barnes, T.~G., III, Moffett, T.~J., \& Slovak, M.~H.\ 1988, \apjs, 66, 43
\bibitem[Barnes et al.(1997)]{ref154} Barnes, T.~G., III, Fernley, J.~A., Frueh, M.~L., Navas, J.~G., Moffett, T.~J., \& Skillen, I.\ 1997, \pasp, 109, 645 
\bibitem[Beavers \& Eitter(1986)]{ref159} Beavers, W.~I., \& Eitter, J.~J.\ 1986, \apjs, 62, 147 
\bibitem[Benedict et al.(2007)]{benedict07} Benedict, G.~F., et al.\ 2007, \aj, 133, 1810 
\bibitem[Berdnikov(1999)]{ref079} Berdnikov, L.~N., 1999, Catalog of Cepheid Photometry and Radial Velocities, version from 1999
\bibitem[Berdnikov \& Ivanov(1986)]{berdnikov86} Berdnikov, L.~N., \& Ivanov, G.~R.\ 1986, \apss, 125, 201 
\bibitem[Berry et al.(2011)]{berry11} Berry, M., Ivezi{\'c}, {\v Z}., Sesar, B., et al.\ 2011, arXiv:1111.4985 
\bibitem[Bersier et al.(1994)]{ref056} Bersier, D., Burki, G., Mayor, M., \& Duquennoy, A.\ 1994, \aaps, 108, 25 
\bibitem[Bird et al.(2009)]{bird09} Bird, J.~C., Stanek, K.~Z., \& Prieto, J.~L.\ 2009, \apj, 695, 874 
\bibitem[Bono et al.(1998)]{bono98} Bono, G., Caputo, F., \& Marconi, M.\ 1998, \apjl, 497, L43 
\bibitem[Bono et al.(2005)]{bono05} Bono, G., Marconi, M., Cassisi, S., Caputo, F., Gieren, W., \& Pietrzynski, G.\ 2005, \apj, 621, 966 
\bibitem[Bono et al.(2008)]{bono08} Bono, G., Caputo, F., Fiorentino, G., Marconi, M., \& Musella, I.\ 2008, \apj, 684, 102 
\bibitem[Bono et al.(2010)]{bono10} Bono, G., Caputo, F., Marconi, M., \& Musella, I.\ 2010, \apj, 715, 277 
\bibitem[Bresolin(2011)]{bresolin11} Bresolin, F.\ 2011, \apj, 729, 56 
\bibitem[Burki et al.(1982)]{burki82} Burki, G., Mayor, M., \& Benz, W.\ 1982, \aap, 109, 258 
\bibitem[Caldwell \& Coulson(1984)]{ref118} Caldwell, J.~A.~R., \& Coulson, I.~M.\ 1984, South African Astronomical Observatory Circular, 8, 1 
\bibitem[Caldwell et al.(1986)]{ref092} Caldwell, J.~A.~R., Coulson, I.~M., Jones, J.~H.~S., Black, C.~A., \& Feast, M.~W.\ 1986, \mnras, 220, 671 
\bibitem[Cardelli et al.(1989)]{cardelli89} Cardelli, J.~A., Clayton, G.~C., \& Mathis, J.~S.\ 1989, \apj, 345, 245 
\bibitem[Castelli \& Kurucz(2004)]{castelli04} Castelli, F., \& Kurucz, R.~L.\ 2004, arXiv:astro-ph/0405087 
\bibitem[Cioni et al.(2000)]{cioni00} Cioni, M.-R.~L., van der Marel, R.~P., Loup, C., \& Habing, H.~J.\ 2000, \aap, 359, 601 
\bibitem[Coulson \& Caldwell(1985)]{ref022} Coulson, I.~M., \& Caldwell, J.~A.~R.\ 1985, South African Astronomical Observatory Circular, 9, 5 
\bibitem[Coulson et al.(1985)]{ref019} Coulson, I.~M., Caldwell, J.~A.~R., \& Gieren, W.~P.\ 1985, \apjs, 57, 595 
\bibitem[Elias et al.(1982)]{elias82} Elias, J.~H., Frogel, J.~A., Matthews, K., \& Neugebauer, G.\ 1982, \aj, 87, 1029 
\bibitem[Falco et al.(1999)]{falco99} Falco, E.~E., Impey, C.~D., Kochanek, C.~S., et al.\ 1999, \apj, 523, 617 
\bibitem[Feast(1967)]{ref125} Feast, M.~W.\ 1967, \mnras, 136, 141
\bibitem[Feast \& Catchpole(1997)]{feast97} Feast, M.~W., \& Catchpole, R.~M.\ 1997, \mnras, 286, L1 
\bibitem[Fernie et al.(1995)]{ref065} Fernie, J.~D., Khoshnevissan, M.~H., \& Seager, S.\ 1995, \aj, 110, 1326 
\bibitem[Fiorentino et al.(2002)]{fiorentino02} Fiorentino, G., Caputo, F., Marconi, M., \& Musella, I.\ 2002, \apj, 576, 402 
\bibitem[Fiorentino et al.(2007)]{fiorentino07} Fiorentino, G., Marconi, M., Musella, I., \& Caputo, F.\ 2007, \aap, 476, 863 
\bibitem[Freedman \& Madore(1990)]{freedman90} Freedman, W.~L., \& Madore, B.~F.\ 1990, \apj, 365, 186 
\bibitem[Freedman \& Madore(2010a)]{freedman10a} Freedman, W.~L., \& Madore, B.~F.\ 2010a, \araa, 48, 673 
\bibitem[Freedman \& Madore(2010b)]{freedman10b} Freedman, W.~L., \& Madore, B.~F.\ 2010b, \apj, 719, 335 
\bibitem[Freedman \& Madore(2011)]{freedman11} Freedman, W.~L., \& Madore, B.~F.\ 2011, \apj, 734, 46 
\bibitem[Freedman et al.(1985)]{ref091} Freedman, W.~L., Grieve, G.~R., \& Madore, B.~F.\ 1985, \apjs, 59, 311 
\bibitem[Freedman et al.(2001)]{freedman01} Freedman, W.~L., et al.\ 2001, \apj, 553, 47 
\bibitem[Freedman et al.(2008)]{freedman08} Freedman, W.~L., Madore, B.~F., Rigby, J., Persson, S.~E., \& Sturch, L.\ 2008, \apj, 679, 71 
\bibitem[Freedman et al.(2008)]{ref176} Freedman, W.~L., Madore, B.~F., Rigby, J., Persson, S.~E., \& Sturch, L.\ 2008, \apj, 679, 71 
\bibitem[Fr{\'e}maux et al.(2006)]{fremaux06} Fr{\'e}maux, J., Kupka, F., Boisson, C., Joly, M., \& Tsymbal, V.\ 2006, \aap, 449, 109 
\bibitem[Gallenne et al.(2011)]{gallenne11} Gallenne, A., Kervella, P., \& M{\'e}rand, A.\ 2011, arXiv:1111.7215
\bibitem[Gerke et al.(2011)]{gerke11} Gerke, J.~R., Kochanek, C.~S., Prieto, J.~L., Stanek, K.~Z., \& Macri, L.~M.\ 2011, arXiv:1103.0549 
\bibitem[Gieren et al.(2005)]{gieren05} Gieren, W., Storm, J., Barnes, T.~G., III, et al.\ 2005, \apj, 627, 224 
\bibitem[Glass(1973)]{glass73} Glass, I.~S.\ 1973, \mnras, 164, 155 
\bibitem[Gordon et al.(2003)]{gordon03} Gordon, K.~D., Clayton, G.~C., Misselt, K.~A., Landolt, A.~U., \& Wolff, M.~J.\ 2003, \apj, 594, 279 
\bibitem[Gorynya et al.(1992)]{ref119a} Gorynya, N.~A., Irsmambetova, T.~R., Rastorgouev, A.~S., \& Samus, N.~N.\ 1992, Soviet Astronomy Letters, 18, 316 
\bibitem[Gorynya et al.(1996)]{ref119b} Gorynya, N.~A., Samus', N.~N., Rastorguev, A.~S., \& Sachkov, M.~E.\ 1996, Astronomy Letters, 22, 175 
\bibitem[Gorynya et al.(1998)]{ref119c} Gorynya, N.~A., Samus', N.~N., Sachkov, M.~E., Rastorguev, A.~S., Glushkova, E.~V., \& Antipin, S.~V.\ 1998, Astronomy Letters, 24, 815 
\bibitem[Gould(1994)]{gould94} Gould, A.\ 1994, \apj, 426, 542 
\bibitem[Grayzeck(1978)]{ref123} Grayzeck, E.~J.\ 1978, \aj, 83, 1390 
\bibitem[Groenewegen(2008)]{ref174} Groenewegen, M.~A.~T.\ 2008, \aap, 488, 25 
\bibitem[Harris(1980)]{ref009} Harris, H.~C.\ 1980, Ph.D.~Thesis,  University of Washington
\bibitem[Harris(1983)]{ref121} Harris, H.~C.\ 1983, \aj, 88, 507 
\bibitem[Hauschildt et al.(1999a)]{hauschildt99a} Hauschildt, P.~H., Allard, F., \& Baron, E.\ 1999, \apj, 512, 377 
\bibitem[Hauschildt et al.(1999b)]{hauschildt99b} Hauschildt, P.~H., Allard, F., Ferguson, J., Baron, E., \& Alexander, D.~R.\ 1999, \apj, 525, 871 
\bibitem[Hendry et al.(1999)]{hendry99} Hendry, M.~A., Tanvir, N.~R., \& Kanbur, S.~M.\ 1999, Harmonizing Cosmic Distance Scales in a Post-HIPPARCOS Era, 167, 192 
\bibitem[Hertzsprung(1926)]{hertzsprung26} Hertzsprung, E.\ 1926, \bain, 3, 115 
\bibitem[Hilditch et al.(2005)]{hilditch05} Hilditch, R.~W., Howarth, I.~D., \& Harries, T.~J.\ 2005, \mnras, 357, 304 
\bibitem[Imbert(1999)]{ref171} Imbert, M.\ 1999, \aaps, 140, 79 
\bibitem[Imbert et al.(1985)]{ref090} Imbert, M., et al.\ 1985, \aaps, 61, 259 
\bibitem[Imbert et al.(1989)]{ref097} Imbert, M., et al.\ 1989, \aaps, 81, 339
\bibitem[Indebetouw et al.(2005)]{indebetouw05} Indebetouw, R., Mathis, J.~S., Babler, B.~L., et al.\ 2005, \apj, 619, 931 
\bibitem[Joy(1937)]{ref129} Joy, A.~H.\ 1937, \apj, 86, 363 
\bibitem[Keller \& Wood(2006)]{keller06} Keller, S.~C., \& Wood, P.~R.\ 2006, \apj, 642, 834 
\bibitem[Kervella et al.(2006)]{kervella06} Kervella, P., M{\'e}rand, A., Perrin, G., \& Coud{\'e} du Foresto, V.\ 2006, \aap, 448, 623
\bibitem[Kiss(1998)]{ref161} Kiss, L.~L.\ 1998, \mnras, 297, 825 
\bibitem[Kiss \& Vink{\'o}(2000)]{ref173} Kiss, L.~L., \& Vink{\'o}, J.\ 2000, \mnras, 314, 420 
\bibitem[Klagyivik \& Szabados(2009)]{klagyivik09} Klagyivik, P., \& Szabados, L.\ 2009, \aap, 504, 959 
\bibitem[Kochanek(1997)]{kochanek97} Kochanek, C.~S.\ 1997, \apj, 491, 13 
\bibitem[Kurucz(1979)]{kurucz79} Kurucz, R.~L.\ 1979, \apjs, 40, 1 
\bibitem[Laney \& Caldwell(2007)]{laney07} Laney, C.~D., \& Caldwell, J.~A.~R.\ 2007, \mnras, 377, 147 
\bibitem[Laney \& Stobie(1986)]{ref093} Laney, C.~D., \& Stobie, R.~S.\ 1986, South African Astronomical Observatory Circular, 10, 51 
\bibitem[Laney \& Stobie(1992)]{ref036} Laney, C.~D., \& Stobie, R.~S.\ 1992, \aaps, 93, 93 
\bibitem[Laney \& Stobie(1993)]{laney93} Laney, C.~D., \& Stobie, R.~S.\ 1993, \mnras, 263, 921 
\bibitem[Lloyd Evans(1968)]{ref128} Lloyd Evans, T.\ 1968, \mnras, 141, 109 
\bibitem[Lloyd Evans(1980)]{ref010} Lloyd Evans, T.\ 1980, South African Astronomical Observatory Circular, 1, 257 
\bibitem[Macri et al.(2006)]{macri06} Macri, L.~M., Stanek, K.~Z., Bersier, D., Greenhill, L.~J., \& Reid, M.~J.\ 2006, \apj, 652, 1133 
\bibitem[Madore(1975)]{ref002} Madore, B.~F.\ 1975, \apjs, 29, 219 
\bibitem[Madore(1982)]{madore82} Madore, B.~F.\ 1982, \apj, 253, 575 
\bibitem[Madore et al.(2009)]{madore09} Madore, B.~F., Freedman, W.~L., Rigby, J., et al.\ 2009, \apj, 695, 988 
\bibitem[Madore \& Freedman(1991)]{madore91} Madore, B.~F., \& Freedman, W.~L.\ 1991, \pasp, 103, 933 
\bibitem[Madore \& Freedman(2009)]{madore_freedman09} Madore, B.~F., \& Freedman, W.~L.\ 2009, \apj, 696, 1498 
\bibitem[Madore \& Freedman(2011)]{madore11} Madore, B.~F., \& Freedman, W.~L.\ 2011, \apj, 744, 132
\bibitem[Marconi et al.(2005)]{marconi05} Marconi, M., Musella, I., \& Fiorentino, G.\ 2005, \apj, 632, 590 
\bibitem[Marengo et al.(2010)]{ref177} Marengo, M., Evans, N.~R., Barmby, P., Bono, G., Welch, D.~L., \& Romaniello, M.\ 2010, \apj, 709, 120 
\bibitem[Marshall et al.(2006)]{marshall06} Marshall, D.~J., Robin, A.~C., Reyl{\'e}, C., Schultheis, M., \& Picaud, S.\ 2006, \aap, 453, 635 
\bibitem[Martin(1980)]{ref088} Martin, W.~L.\ 1980, South African Astronomical Observatory Circular, 1, 172 
\bibitem[Martin \& Warren(1979)]{ref086} Martin, W.~L., \& Warren, P.~R.\ 1979, South African Astronomical Observatory Circular, 1, 98 
\bibitem[Mathewson et al.(1988)]{ref149} Mathewson, D.~S., Ford, V.~L., \& Visvanathan, N.\ 1988, \apj, 333, 617
\bibitem[McGonegal et al.(1982)]{mcgonegal82} McGonegal, R., McAlary, C.~W., Madore, B.~F., \& McLaren, R.~A.\ 1982, \apjl, 257, L33 
\bibitem[Metzger et al.(1991)]{ref034} Metzger, M.~R., Caldwell, J.~A.~R., McCarthy, J.~K., \& Schechter, P.~J.\ 1991, \apjs, 76, 803 
\bibitem[Metzger et al.(1992)]{ref042} Metzger, M.~R., Caldwell, J.~A.~R., \& Schechter, P.~L.\ 1992, \aj, 103, 529 
\bibitem[Misselt et al.(1999)]{misselt99} Misselt, K.~A., Clayton, G.~C., \& Gordon, K.~D.\ 1999, \apj, 515, 128 
\bibitem[Mochejska et al.(2000)]{mochejska00} Mochejska, B.~J., Macri, L.~M., Sasselov, D.~D., \& Stanek, K.~Z.\ 2000, \aj, 120, 810 
\bibitem[Moffett \& Barnes(1984)]{ref018} Moffett, T.~J., \& Barnes, T.~G., III 1984, \apjs, 55, 389 
\bibitem[Moffett et al.(1998)]{ref145} Moffett, T.~J., Gieren, W.~P., Barnes, T.~G., III, \& Gomez, M.\ 1998, \apjs, 117, 135 
\bibitem[Motta et al.(2002)]{motta02} Motta, V., Mediavilla, E., Mu{\~n}oz, J.~A., et al.\ 2002, \apj, 574, 719 
\bibitem[Nardetto et al.(2004)]{nardetto04} Nardetto, N., Fokin, A., Mourard, D., Mathias, P., Kervella, P., \& Bersier, D.\ 2004, \aap, 428, 131 
\bibitem[Nardetto et al.(2009)]{ref169} Nardetto, N., Gieren, W., Kervella, P., Fouqu{\'e}, P., Storm, J., Pietrzynski, G., Mourard, D., \& Queloz, D.\ 2009, \aap, 502, 951 
\bibitem[Ngeow et al.(2003)]{ngeow03} Ngeow, C.-C., Kanbur, S.~M., Nikolaev, S., Tanvir, N.~R., \& Hendry, M.~A.\ 2003, \apj, 586, 959 
\bibitem[Ngeow et al.(2010)]{ngeow10} Ngeow, C.-C., Ita, Y., Kanbur, S.~M., et al.\ 2010, \mnras, 408, 983 
\bibitem[Ngeow \& Kanbur(2005)]{ngeow05} Ngeow, C.-C., \& Kanbur, S.~M.\ 2005, \mnras, 360, 1033 
\bibitem[Ngeow \& Kanbur(2010)]{ref178} Ngeow, C.-C., \& Kanbur, S.~M.\ 2010, \apj, 720, 626 
\bibitem[Ngeow et al.(2011)]{ngeow11} Ngeow, C.-C., Marconi, M., Musella, I., Cignoni, M., \& Kanbur, S.~M.\ 2011, arXiv:1111.1791 
\bibitem[Nikolaev et al.(2004)]{nikolaev04} Nikolaev, S., Drake, A.~J., Keller, S.~C., et al.\ 2004, \apj, 601, 260 
\bibitem[Nishiyama et al.(2009)]{nishiyama09} Nishiyama, S., Tamura, M., Hatano, H., et al.\ 2009, \apj, 696, 1407
\bibitem[Oosterhoff(1960)]{ref162} Oosterhoff, P.~T.\ 1960, \bain, 15, 199 
\bibitem[Paczy{\'n}ski \& Pindor(2000)]{paczynski00} Paczy{\'n}ski, B., \& Pindor, B.\ 2000, \apjl, 533, L103 
\bibitem[Pel(1976)]{ref004} Pel, J.~W.\ 1976, \aaps, 24, 413 
\bibitem[Perryman \& ESA(1997)]{ref163} Perryman, M.~A.~C., \& ESA 1997, ESA Special Publication, 1200  
\bibitem[Persson et al.(2004)]{ref167} Persson, S.~E., Madore, B.~F., Krzemi{\'n}ski, W., Freedman, W.~L., Roth, M., \& Murphy, D.~C.\ 2004, \aj, 128, 2239 
\bibitem[Pont et al.(1994)]{ref055} Pont, F., Burki, G., \& Mayor, M.\ 1994, \aaps, 105, 165 
\bibitem[Pont et al.(1997)]{ref156} Pont, F., Queloz, D., Bratschi, P., \& Mayor, M.\ 1997, \aap, 318, 416 
\bibitem[Press et al.(1992)]{press92} Press, W.~H., Teukolsky, S.~A., Vetterling, W.~T., \& Flannery, B.~P.\ 1992, Cambridge: University Press, 2nd ed.
\bibitem[Riess et al.(2009a)]{riess09a} Riess, A.~G., Macri, L., Casertano, S., et al.\ 2009a, \apj, 699, 539 
\bibitem[Riess et al.(2009b)]{riess09b} Riess, A.~G., Macri, L., Li, W., et al.\ 2009b, \apjs, 183, 109 
\bibitem[Riess et al.(2011a)]{riess11a} Riess, A.~G., Macri, L., Casertano, S., et al.\ 2011, \apj, 730, 119 
\bibitem[Riess et al.(2011b)]{riess11b} Riess, A.~G., Fliri, J., \& Valls-Gabaud, D.\ 2011, arXiv:1110.3769 
\bibitem[Romaniello et al.(2008)]{romaniello08} Romaniello, M., Primas, F., Mottini, M., et al.\ 2008, \aap, 488, 731 
\bibitem[Rom{\'a}n-Z{\'u}{\~n}iga et al.(2007)]{roman07} Rom{\'a}n-Z{\'u}{\~n}iga, C.~G., Lada, C.~J., Muench, A., \& Alves, J.~F.\ 2007, \apj, 664, 357 
\bibitem[Sakai et al.(2004)]{sakai04} Sakai, S., Ferrarese, L., Kennicutt, R.~C., Jr., \& Saha, A.\ 2004, \apj, 608, 42 
\bibitem[Sandage \& Tammann(1971)]{sandage71} Sandage, A., \& Tammann, G.~A.\ 1971, \apj, 167, 293 
\bibitem[Sandage et al.(2004)]{sandage04} Sandage, A., Tammann, G.~A., \& Reindl, B.\ 2004, \aap, 424, 43 
\bibitem[Sasselov et al.(1997)]{sasselov97} Sasselov, D.~D., Beaulieu, J.~P., Renault, C., et al.\ 1997, \aap, 324, 471 
\bibitem[Schechter et al.(1992)]{ref043} Schechter, P.~L., Avruch, I.~M., Caldwell, J.~A.~R., \& Keane, M.~J.\ 1992, \aj, 104, 1930 
\bibitem[Schmidt et al.(1995)]{ref064} Schmidt, E.~G., Chab, J.~R., \& Reiswig, D.~E.\ 1995, \aj, 109, 1239 
\bibitem[Scowcroft et al.(2011)]{scowcroft11} Scowcroft, V., Freedman, W., Madore, B.~F., et al.\ 2011, arXiv:1108.4672 
\bibitem[Sebo \& Wood(1995)]{ref102} Sebo, K.~M., \& Wood, P.~R.\ 1995, \apj, 449, 164
\bibitem[Shappee \& Stanek(2011)]{shappee11} Shappee, B.~J., \& Stanek, K.~Z.\ 2011, \apj, 733, 124 
\bibitem[Simon \& Lee(1981)]{simonlee81} Simon, N.~R., \& Lee, A.~S.\ 1981, \apj, 248, 291 
\bibitem[Soszy{\'n}ski et al.(2008)]{ref166} Soszy{\'n}ski, I., et al.\ 2008, \actaa, 58, 163 
\bibitem[Soszy{\'n}ski et al.(2010)]{ref175} Soszy{\'n}ski, I., et al.\ 2010, \actaa, 60, 17 
\bibitem[Stetson(1996)]{stetson96} Stetson, P.~B.\ 1996, \pasp, 108, 851 
\bibitem[Stibbs(1955)]{ref126} Stibbs, D.~W.~N.\ 1955, \mnras, 115, 363
\bibitem[Storm et al.(2004)]{ref172} Storm, J., Carney, B.~W., Gieren, W.~P., Fouqu{\'e}, P., Latham, D.~W., \& Fry, A.~M.\ 2004, \aap, 415, 531 
\bibitem[Storm et al.(2011a)]{storm11a}  Storm, J., Gieren, W., Fouqu{\'e}, P., et al.\ 2011, \aap, 534, A94
\bibitem[Storm et al.(2011b)]{storm11b} Storm, J., Gieren, W., Fouqu{\'e}, P., et al.\ 2011, \aap, 534, A95 
\bibitem[Stothers(1988)]{stothers88} Stothers, R.~B.\ 1988, \apj, 329, 712 
\bibitem[Szabados(1981)]{ref014} Szabados, L.\ 1981, Commmunications of the Konkoly Observatory Hungary, 77, 1
\bibitem[Szabados \& Pont(1998)]{ref157} Szabados, L., \& Pont, F.\ 1998, \aaps, 133, 51 
\bibitem[Szabados \& Klagyivik(2011)]{szabados11} Szabados, L., \& Klagyivik, P.\ 2011, arXiv:1112.0115 
\bibitem[Szab{\'o} et al.(2007)]{szabo07} Szab{\'o}, R., Buchler, J.~R., \& Bartee, J.\ 2007, \apj, 667, 1150 
\bibitem[Tanvir et al.(2005)]{tanvir05} Tanvir, N.~R., Hendry, M.~A., Watkins, A., et al.\ 2005, \mnras, 363, 749 
\bibitem[Udalski et al.(1999)]{udalski99} Udalski, A., Szymanski, M., Kubiak, M., Pietrzynski, G., Soszynski, I., Wozniak, P., \& Zebrun, K.\ 1999, \actaa, 49, 201 
\bibitem[Valle et al.(2009)]{valle09} Valle, G., Marconi, M., Degl'Innocenti, S., \& Prada Moroni, P.~G.\ 2009, \aap, 507, 1541 
\bibitem[van Genderen(1978)]{vangenderen78} van Genderen, A.~M.\ 1978, \aap, 65, 147 
\bibitem[van Genderen(1983)]{ref089} van Genderen, A.~M.\ 1983, \aaps, 52, 423 
\bibitem[Wallerstein \& Helfer(1966)]{wallerstein66} Wallerstein, G., \& Helfer, H.~L.\ 1966, \aj, 71, 350 
\bibitem[Walraven et al.(1964)]{ref001} Walraven, J.~H., Tinbergen, J., \& Walraven, T.\ 1964, \bain, 17, 520
\bibitem[Walraven \& Walraven(1960)]{walraven60} Walraven, T., \& Walraven, J.~H.\ 1960, \bain, 15, 67 
\bibitem[Wang(2008)]{wang08} Wang, Y.\ 2008, \prd, 77, 123525 
\bibitem[Welch et al.(1984)]{ref017} Welch, D.~L., Wieland, F., McAlary, C.~W., McGonegal, R., Madore, B.~F., McLaren, R.~A., \& Neugebauer, G.\ 1984, \apjs, 54, 547 
\bibitem[Welch et al.(1987)]{ref095} Welch, D.~L., McLaren, R.~A., Madore, B.~F., \& McAlary, C.~W.\ 1987, \apj, 321, 162 
\bibitem[Welch et al.(1993)]{ref100} Welch, D.~L., Mateo, M., Olszewski, E.~W., Fischer, P., \& Takamiya, M.\ 1993, \aj, 105, 146 
\bibitem[Wesselink(1946)]{wesselink46} Wesselink, A.~J.\ 1946, \bain, 10, 91 
\bibitem[Wilson et al.(1989)]{ref032} Wilson, T.~D., Carter, M.~W., Barnes, T.~G., III, van Citters, G.~W., Jr., \& Moffett, T.~J.\ 1989, \apjs, 69, 951 
\bibitem[Wozniak \& Stanek(1996)]{wozniak96} Wozniak, P.~R., \& Stanek, K.~Z.\ 1996, \apj, 464, 233 
\bibitem[Yoachim et al.(2009)]{yoachim09} Yoachim, P., McCommas, L.~P., Dalcanton, J.~J., \& Williams, B.~F.\ 2009, \aj, 137, 4697 
\end{thebibliography}
\end{document}